\documentclass[twocolumn]{aastex61}  
\submitjournal{ApJS}
\usepackage{graphicx}
\usepackage{color}
\usepackage{amsmath}
\usepackage{array}
\usepackage{txfonts}
\begin{document}

   \title{Stellar flares observed in long cadence data from the Kepler mission}

   \correspondingauthor{Tom Van Doorsselaere}
   \email{tom.vandoorsselaere@kuleuven.be}
   \author[0000-0001-9628-4113]{Tom Van Doorsselaere}
   \affil{Centre for mathematical Plasma Asytrophysics, 
   Mathematics Department, KU~Leuven,
   Celestijnenlaan 200B bus 2400, 3001 Leuven, Belgium}
   \author{Hoda Shariati}
   \affil{Instituut voor Sterrenkunde, KU Leuven, Celestijnenlaan 200D, 3001 Leuven, Belgium}
   \author{Jonas Debosscher}
   \affil{Instituut voor Sterrenkunde, KU Leuven, Celestijnenlaan 200D, 3001 Leuven, Belgium}
   \affil{Royal Belgian Institute for Space Aeronomy, Ringlaan-3-Avenue Circulaire, B-1180 Brussels, Belgium}
   \date{Received xx; accepted xx}

  \begin{abstract}
   {We aim to perform a statistical study of stellar flares observed by Kepler. We want to study the flare amplitude, duration, energy and occurrence rates, and how they are related to the spectral type and rotation period.}
   {To that end, we have developed an automated flare detection and characterisation algorithm. We have harvested the stellar parameters from the Kepler input catalogue and the rotation periods from \citet{mcquillan2014}.}
   {We find several new candidate A stars showing flaring activity. Moreover, we find 653 giants with flares. From the statistical distribution of flare properties, we find that the flare amplitude distribution has a similar behaviour between F+G-types and K+M-types. The flare duration and flare energy seem to be grouped between G+K+M-types vs. F-types and giants. We also detect a tail of stars with high flare occurrence rates across all spectral types (but most prominent in the late spectral types), and this is compatible with the existence of ``flare stars''. Finally, we have found a strong correlation of the flare occurrence rate and the flare amplitude with the stellar rotation period: a quickly rotating star is more likely to flare often, and has a higher chance to generate large flares.}
  \end{abstract}

%

\section{Introduction}
It is well known that not only the Sun is showing magnetic activity in the form of flares \citep[e.g.][]{kowalski2010,hawley2014}. On the Sun, it is believed that solar flares are caused by magnetic reconnection \citep[e.g.][]{sun2015}. Thus, it is a straightforward assumption that stellar flares are also caused by magnetic reconnection of coronal structures. Hence, the presence of flares can be used as a proxy for the presence of a stellar corona.\par
Before space photometry, stellar flare studies were focused on active flare stars, in order to increase the chance of capturing a flare during the limited telescope observing time. Some early statistical studies exist \citep{shakhovskaia1989,kowalski2013}, but they are rare. However, with the Kepler mission, it became possible to systematically study stellar flares. The first to use the Kepler data for this purpose were \citet{walkowicz2011}, and there they limited the sample to cool dwarfs. To study superflares, \citet{maehara2012, shibayama2013, candelaresi2014, maehara2015} selected a sample of G-type stars or later, to obtain power laws for flare amplitudes per star. Moreover, several case studies of flare stars using Kepler data have been published \citep[e.g.][]{ramsay2013,lurie2015}, and those studies are continued with K2 as well \citep{ramsay2014}.
\par
\citet{balona2012,balona2013} used visual inspection of the light curves to identify flares in a much broader (in terms of spectral type) sample of Kepler observations. They found that even some A stars show flaring activity. This is unexpected from stellar evolution theory, because these stars are not believed to have an outer convection zone. Due to the dynamo effect, the latter is considered to be a crucial ingredient for the star to show magnetic activity, such as flares. Thus, it is plausible to ascribe the flares on A stars to cool companions. However, it was argued by \citet{balona2012,balona2015b} that this could not be the case, because then the flare amplitude would be unusually large for these cool companions. Still, \citet{pedersen2017} studied the list of A stars of \citet{balona2013} in great detail. They found that several cases could be explained by contaminated pixel data or found that several of the flaring A stars were in a binary (implying that the flare is originating from the companion). Yet, not all flaring A stars could be excluded, and thus it remains inconclusive if A stars can flare or not.\par
More recently, \citet{pitkin2014,davenport2016} developed automated algorithms to process the large Kepler database, and study flares in a statistical sense. Still, several aspects are missing from those studies. For example, it was previously shown that X-ray luminosity (presumably from the stellar corona) scales with the rotation period of the star \citep{wright2011}, but this aspect was not studied in the previous statistical studies on stellar flares. \par
With the high-quality data from Kepler, even seismology of stellar flares can be attempted using quasi-periodic pulsations in stellar flares \citep[for a review, see][]{vd2016}. Observations of stellar flare oscillations were reported by \citet{mathioudakis2003,mitra-kraev2005,welsh2006,anfinogentov2013,srivastava2013,balona2015}, with even multi-periodic events \citep{pugh2015} being detected now. Given the similarity with quasi-periodic pulsations in solar flares \citep{cho2016}, a lot of potential exists for remote sensing of stellar coronae and their magnetic fields. \par
In this paper, we develop a new automated detection and characterisation method for flares in the Kepler mission data. We perform a statistical study of the stellar flares. In particular, we study the dependence of the flare occurrence rate, flare duration and flare amplitude on stellar spectral type and stellar rotation periods. 

\section{Detection algorithm}
\label{sec:methods}
\subsection{Preprocessing}
We use the raw light curves from the Kepler mission during quarter 15 (Q15 for short). We focus on the long cadence, which has a time cadence of approximately $\Delta t=29\mathrm{min}$. Each time series consists of different segments between data gaps and intensity discontinuities, which are instrumental in nature. For each of these segments, we fit a 3rd order polynomial to remove the instrumental effects. We have removed the detected flare candidates near these discontinuities, to ensure that our flare sample is not influenced by these instrumental effects.\par

After this detrending, we have prewhitened the time series, following the procedure outlined in \citet{degroote2009}. We have computed the Lomb-Scargle periodogram of the data to determine the most significant frequency peak in the interval $f=[0.1\mbox{d}^{-1}, 24.5\mbox{d}^{-1}]$. If this peak is statistically significant (with $\mbox{S/N}>4$ using the white noise approximation), then a sine with this frequency is fitted to the data, and then removed. This procedure is repeated until no significant peaks are found, or at most 100 frequencies have been removed from the light curve. This prewhitening procedure removes most of the regular periodic effects, such as intensity variations caused by stellar pulsations or stellar spots. However, the prewhitening does not do so well for eclipsing binaries, and thus any flare events on known eclipsing binaries are removed from the database (using the information at \url{http://keplerebs.villanova.edu}). \par
\begin{figure}
	\includegraphics[width=\linewidth]{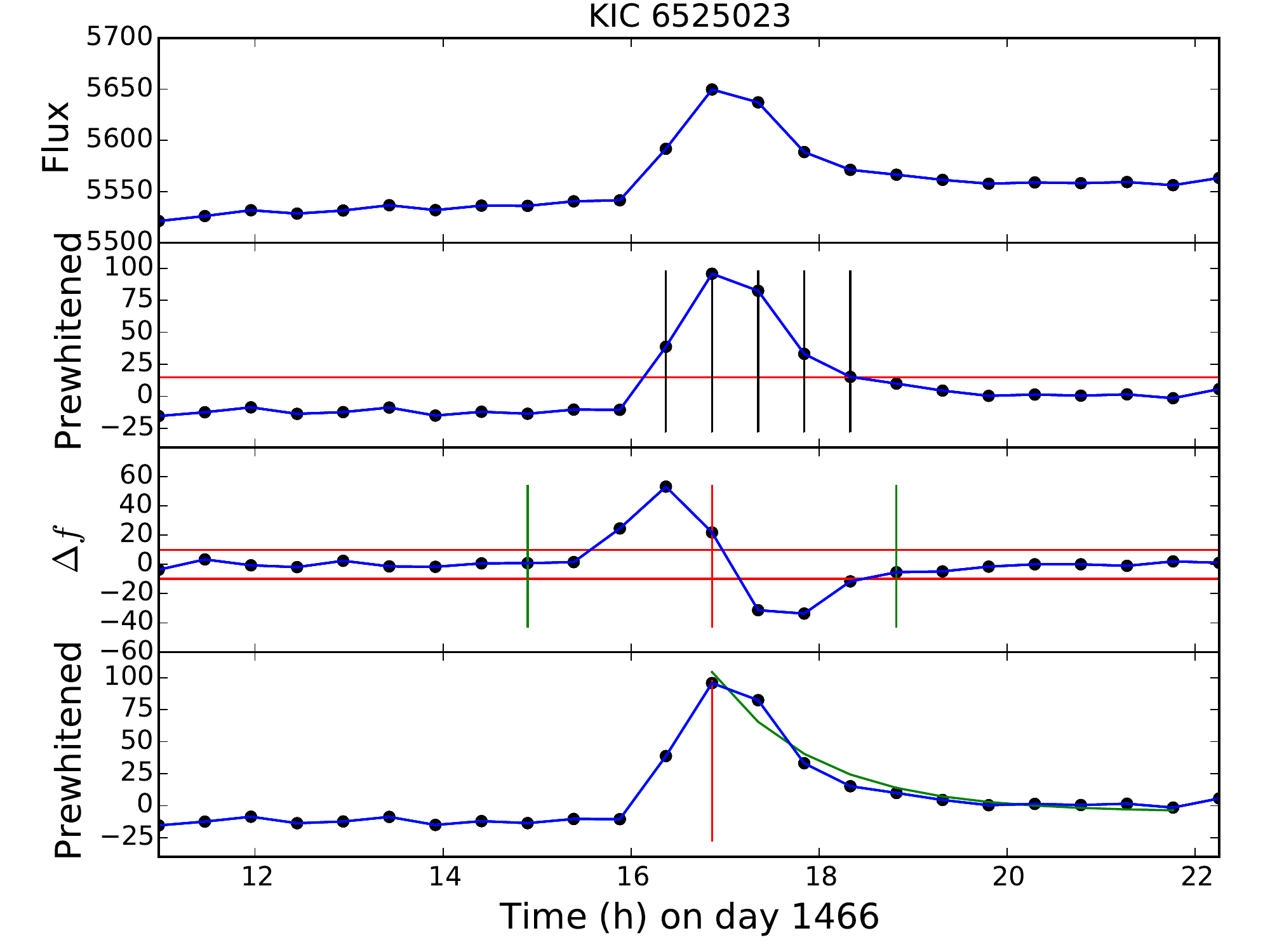}
	\caption{Illustration of the detection method. The top panel shows the raw data from the Kepler database (black dots) for KIC 6525023, as a function of time on day 1466 of the Kepler mission. The second panel shows the detrended light curve $f(t)$ (black dots) with the employed threshold (red line). The points above the threshold are highlighted with vertical black lines. The third panel shows the central difference $\Delta f$ of the intensity signal (black dots), the thresholds (horizontal red lines), and the flare window (vertical green lines). The bottom panel shows the detrended light curve (black dots), with the flare peak indicated (vertical red line) and the fitted exponential decay (green line).}
	\label{fig:detection}
\end{figure}

\subsection{Thresholding}
The detrended and prewhitened light curve is then subjected to a double threshold method. We have first computed the standard deviation of the central difference of the time series $\Delta f$ ($(f_{i+1}-f_{i-1})/2$ at time $t_i$, which estimates its slope). After making an initial estimate of the standard deviation $\sigma_{\Delta f}$, we have made this estimate more robust by removing once all outliers above $6\sigma_{\Delta f}$ from the central difference time series. \\
Then we check if the flare intensity is sufficiently higher than the mean intensity of the star: the threshold for detection of a flare is set at $4.5\sigma_{\Delta f}$, which statistically corresponds to $4.5 \sigma_f/\sqrt{2}$, where $\sigma_f$ is the standard deviation of the noise of the detrended and prewhitened flux $f(t)$. \\
The second threshold is in the central difference of the flux $\Delta f(t)$, in order to check that the increase in intensity was sufficiently rapid, following the FRED profile (Fast Rise Exponential Decay). The running difference threshold is taken as $3\sigma_{\Delta f}$. Additionally, we check if an interval of 4 data points on either side of the flare maximum at time $t_0$ contains a maximum in the flare slope left of the peak ($[t_0-4\Delta t, t_0]$), and a minimum in the slope right of the peak ($[t_0,t_0+4\Delta t]$). This last criterion ensures that the flux changes are rapid enough. \\
\subsection{Parametrization}
After the detection, we perform further filtration of the detected flare candidates. We want to study decay times of flares, and therefore we fit the detected flares with an exponentially decaying function, following \citet{anfinogentov2013,pugh2016}:
\begin{equation}g(t-t_0)=a\exp{(-k(t-t_0))}+b.\label{eq:fit}\end{equation}
To do this, we introduce three more constraints on the flare detection. First, we determine the length of the time series to be fitted. The start of the time series is the time of the peak of the flare $t_0$. The time $t_\mathrm{e}$ is found as the time  for which the gradient $\Delta f$ is closest to zero, and has to be within the 5 points right of the last point where the flux $f$ exceeds the threshold. For the fitting, 4 more data points are added to the time series, and the fit is thus performed between $[t_0,t_e+4\Delta t]$. \\
Then, there are three possible reasons to reject the flare: (1) there are multiple local maxima in flux during the flare between times $[t_0, t_\mathrm{e}]$ and a FRED profile is thus not a good match, or (2) the fit with the exponential function $g$ fails, or (3) the flare has a negative amplitude $a\leq 0$ or the amplitude is smaller than the background $a\leq b$.\\
It is worth noting that there is no restriction on the duration of the flare $1/k$, and thus also short duration flares ($1/k<30\mbox{min}$) are retained in the results. However, we have checked the results in Sec.~\ref{sec:results} by excluding those short flares. We find that all incidence rates are decreased with 0.50\%, except for the giants (where the incidence rate remains nearly constant). This tells us that the short duration flares are uniformly distributed over all spectral types, except for the giants. Moreover, the distributions of flare amplitudes and energies (see Sec.~\ref{sec:amplitude}) are not modified by the exclusion of short duration flares. 

\section{Results}
\label{sec:results}
\subsection{General interpretation} 
\label{sec:general}
\begin{figure}
	\includegraphics[width=\linewidth]{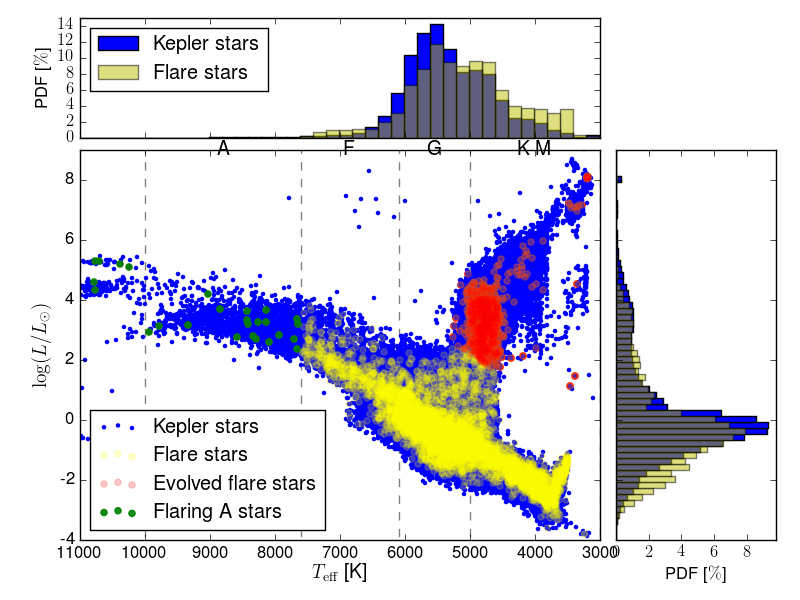}
	\caption{The HR diagram with the detected flare stars indicated. The central panel has stellar temperature as horizontal axis, while the vertical axis shows the luminosity. Kepler stars are shown with blue dots. Yellow, green and red dots indicate stars with detected flares, yellow dots are close to the main sequence, green dots are A stars with flares, and red dots are giants. The top and side panel show the histogram of the Kepler stars (in blue), and flare stars (in yellow).}
	\label{fig:hr}
\end{figure}
We have constructed a Hertzsprung-Russell diagram in Fig.~\ref{fig:hr}. To this end, we have used the information in the Kepler Input Catalogue (KIC, v10 downloaded from \url{https://archive.stsci.edu/kepler/kic10/search.php}), where we have extracted the temperature $T_\mathrm{eff}$ and used the stellar radius $R/R_\odot$ to estimate the luminosity $L$ (in units of the solar luminosity $L_\odot$) by 
\begin{equation}
	L=\log{\left\lbrace\left(\frac{R}{R_\odot}\right)^2 \left(\frac{T_\mathrm{eff}}{5780\mathrm{K}}\right)^4\right\rbrace}.
\end{equation}
To estimate the spectral type, we have taken an interval of [3000K, 5000K] as K and M stellar types, [5000K, 6100K] as G stars, [6100K, 7600K] as F stars, and [7600K, 10000K] as A stars. These temperature intervals are shown by vertical dashed lines in Fig.~\ref{fig:hr}.\\ 
The stars and detected flares are listed in the additional material to this paper (\texttt{vandoorsselaere\_et\_al\_2017\_q15\_flares\_sorted.txt}). This file contains a line for each detected flare candidate with the Kepler ID of the object, flare start time, the effective temperature $T_\mathrm{eff}$, stellar radius $R/R_\odot$ and $\log{g}$. \par
In Fig.~\ref{fig:hr}, the Kepler stars are shown with blue dots, while the flare stars are shown with red, green or yellow dots. The top and right panel show the normalised histograms of the distribution of Kepler and flare stars over temperature and luminosity. It is clear that the flare distribution (in yellow) is more concentrated towards the lower end of luminosity and temperature. This confirms the finding of \citet{davenport2016} that the flare star occurrence rate increases towards later spectral types. \par
In total, we have detected 16850 flares on 6662 stars out of a total of 188837 in the Kepler field of view during Q15. This translates in 3.5\% of the stars being flare stars. This number is higher than the percentages mentioned in \citet{walkowicz2011,balona2012, davenport2016}. Especially the latter work is very relevant, because they also use an automated detection method as we do. However, \citet{davenport2016} have removed all flare star candidates with less than 25 flares during the entire Kepler observations. We have kept all flare stars, even if they only show one flare, because we want to avoid a selection effect in the sample.  \par
\begin{table}
	\caption{Flare star incidence over the HR diagram. The first column indicates the stellar type, the second column indicates the number of objects in this type, the third column shows how many have at least one flare, and the fourth column is the ratio of column 3 by column 2 to get the incidence of flare stars for the given type. }
	\label{tab:overview}
	\begin{tabular}{llll}
		\hline
		Stellar type & \# objects & \# flare stars & Incidence\\
		\hline \hline
		A+B & 2141 & 28 & 1.31\%\\
		F & 22107 & 708 & 3.20\%\\
		G & 116178 & 3365 & 2.90\%\\
		K+M & 48411 & 2556 & 5.28\%\\
		giants & 22837 & 653 &2.86\% \\
		\hline
	\end{tabular}
\end{table}
In Table~\ref{tab:overview}, we give an overview of the flare star incidence for each spectral type. A star can be counted both as a giant and as belonging to its spectral type. These incidence rates are relatively close to the values found in \citet{davenport2016}. Also in this table, it is confirmed that the flare star incidence is twice as high for K and M stellar types, compared to F and G stellar types. Our flare star incidences seem to be different from the statistics by \citet{balona2015b}: he found a much higher incidence rate of 12\% for K and M-type stars, and a reduced incidence rate of 1.2\% for F-type stars, but does find a similar incidence rate for the G-type stars. The differences between our study and \citet{balona2015b} could be caused by the much smaller sample size of the latter work.

\subsection{False detections \& Detection errors}
\label{sec:errors}
\begin{figure}
	\includegraphics[width=1.1\linewidth]{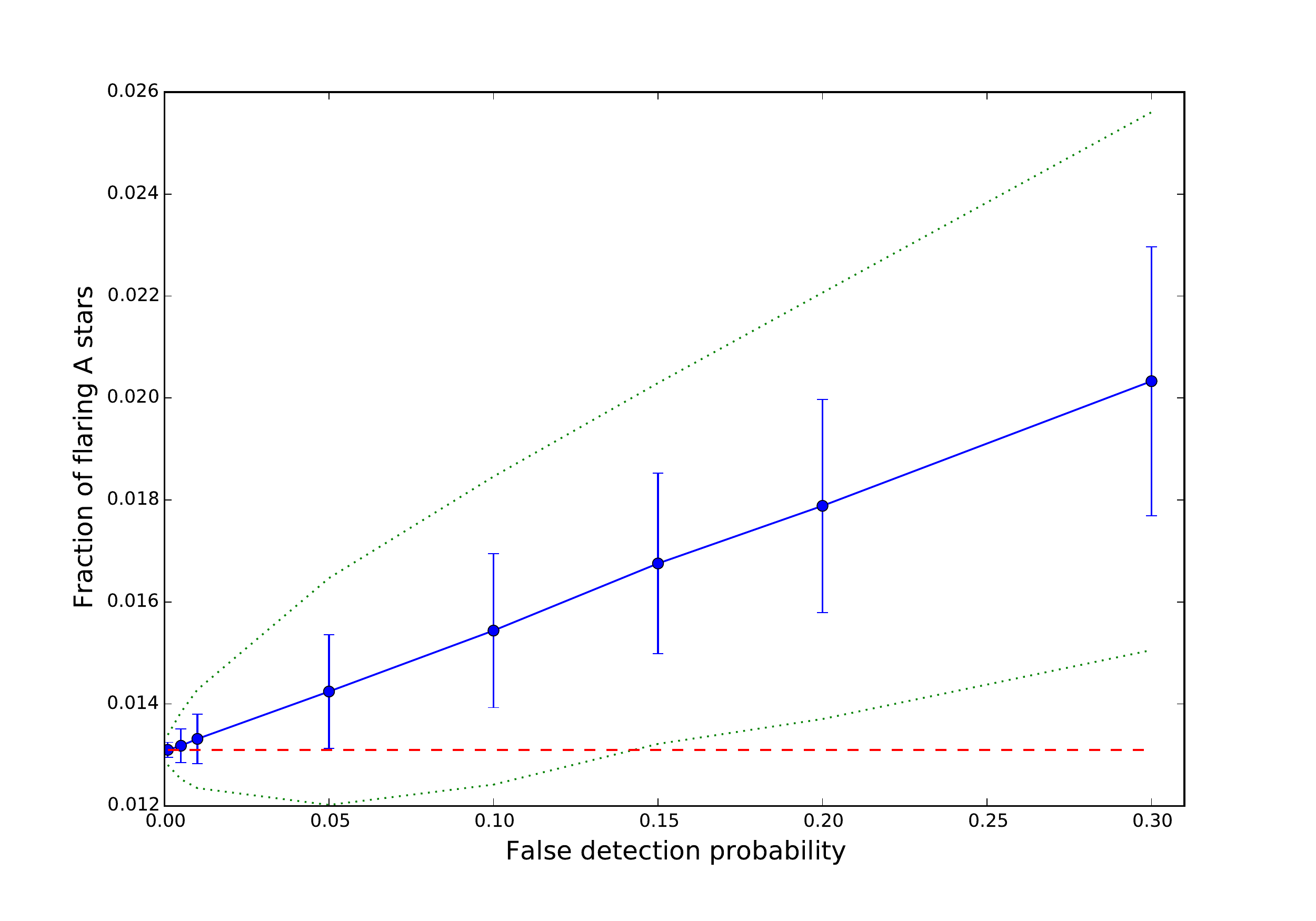}
	\caption{The fraction of A stars that show flaring as a function of the false detection probability. The blue dots show the mean fraction of the Monte Carlo results, while the blue bars show the 1-$\sigma$ errors. The green dotted lines show the 2-$\sigma$ uncertainty strip. The red dashed line shows the value with a false detection probability of 0 (i.e. the algorithm is perfect), and is also given in table~\ref{tab:overview}.}
	\label{fig:errors}
\end{figure}
It is important to quantify the number of false detections and detection errors of our algorithm, because this will likely influence the statistics of the stellar flare detections. The reasons for false detections could be twofold: (1) the algorithm may not function as expected, or (2) there are signals coming from nearby stars which fall in the Kepler pixel mask. Note that the first was circumvented by \citet{davenport2016} by excluding stars with fewer than 25 flares, while the latter was avoided by \citet{maehara2012} by eliminating stars which have another star within 12 arcsec of the target. \par
To assess the false detection or detection errors, we have performed Monte Carlo simulations. To do this, we have first fixed a value for the possible false detection probability $p_\mathrm{false}$ (with values between 0.1\% to 30\%, see horizontal axis of Fig.~\ref{fig:errors}). Then, we have randomly chosen $p_\mathrm{false}N_\mathrm{flare\ star}$ out of the fixed number of detected flare stars $N_\mathrm{flare\ star}$. For each of these randomly chosen flare stars, we have replaced their characterisation from the Kepler Input Catalogue by a random value from another star in the field. This is then mimicking the fact that the flare could occur on a background star (within the pixel mask), for which the spectral type, temperature and radius distribution is the same as the distribution of Kepler (non-flaring) stars.  Then, we have repeated this process 1000-3000 times for a different random selection of stars (using the Monte Carlo spirit). \par
Fig.~\ref{fig:errors} then shows the mean (with blue dots) of the obtained distribution for the fraction of flaring A stars, and the 1-$\sigma$ error bars (given by the square root of the variance of the distribution). The green dotted lines show the 2-$\sigma$ interval around each mean. It is clear that the mean is increasing for a larger false detection probability $p_\mathrm{false}$, and this is expected, because the mean should evolve to the global mean (over all spectral types). In fact, when $p_\mathrm{false}=1$, it should almost go to the global average of 3.5\% of flare stars in the population of stars. 
The red dashed line (in Fig.~\ref{fig:errors}) shows the fraction of the flaring A stars which was found for the total population (see Table~\ref{tab:overview}). Since it makes physical sense that the fraction is smaller than for the average field stars (comprised mostly of G types), it is probably a good assumption that the obtained value of 1.31\% must fall within the 2-$\sigma$ uncertainty strip of the Monte Carlo simulation. This allows us to estimate our false detection probability of being at most 15\%. \par

\subsection{Active A-stars}
\label{sec:astars}
As described in the introduction, flares were surprisingly found on A stars by \citet{balona2012}. In Fig.~\ref{fig:hr}, it is clear that some flares were found in stars with temperature hotter than 7600K, which we classify as A stars (indicated with green dots). The stellar objects and their effective temperature are listed in Table~\ref{tab:astars}. As can be seen, some temperatures even go higher than 10000K (and are possibly B or even O stars), but given the uncertainty and bias on temperatures in the data from the Kepler Input Catalogue it is safer to assume that these are A stars as well. The table lists in the right column possible alternative explanations for the flares as given by \citet{pedersen2017}. However, our study finds 24 new A stars which show flaring activity. Thus, we add to the body of evidence that there may be some A type stars which show magnetic activity. 
\begin{table}
	\caption{List of A stars with a flare detected. The left column shows the stellar Kepler ID, the middle column shows the stellar temperature $T_\mathrm{eff}$, and the third column indicates which objects are new here, but also lists unreliable flares (see Sec.~\ref{sec:astars}). The right column indicates the explanations as listed in \citet{pedersen2017} for object that were previously found.}
	\label{tab:astars}
	\begin{tabular}{rrll}
		\hline
		Kepler ID & $T_\mathrm{eff}$ (K) & Comment & \citet{pedersen2017} \\
		\hline
		\hline
1294756 & 8411.0 & New &\\
1430353 & 10765.0 & New & \\
& & Unreliable & \\
4547333 & 10769.0 & New & \\
& & Doubt & \\
4573879 & 8288.0 & New & \\
5113797 & 8139.0 & & Overlapping neighbour? \\
5201872 & 7937.0 & & Cool companion\\
& & & Overlapping neighbour?  \\
5273195 & 8588.0 & New & \\
5284647 & 9777.0 & New & \\
5632093 & 8085.0 & New & \\
5879187 & 7668.0 & New & \\
& & Doubt & \\
5905878 & 8337.0 & New & \\
6954726 & 16764.0 & New & \\
& & Unreliable & \\
7097723 & 8438.0 & & Contamination\\
& & & Cool companion?\\
7523115 & 8438.0 & New & \\
7599132 & 10251.0 & New & \\
8044889 & 7653.0 & & Cool companion\\
8129631 & 9946.0 & New & \\
& & Doubt & \\
8142623 & 9332.0 & New & \\
8214398 & 8848.0 & New & \\
8264075 & 7654.0 & New & \\
& & Unreliable & \\
8515910 & 8143.0 & New & \\
8881883 & 10710.0 & New & \\
10149211 & 10785.0 & New & \\
10593239 & 8259.0 & New & \\
10974032 & 9038.0 & New & \\
11293898 & 15072.0 & New & \\
& & Unreliable & \\
11912716 & 10386.0 & New & \\
11919968 & 7707.0 & New & \\
		\hline
	\end{tabular}
\end{table}
\par
In the appendix~\ref{sec:aflares}, we show the flare light curves for the A-stars. An uncritical look will reveal Fig. \ref{fig:1430353}, \ref{fig:6954726}, \ref{fig:8264075}, \ref{fig:11293898a}, \ref{fig:11293898b} as being unreliable flares (indicated with ``Unreliable'' in the third column in table~\ref{tab:astars}).  Furthermore, an even more critical look will also cast doubt on the flares in Figs.~\ref{fig:4547333}, \ref{fig:5879187a}, \ref{fig:5879187b}, \ref{fig:8129631}, because the stellar variability has the same time scales as the detected flares (indicated with ``Doubt'' in the third column in table~\ref{tab:astars}).\par
Taking only the ``unreliable'' flare stars as misdetections, we have misdetected 10 flares out of 61 flares, amounting to 16\% of the detected flares. With this assumption, the false detection probability of flaring A-stars is thus 4 out of 28, which is 14\%. This value is close to our misdetection probability as estimated in Sec.~\ref{sec:errors}. However, when also taking into account the A-stars in ``doubt'', the misdetection probability goes as high as 25\%.\par
One may also argue that the A-stars disqualified by \citet{pedersen2017} should not be listed in table~\ref{tab:astars}. However, the automated detection algorithm {\em should} detect these stars: indeed, they appear as A-stars with flares and are thus correctly detected by our algorithm \citep[despite the flares most likely originating on a cool companion or neighbour,][]{pedersen2017}.

\subsection{Flaring giants}
Most of the flaring stars are found on (or near) the main sequence. On the other hand, several objects have a temperature and luminosity located in the red giant branch or even the asymptotic giant branch. A few of such objects were also previously detected by \citet{balona2015b}. These giant stars with flares are shown by red dots in Fig.~\ref{fig:hr}. To be classified as giants, the flare stars have to satisfy the following relation:
\begin{equation}
	\frac{L-1}{1.5}>\frac{T_\mathrm{eff} -4200K}{1000K}.
	\label{eq:giant}
\end{equation}
This relation was inspired by the shape of the HR diagram, and serves to separate the giant branch from the main sequence. In total, we find 653 giants with flaring activity. \par 
As for A stars, it is equally unexpected that giant stars have strong magnetic fields. During their evolution on the main sequence, stars spin down. Moreover, their increase in size would lead to a decreased surface magnetic field (because of the magnetic field dependence on distance). As a result, their magnetic field is expected to be weaker \citep{simon1989}. This was confirmed observationally by \citet{konstantinova2008}, who found a magnetic field of 5G and 15G for a rapidly rotating giant. However, outbursts which were attributed to stellar activity were observed on Mira A by \citet{karovska2005,vlemmings2015},  \citet{harper2013} found chromospheric emission on giants, and \citet{gaulme2014} found light curve variations compatible with stellar spots on giants.\par
Given the numerous evolved objects with flares revealed by our analysis, it becomes clear that magnetic activity is ubiquitous on the red giant branch. Moreover, the flare star incidence for giants (as given in Table~\ref{tab:overview}) is similar to that of F and G stars. This value of the flare star incidence among giants may be compatible with two scenarios for the explanation of flares on these stars: (1) The flares are actually attributable to a companion of K or M type. The incidence rate among giants is around half of that of K and M stars, which would then require a cool companion to half of the flaring giants. (2) Alternatively, one could argue that this value of the flare star incidence among giants is compatible with the flare star incidence in their progenitors, i.e. the F and G stars. This could mean that the magnetic activity of the stellar atmosphere is more or less conserved during its evolution from the main sequence to the red giant branch. \par
One may also wonder about the nature of the flares. Perhaps these flares are not caused by the same mechanism as on late type main sequence stars, where it is believed an internal dynamo leads to complex surface magnetic fields, reconnection and flares. The giants may have a different mechanism that operates to create these outbursts. There may be an enhanced surface dynamo \citep[e.g.][]{amari2015}, or the extended magnetic loops could reconnect due to centrifugal forces. 

\subsection{Flare amplitudes}
\label{sec:amplitude}
\begin{figure*}
	\includegraphics[width=\linewidth]{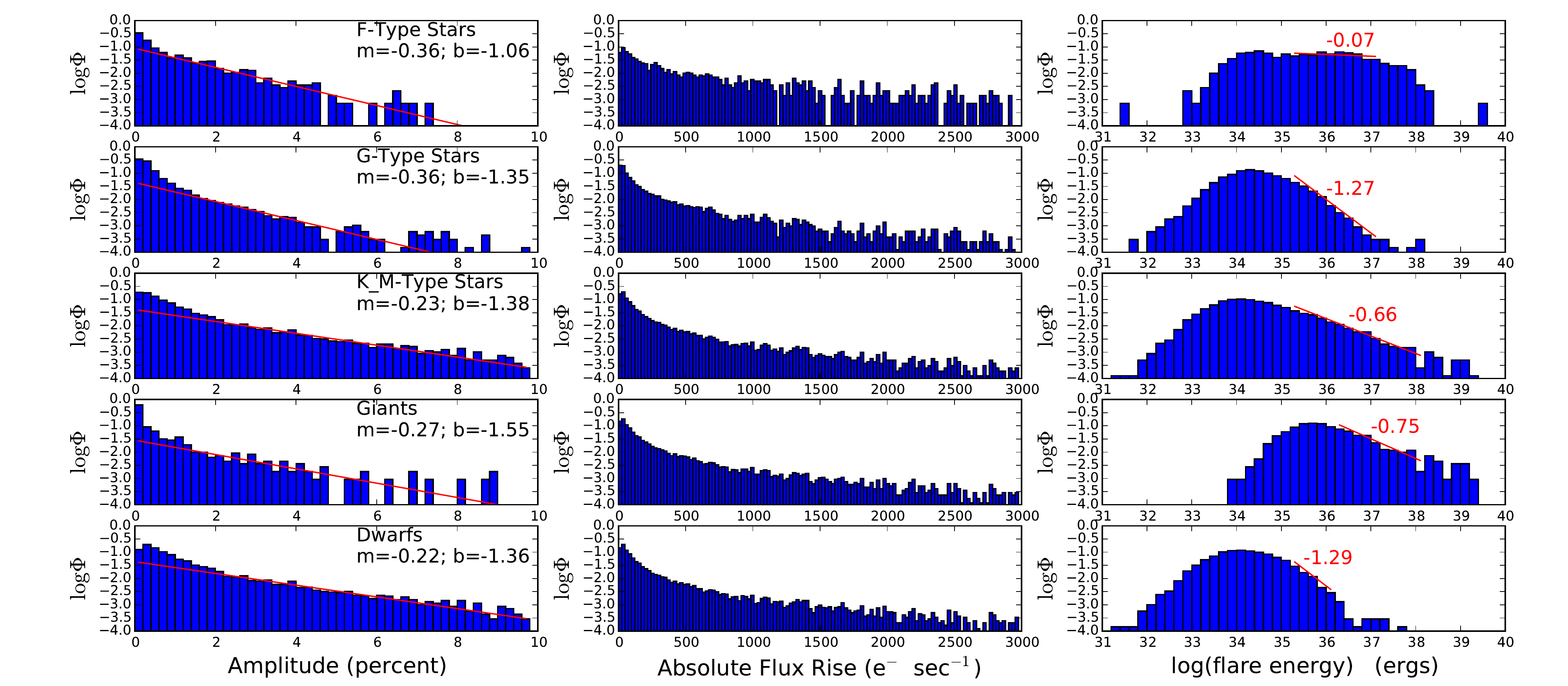}
	\caption{Histogram of flare amplitude and energy per spectral type. The vertical axis is the logarithm of the histogram count. The horizontal axis is the relative flux increase in the left figure, and the absolute flux increase in the middle figure, and the logarithm of the flare energy (in ergs) in the right panel. The left panel shows the exponential fit between 2\% and 4\% with a red line, while the right panel shows the fit only in the fitting range.}
	\label{fig:amplitude}
\end{figure*}
In Fig.~\ref{fig:amplitude}, we show the (logarithm base-10 of the) histograms of the flare amplitude and energy for different stellar categories. The absolute amplitudes (middle panels) are calculated by the flare amplitude minus the pre-flare stellar intensity, and the relative amplitude normalises the absolute flare amplitude by the pre-flare stellar intensity. The flare energy $E_{\mbox{flare}}$ is calculated by using Eq.~S5 from \citet{maehara2012} \citep[equivalent to Eq.~6 of][when assuming the exponential intensity profile of Eq.~\ref{eq:fit}]{shibayama2013}:\begin{equation}E_{\mbox{flare}}=\frac{a}{k}LL_\odot,\end{equation} with $L_\odot$ given in ergs, and its distribution is shown in the right panels. \par
The total population of flaring stars is subdivided in spectral types using the temperatures listed in the Kepler input catalogue (as explained in Sec.~\ref{sec:general}). We do not show the results of A-type stars or hotter, because there is only a limited number of stars in that category (see Sec.~\ref{sec:astars}). Moreover, the K+M-types (third row) have been split between giant stars and dwarf stars (using Eq.~\ref{eq:giant}). \\ The horizontal scales of the histogram have been limited to 10\% and 3000 e$^-$sec$^{-1}$, because some flares with much higher amplitudes have been detected, but these events disallow a clear comparison between the different spectral types. \\ The histogram of the relative amplitude has been fitted with an exponential distribution ($10^{mx}$), motivated by its appearance. This is done by fitting a straight line $y=m x+b$ (with $y$ the logarithm base-10 of the histogram and $x$ the relative amplitudes) by a regression between the amplitudes $x=2\%$ and $x=4\%$ to avoid the influence of the tail (which is presumably dominated by small-number statistics). The fitted values of $m$ and $b$ are shown in the key.\par
From the histograms and their fits, it is clear that the F and G-type stars show the same behaviour in the relative amplitude, and even an identical slope is found from the fit. This is an indication that the magnetic phenomena on these types of stars have the same underlying physical mechanism, because the relative flare amplitude is equally distributed. However, for the flare energy, a widely different behaviour is found. The F-stars show a flat energy distribution, while the G-type stars show a steeply decreasing energy distribution. \par
For the K and M-type stars, the histogram for the relative amplitude seems to be different. There seem to be more flares with a high amplitude, and they form a longer tail in the histogram. This is also confirmed by the slope fit of -0.23, which is shallower than the early-type stars (F and G). We have also split the K and M-type stars between giant and dwarf stars, and show their histograms separately. Given the smaller number of flaring giants, the distribution of the amplitudes of flares on dwarfs is very close to the distribution of the K and M-type stars. Even the fitted slope is not changed very much. However, the distribution of the amplitudes of flares on giants is quite different from the distribution of the dwarf stars, because the heavy tail of large-amplitude flares is missing from the distribution. This is also confirmed by the fitted slope; it is steeper than the slope of the dwarf amplitude distribution. \\
From these results, we may tentatively conclude that the magnetic mechanism for flares  is the same on F and G-type stars, but differs for dwarf stars (K and M-type). For giants, it is unclear what magnetic mechanism they could use for generating flares. On the one hand, it could be that there is yet another mechanism at work to generate flares on giants. On the other hand, it could be that there are two mixed populations of giants, with each population generating flares with the magnetic mechanism of F and G-types or that of the dwarfs. One of the two populations could constitute of giants who were already flaring during their life on the main sequence, and the other population may generate flares with a similar mechanism as the dwarf stars. \par
From the flare energy distributions (right panel, Fig.~\ref{fig:amplitude}), we see that the low energy end is strongly influenced by observational bias, because only the stronger flares are detected. It also shows that the flare detection routine (which is based on thresholds) is influenced by the background brightness of the star, because the peak of the flare energy distribution depends on the stellar type. In particular, the flare energy distribution for the giants (and to a lesser extent for the F-stars) is shifted towards higher energies. This is because only the very strongest flares are detected against the much stronger emission from the giants (due to their larger radius).\\ The high energy tail is governed by an apparent power law, which is physical in nature. The fits to the power law (with $10^{mx}$) are shown with red lines in the fitting regions, and their slopes are indicated. The F-type stars have a very shallow distribution, but the power law slope of the G-types and dwarfs is consistent. The giants have a slope between the flat F-type distribution and the G-type and K+M-type distributions. As we have inferred from the amplitudes as well, it could point in the direction of two flare generation mechanisms at work in flaring giant stars. 

\subsection{Flare duration}
\begin{figure}
	\includegraphics[width=\linewidth]{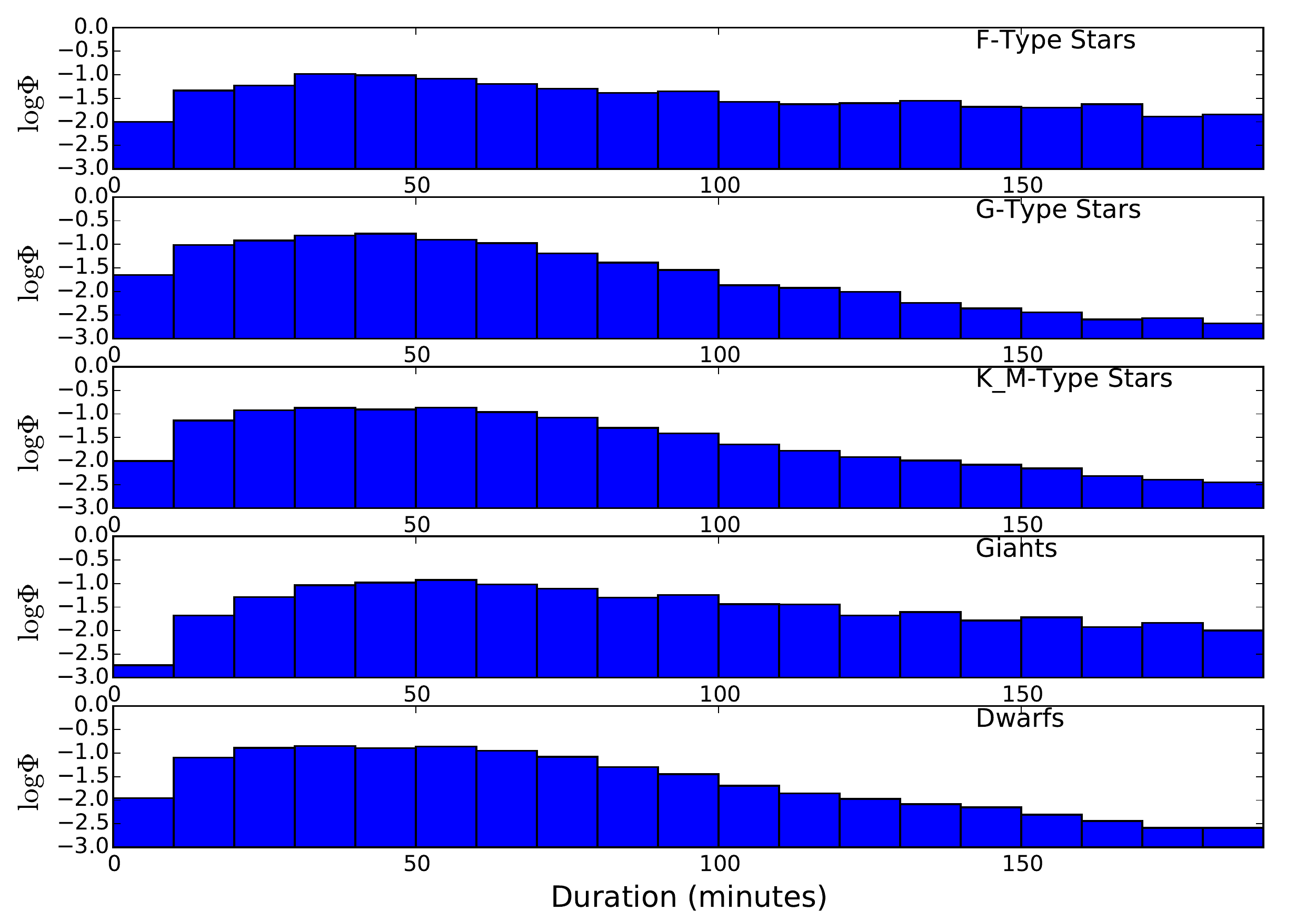}
	\caption{Histogram of flare duration per spectral type. The vertical axis is the logarithm of the histogram count. The horizontal axis is the duration of the flare in minutes, taken as $1/k$ (Eq.~\ref{eq:fit}).}
	\label{fig:duration}
\end{figure}
In Fig.~\ref{fig:duration}, we show the histograms for the duration of the flares, split up between stellar type similarly as in Sec.~\ref{sec:amplitude}. The duration is calculated as $1/k$, from the fit with Eq.~\ref{eq:fit}. It is clear that the histograms do not follow an exponential distribution, as seemed to be the case for the amplitude distribution (Fig.~\ref{fig:amplitude}). The reason is clear: we have used long-cadence data from the Kepler mission, and thus we have only a small chance of observing flares with a duration shorter than 30 minutes. Moreover, flare detections of only one data point have been rejected, further limiting our detection capabilities in that range. Thus, the location of the peak of the distribution is mainly determined by the cadence time of the telescope and our detection algorithm. Still, it seems that the peak of the histogram for the duration of flares on giants is shifted to the right with respect to the other histograms. One may wonder if this fact contains any physics.\par
From the graph, it seems that now the G and dwarf K+M-type stars have a similar behaviour for the flare duration. This is surprising, because the previous section found that the amplitude of flares on G stars behave more like the amplitudes of flares on F stars. However, now it is clear that the flares last much longer on F stars than on G stars. So, it seems that the separation between the two flare generation mechanisms that we postulated in Sec.~\ref{sec:amplitude} is not so clear-cut. It could be that the transition between the two mechanisms does not coincide with the chosen boundary of spectral types, or that there is an overlapping region in the HR diagram, where the two flare mechanisms occur simultaneously. \par
The distribution of duration of flares on giants follows closely the distribution of F-type stars, with a heavy tail towards longer duration. Thus flares on giants have statistically a longer duration than flares on dwarf K+M stars. This could point in the direction that their flaring mechanism has remained the same as during their main-sequence life as F star. However, such a conclusion is not supported by the difference between F stars and giants in Fig.~\ref{fig:amplitude}. \par
The measured duration of the flares may be heavily influenced by the thresholds we chose for the detection. So, some word of caution should be added that the results of this section may change for different detection algorithms and this could be tested by other teams \citep[e.g.][]{davenport2016}.

\subsection{Flare occurrence rate}
\begin{figure}
	\includegraphics[width=\linewidth]{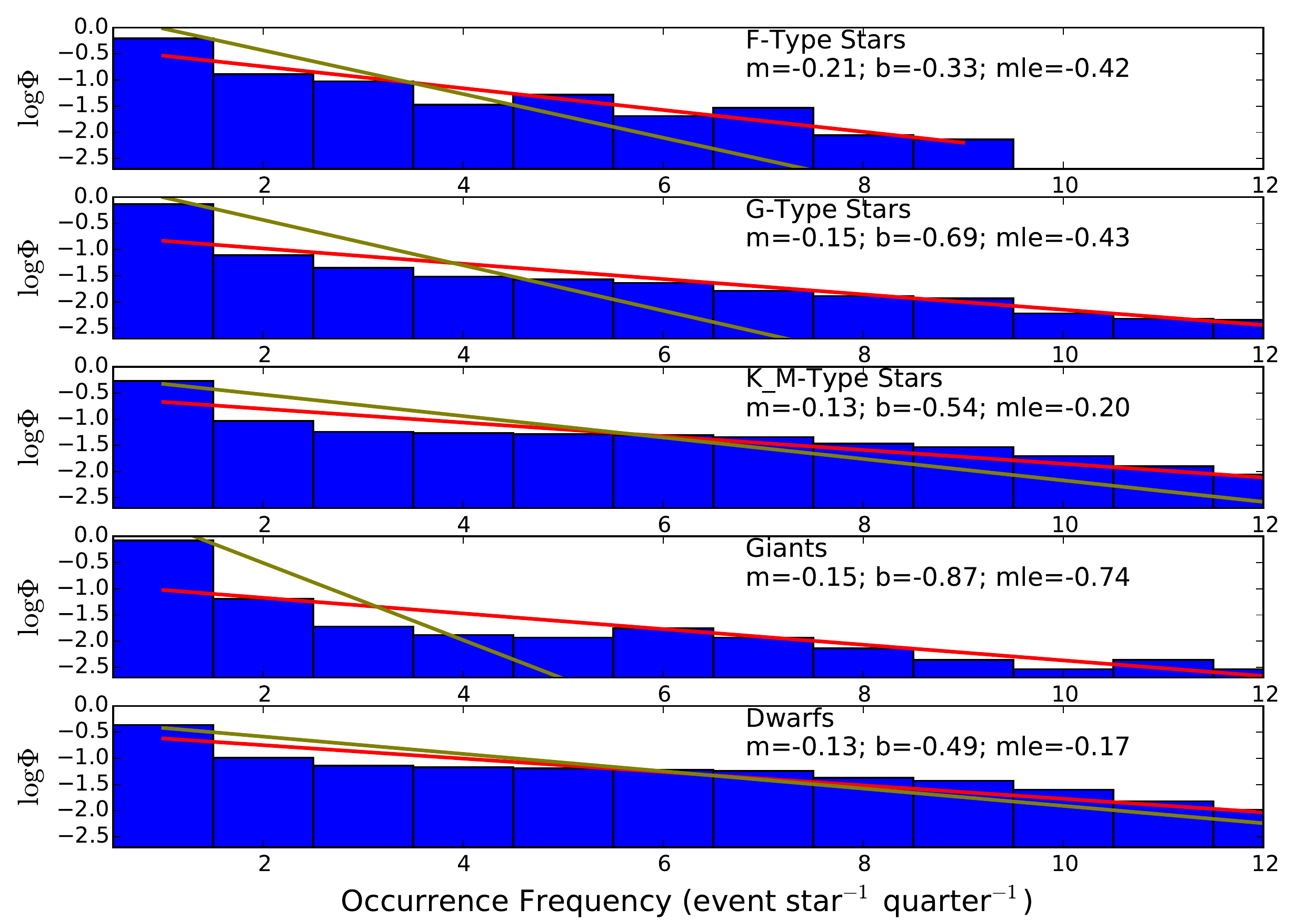}
	\caption{Histogram of flare occurrence rate per spectral type. The vertical axis is the logarithm of the histogram count. The horizontal axis is the flare occurrence rate in units of flares per star per quarter. The key contains the fitting parameters of a straight line (shown in red), and also the MLE for the exponential parameter (shown in olive).}
	\label{fig:occur}
\end{figure}
Fig.~\ref{fig:occur} shows the histogram for the occurrence rates of flares per object. Since most of the histograms seem close to an exponential distribution ($10^{mx}$, $x$ is the occurrence rate here), we have fitted the histogram with a straight line $y=mx+b$ (with $y$ the logarithm of the histogram counts). The parameters of this fit have been displayed in the key on each graph. We have also fitted the distributions by using the Maximum Likelihood Estimator (MLE) for $m$, assuming that the offset-exponential distribution starts from a value $\theta$. Following \citet{johnsonkotz}, the MLE estimator for the shifted exponential distribution is given by \begin{equation}\mbox{MLE}_\theta=\frac{n}{\sum_{i=1}^{n} x_i - n \theta}=\frac{1}{\mu_S - \theta},\label{eq:distribution}\end{equation} with $S=\left\lbrace x_i | i=1,\ldots,n\right\rbrace$ the $n$ samples of the data, and $\mu_S$ is the sample's mean. This is analogous to the width estimator of the Laplace distribution.\\ Thus, MLE$_1\cdot(\log_{10}{e})$ is the MLE for $m$, assuming a minimum occurrence rate of 1 event per star per quarter, because we did not include stars without flares. In the figure, we have printed the obtained value MLE$_1\cdot(\log_{10}{e})$, which is the estimator for $m$. Any further mentions of MLE values always includes the $\log_{10}{e}$ factor, to relate to the $10^{mx}$ proposed distribution.\\ On the one hand, the MLE$_1$ value is dominated by the bulk of the distribution and therefore contains information on the stars which do not flare very often. On the other hand, the fitting is more heavily influenced by the tail of the distribution and thus contains information about the most frequently flaring stars. \par
From the MLE values in the keys of Fig.~\ref{fig:occur}, it seems that the K+M type stars have a higher flare occurrence rate, followed by F and G, and that the giants have the lowest flare occurrence rate. If K+M stars are flaring, they flare on average 3.1 times per quarter, flaring F and G stars are flaring on average 2.0 times per quarter and flaring giants flare on average 1.6 times per quarter.\\ 
However, from the fit of the slope $m$ to the histogram, it seems that all stellar types have a similar tail of flare stars which have a high flare occurrence rate. Thus, the colloquially used term ``flare stars'' seems to make sense with our statistics: each stellar type has a fraction of flare stars. Such flare stars are more frequently occurring in K and M-type stars, but exist for earlier types as well. They are rarest among the giant class.

\subsection{Relation with rotation period}
Because \citet{candelaresi2014} and \citet{davenport2016} found a strong correlation between the flare occurrence rate, luminosity and rotation period, we also investigate the statistical connection between stellar flares and stellar rotation period. To that end, we have used the data of \citet{mcquillan2014}, who have found the stellar rotation from a period analysis of light curve variations due to stellar spots. We have cross-referenced our flare stars with the data available in \citet{mcquillan2014catalog}. In total, we obtained the rotation period for 3033 of our flare stars from that catalogue. We note that none of the flaring giants have reported rotation periods, because they were excluded from \citet{mcquillan2014}. \\
For each star, we estimate the Rossby number Ro by Ro$=p_\mathrm{rot}/\tau$, where $p_\mathrm{rot}$ is the rotation period \citep[taken from][]{mcquillan2014catalog}, and $\tau$ is the convective turnover time. The latter is estimated with formula 11 from \citet{wright2011}, following the procedure in \citet{davenport2016} but using the estimated stellar masses from the KIC. \par
\begin{table}
	\caption{An overview of flare star incidence in the database of \citet{mcquillan2014catalog}. The left column shows the range of the logarithm of the Rossby number $R_o$, the second column shows the number of stars $N_\mathrm{stars}$ in the \citet{mcquillan2014catalog} database, the third column the number of flare stars $N_\mathrm{flare}$ \citep[as detected in this work and cross-referenced with][]{mcquillan2014catalog}, and the fourth column is the flare star incidence rate per rotation bin. The bottom graph shows the flare star incidence rate as a function of Rossby number Ro.}
	\label{tab:rotation}
	\centering
	\begin{tabular}{cccc}
		\hline
		$\log{\mbox{Ro}}$ & $N_\mathrm{stars}$ & $N_\mathrm{flare}$ & \% \\ \hline \hline
		-2.5 -- -1.9  &  26  &  18  &  69.2$\pm$ 10.9 \%\\
-1.9 -- -1.6  &  174  &  88  &  50.6$\pm$ 5.3 \%\\
-1.6 -- -1.4  &  188  &  103  &  54.8$\pm$ 4.9 \%\\
-1.4 -- -1.2  &  301  &  184  &  61.1$\pm$ 3.6 \%\\
-1.2 -- -1  &  412  &  219  &  53.2$\pm$ 3.4 \%\\
-1 -- -0.8  &  655  &  278  &  42.4$\pm$ 3.0 \%\\
-0.8 -- -0.6  &  1074  &  339  &  31.6$\pm$ 2.5 \%\\
-0.6 -- -0.4  &  2205  &  535  &  24.3$\pm$ 1.9 \%\\
-0.4 -- -0.2  &  5492  &  499  &  9.1$\pm$ 1.3 \%\\
-0.2 -- 0  &  8254  &  378  &  4.6$\pm$ 1.1 \%\\
0 -- 0.2  &  11404  &  312  &  2.7$\pm$ 0.9 \%\\
0.2 -- 0.4  &  3541  &  72  &  2.0$\pm$ 1.7 \%\\
0.4 -- 0.6  &  260  &  8  &  3.1$\pm$ 6.1 \%\\
		\hline \\[-.5ex]
	\end{tabular}
	\centerline{\includegraphics[width=\linewidth]{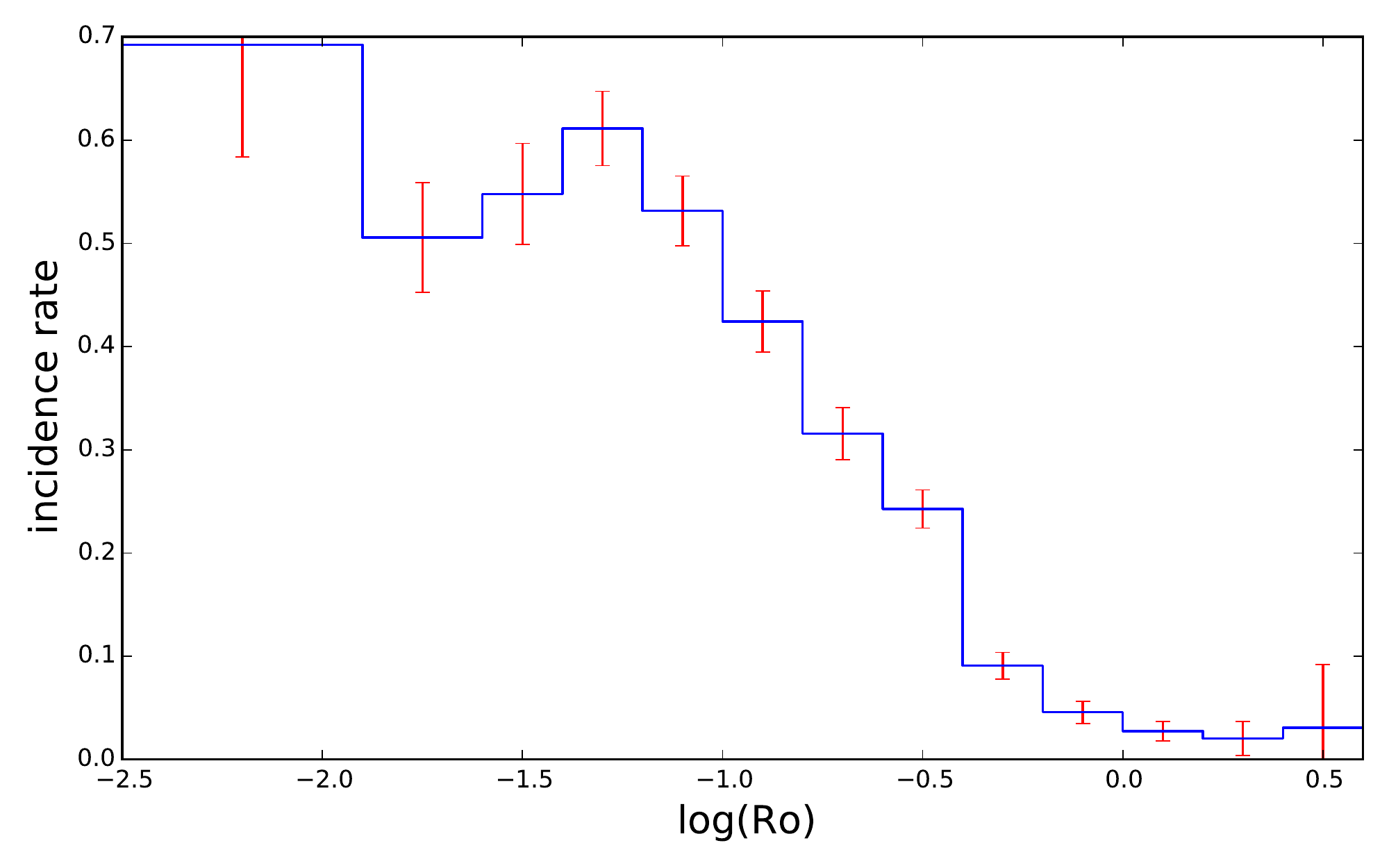}}
\end{table}
Table~\ref{tab:rotation} shows the number of flare stars for each bin, and the flare star incidence rate. Here the standard deviation to the incidence rate $p$ is estimated as $\sqrt{p(1-p)/N_\mathrm{stars}}$. The incidence rates are also shown as a function of the Rossby number in the graph under Table~\ref{tab:rotation}. It is obvious that the stars with a short rotation period are much more often also flare stars, in comparison with stars with a long rotation period. However, for short rotation periods (low Rossby numbers), there is again a drop or saturation from the maximum at $\log{\mbox{Ro}}\sim-1.2$. This seems to compare well to the results of \citet{candelaresi2014}, where the maximum occurs around $\log{\mbox{Ro}}\sim -1$ (see their Fig. 2). However, they plotted flare rate, instead of our occurrence rate. It also agrees with the saturation found by \citet{davenport2016}.\\
From the graph in Table~\ref{tab:rotation}, it seems that flare star incidence rate tails off to a constant value from a Rossby number of 10 onwards, to a value of about 3\%. An interesting question is: does the Sun fall in the 3\% of active stars? Or would we not observe it as a flaring star with our detection algorithm? In any case, these reported incidence rates of 3\% is an overestimate of the real occurrence rate, because the rotation period is measured by the appearance of stellar spots, thus subselecting a sample of only magnetically active stars. \par
\begin{figure*}
	\centerline{
		\includegraphics[width=.45\linewidth]{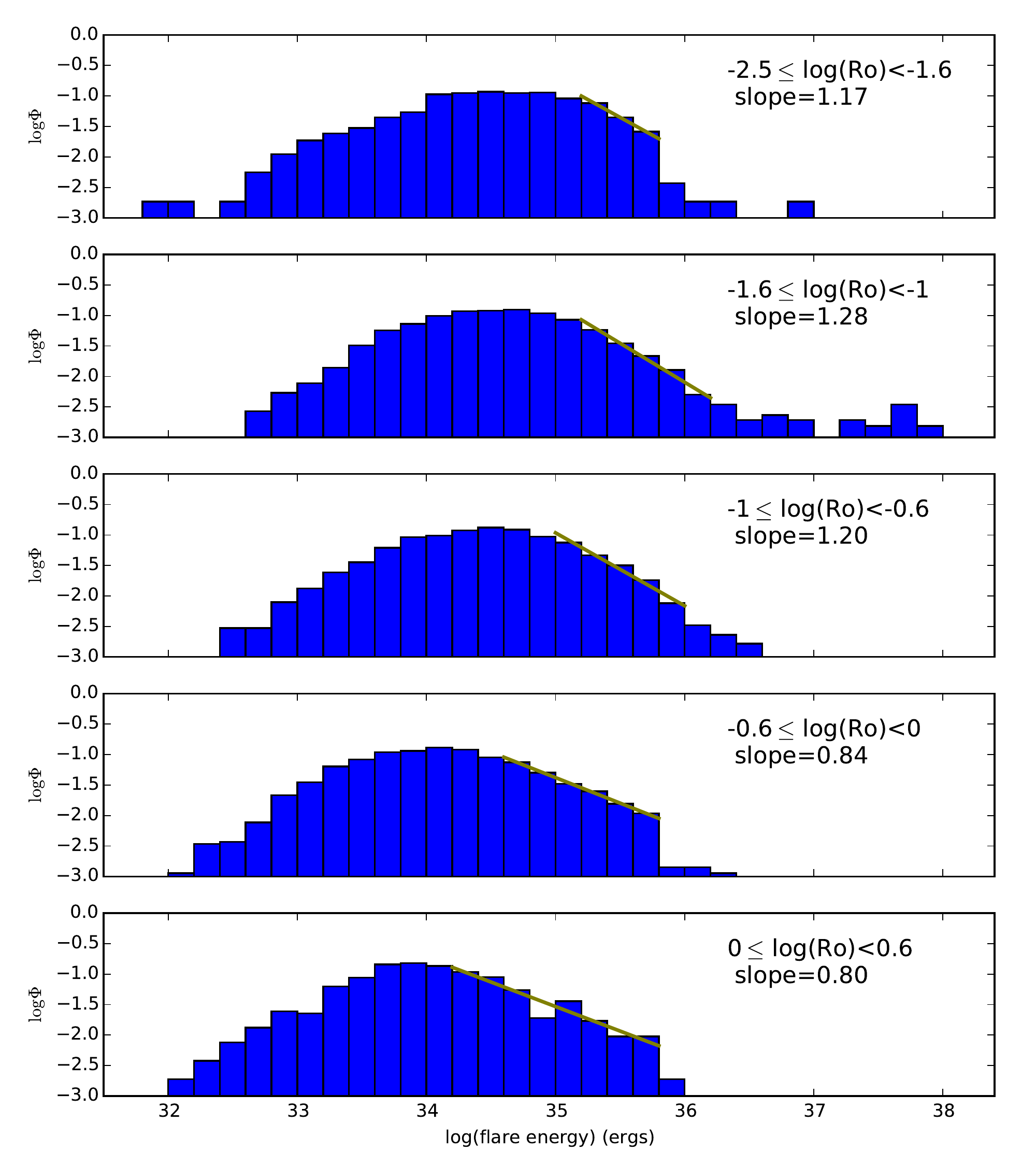}
		\includegraphics[width=.45\linewidth]{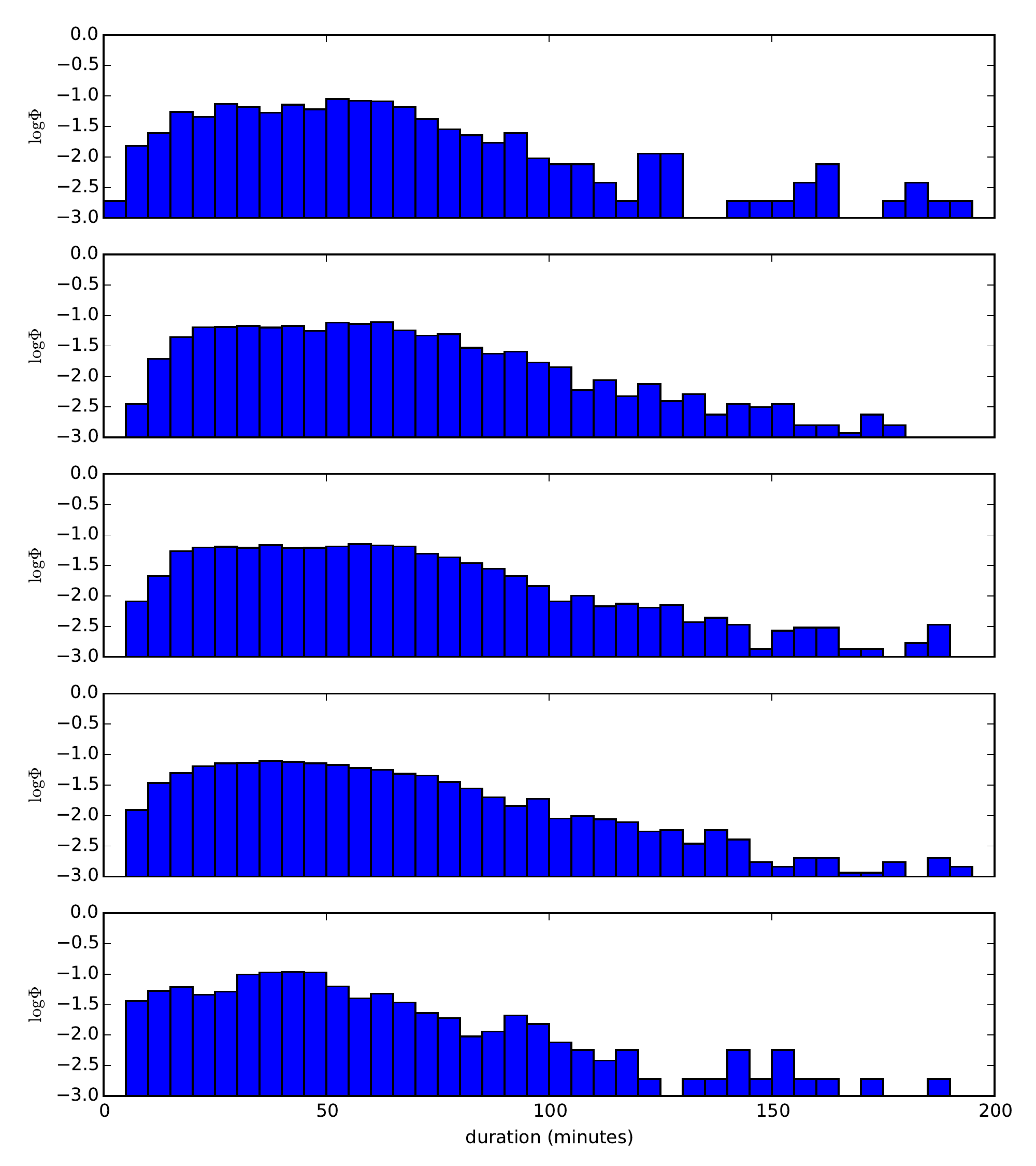}
	}
	\centerline{
		\includegraphics[width=.45\linewidth]{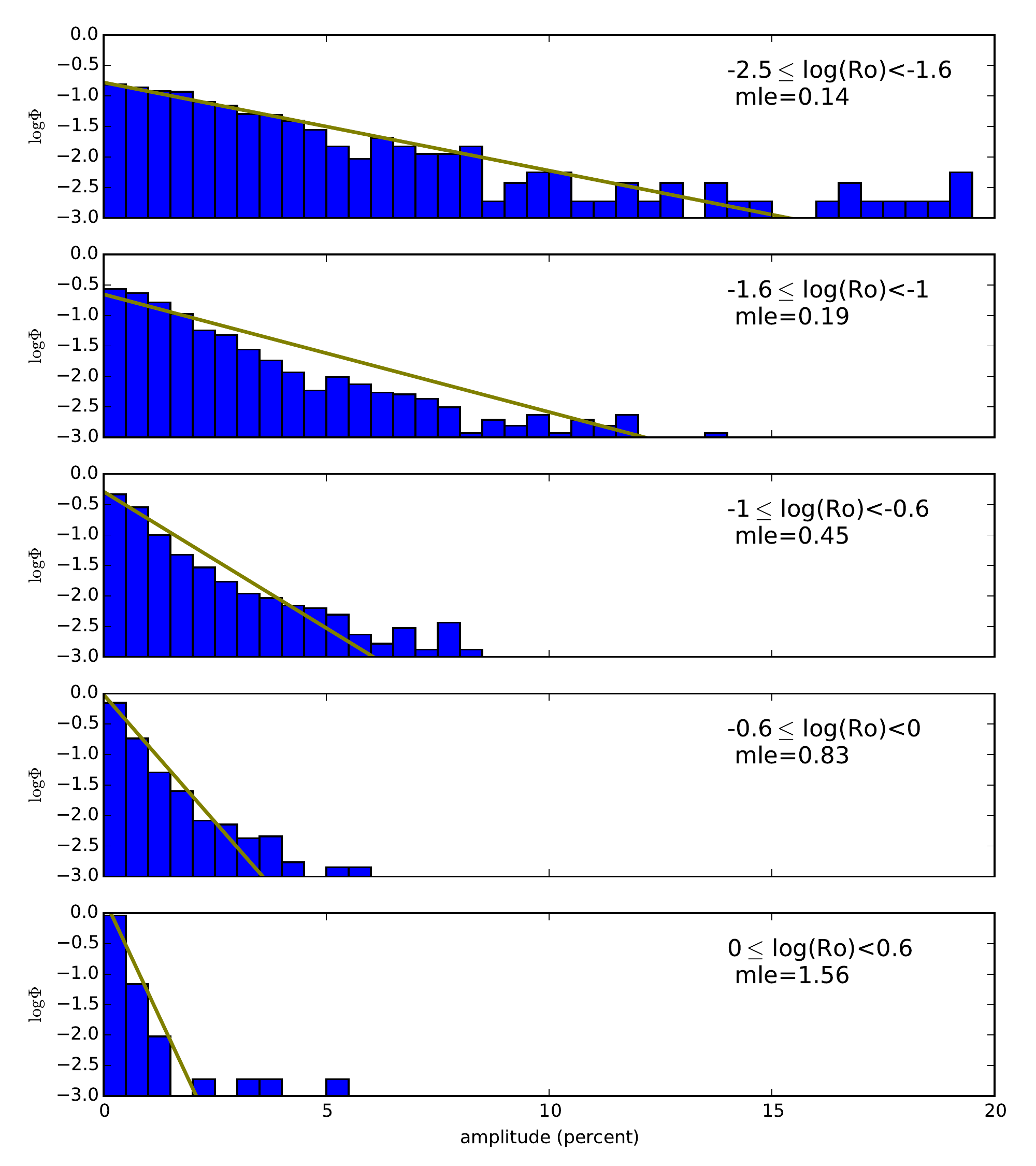}
		\includegraphics[width=.45\linewidth]{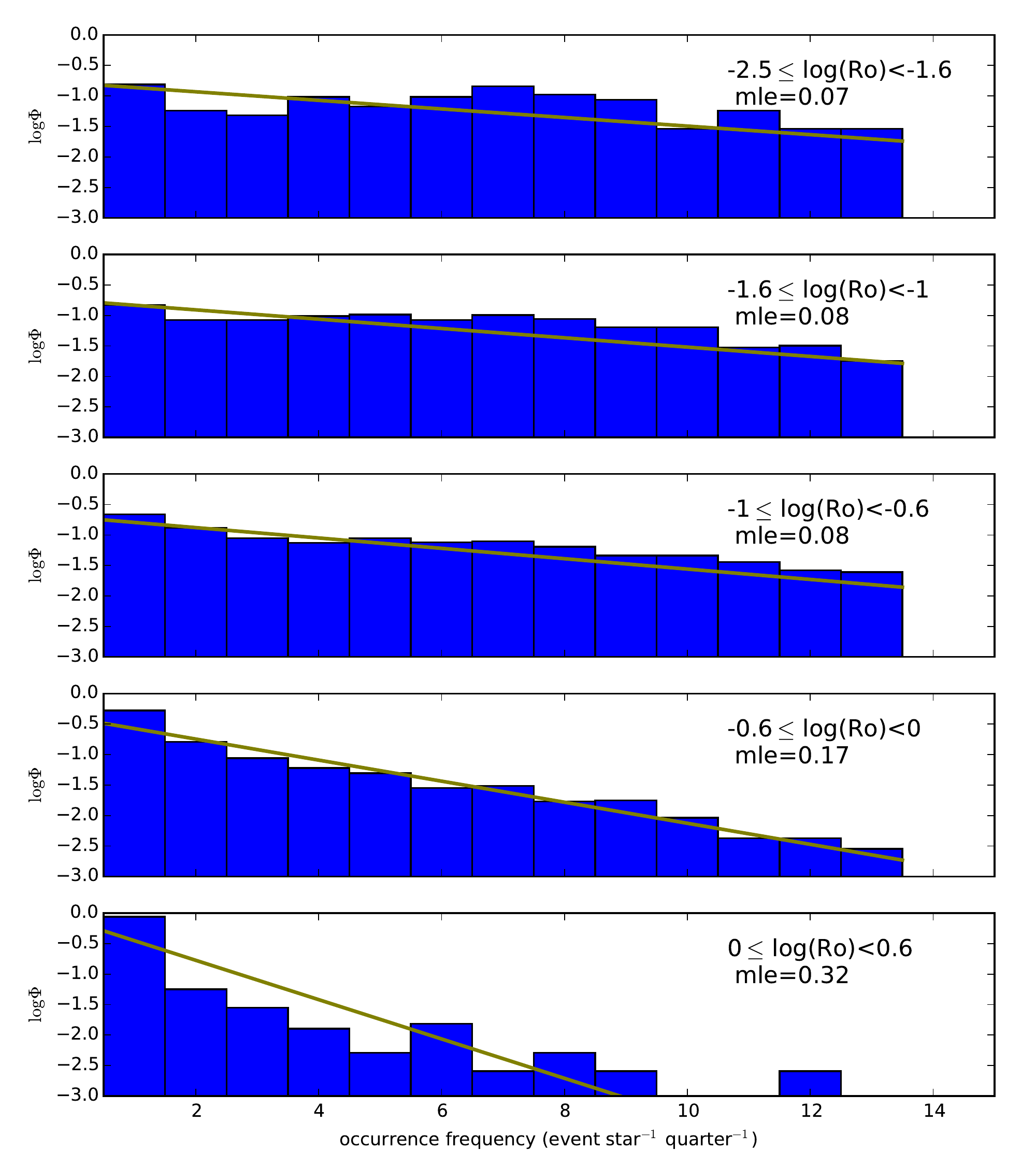}
	}
	\centerline{
		\includegraphics[width=.45\linewidth]{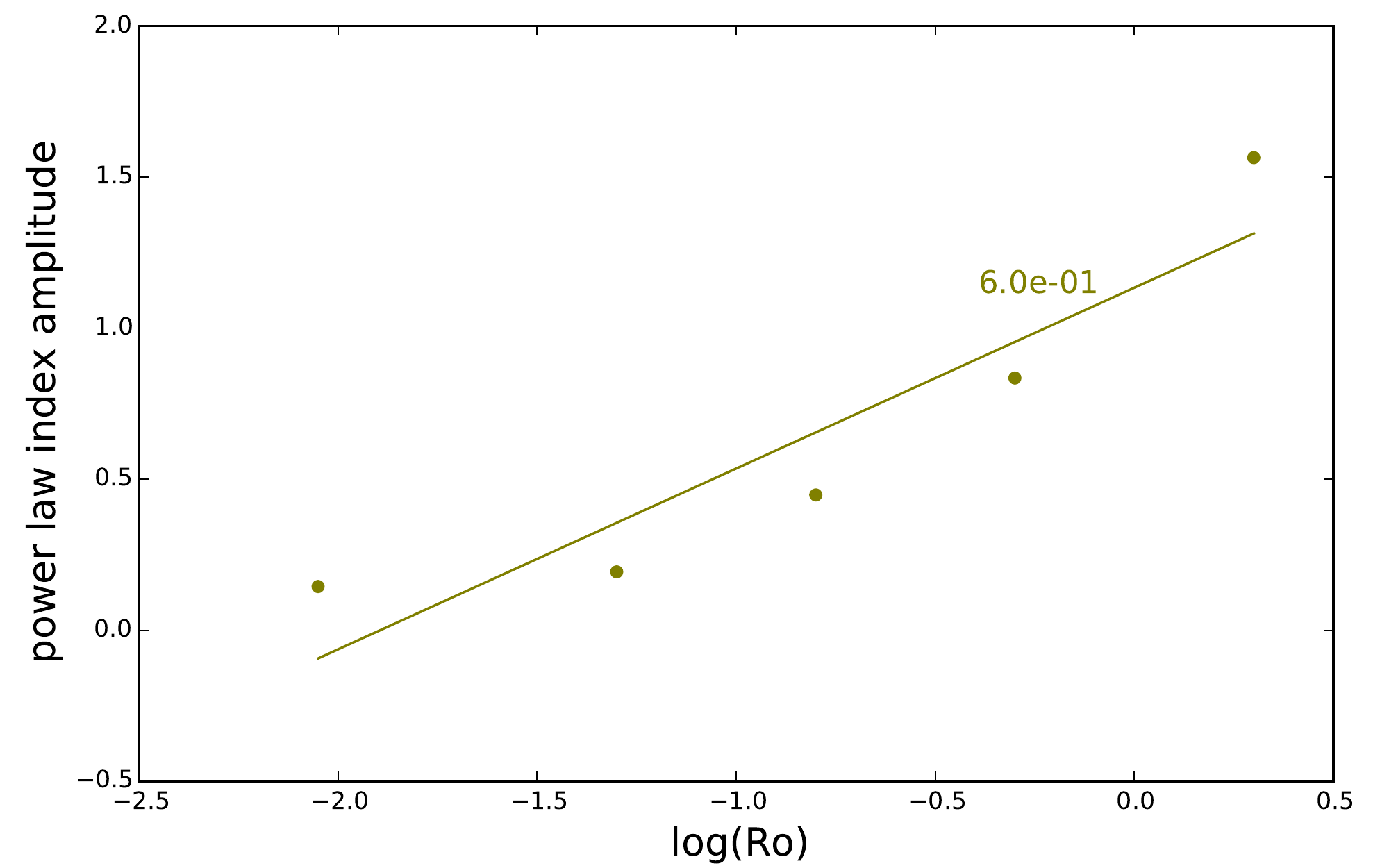}
		\includegraphics[width=.45\linewidth]{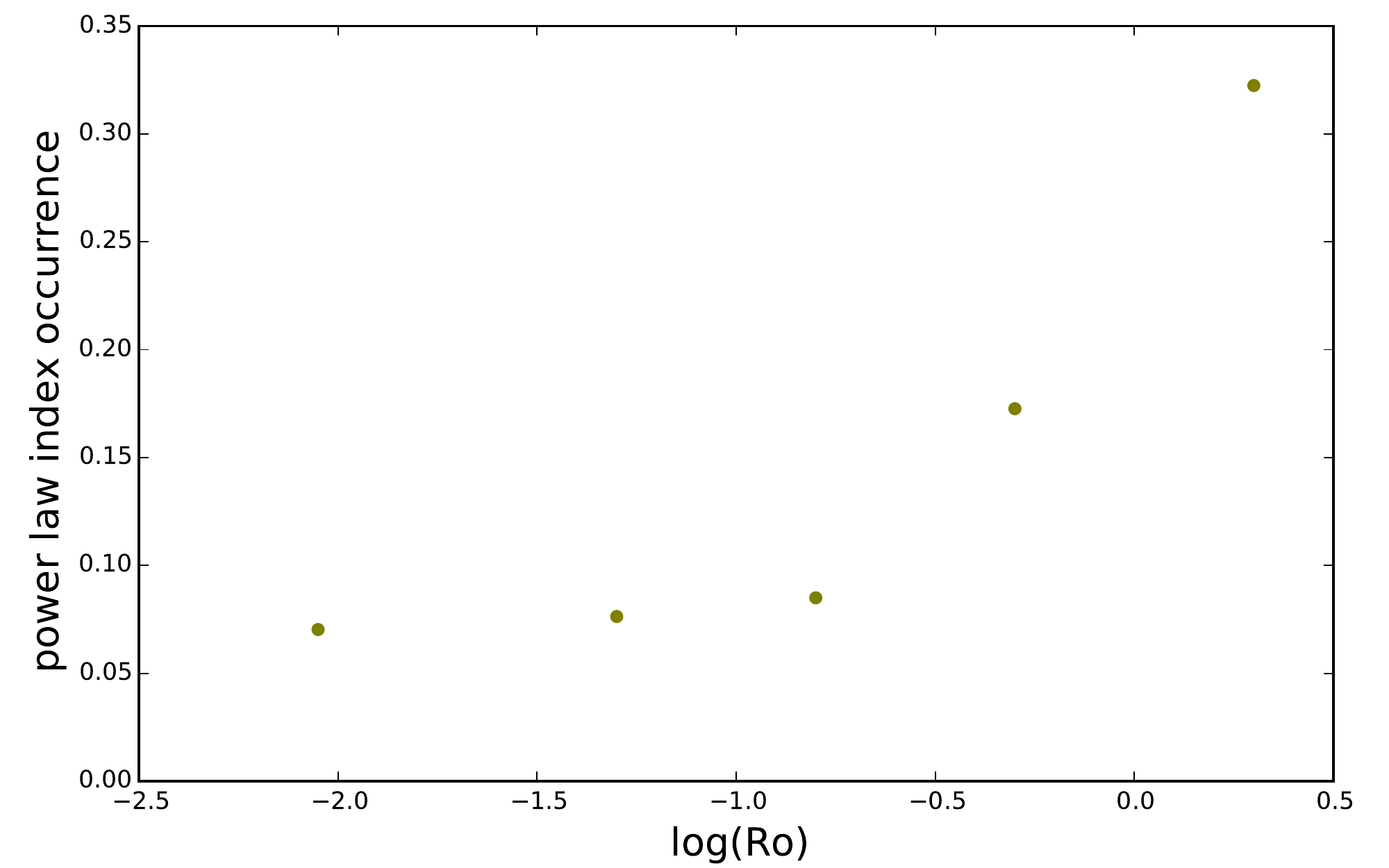}
	}	
	\caption{The top row shows the (logarithm of the) histogram of the flare energy (top left) and flare duration (top right), split up per bin in Rossby number. The second row shows the histogram of flare amplitude (middle left) and flare rate (middle right). The olive line shows the distribution given by the least-squares fit (top left), the MLE (middle left) and MLE$_1$ (middle right). The key contains the parameters of the least-squares fit and the MLE/MLE$_1$. The bottom row shows the MLE/MLE$_1$ estimator (olive dots) per Rossby bin, and the linear least-squares fit to those values in the bottom left panel. The indicated value is the slope of the linear least-squares fit.}
	\label{fig:rotation}
\end{figure*}
In Fig.~\ref{fig:rotation} (top rows), we show the histograms for flare energy, flare duration, flare amplitude, and flare occurrence rate per bin of Rossby numbers (roughly corresponding to left of the incidence rate peak, the peak, two in the tail, and one for the slow rotators). The graphs show the distribution obtained using the least-squares fit or MLE/MLE$_1$ estimator (Eq.~\ref{eq:distribution}) as the olive line. The values of the MLE or slope of the least-squares fit are indicated in the key of the figure. In the bottom row, we show the values of MLE/MLE$_1$ as a function of the Rossby number.\par
The graph for the occurrence frequency (Fig.~\ref{fig:rotation}, middle right panel) confirms the behaviour of Table~\ref{tab:rotation}. If a star is rapidly rotating, it is not only highly likely to flare, but it is also more likely that it will flare more often than stars which are not fast rotators. Here as well, the saturation past $\log{\mbox{Ro}}\sim -1.2$ is apparent. In the middle left panel of Fig.~\ref{fig:rotation}, we see that stellar rotation periods also have a strong influence on the flare amplitude. If a star is rotating quickly, it has a higher chance of creating a big flare. \\
Thus, the effect of rotation is twofold: quickly rotating stars have a higher chance that a flare is occurring (or that flares occur more often), and flares that occur have a higher chance of being strong flares. \\
On the other hand, the graphs for the flare energy (top left panel) show that the power law slope does not depend on the Rossby number. Thus, this confirms the results of \citet{candelaresi2014} who also found that the slope for the flare energy distribution is nearly constant. This was recently confirmed with LAMOST by \citet{chang2017}.\par
In the bottom panels of Fig.~\ref{fig:rotation}, we show how the MLEs of the amplitude distribution and occurrence rate distribution vary with the rotation period. A very clear correlation is found for the slower rotators, allowing us to accurately predict flare amplitude and flare occurrence rate distributions when a stellar rotation period is given. Perhaps such a scaling would modify the estimates for the occurrence of superflares on our Sun \citep{shibata2013,notsu2015}.

\section{Conclusions}
\label{sec:conclusions}
In this paper, we have developed a new method for automated flare detection, using thresholds in the intensity increase, the increase in the running difference and the flare duration. We have applied it to the long-cadence data from quarter 15 of the Kepler mission, and made the flare detections available to the wider community (\texttt{vandoorsselaere\_et\_al\_2017\_q15\_flares\_sorted.txt}) for further study. \par
In Sec.~\ref{sec:results}, we have reported the discovery of several new flaring A-stars, which had not been found by previous studies. From stellar evolution models, it is unexpected that A-stars would flare, because they have a small or no surface convection zone, and thus a magnetic dynamo could not operate. Even though \citet{pedersen2017} discarded some flare detections in A-stars, our paper adds further to the evidence of \citet{balona2012}, and these stars should be studied in detail in order to verify the stellar evolution theory. \par
\citet{balona2015b} had found a few cases of flaring giants. We have extended the sample of flaring giants by 653 objects. It is unexpected that evolved stars would have strong magnetic fields, because they have spun down over their evolution. Moreover, the sheer size of these stars would also decrease the strength of magnetic fields generated in the core. Thus, observations of stellar flares on giants could be evidence that a surface dynamo is operating efficiently there. The giants show no rotational modulation, indicating that starspots do not cover a significant amount of the surface area. \par
We have then embarked on a statistical study of the flares in our sample. Regarding the flare amplitude, we have found that flares on F and G-type stars behave similarly, and are to be contrasted with K and M-type stars (including giants). This could be an argument that the magnetic activity is just determined by the location in the HR-diagram, and that two causes for magnetic activity exist.\\
On the other hand, the flare duration could be grouped between F stars and giants on the one side, and G, K and M stars on the other side. This then seems to suggest that the flare activity is determined by the initial stellar mass, and is kept while flare stars move from the main sequence to the giant branch. From the flare energy, we find the same power law for the G, K and M dwarfs. \\
From the statistics of the flare occurrence rates, we have found that all spectral classes have a tail of stars which are frequently flaring. We could name these stars ``flare stars''. Such a tail of frequently flaring stars is more prominent for K and M-types, but also easily found for earlier types. Flare stars are less frequently observed for giant stars.\par
We have cross-referenced our flaring stars with the rotation periods found in \citet{mcquillan2014catalog}. By splitting the flare stars by Rossby number, it became very clear that the rotation period has a very large influence on stellar flares. A rapidly rotating star has a higher chance to flare, has a higher chance to flare more often, and the flares on rapidly rotating stars are often stronger than on slowly rotating stars. This matches earlier findings that the X-ray luminosity of stars is correlated with its rotation period \citep{wright2011}, and thus flaring activity is also a good predictor of rotation period and X-ray luminosity. We confirm the findings of \citet{candelaresi2014} and \citet{davenport2016} that there is a saturation of the dynamo for very rapidly rotating stars, or even a decrease in flaring activity for lower Rossby numbers. It remains to be seen how the chance for a superflare on our Sun is influenced, when incorporating our statistical relation between flare amplitude and flare occurrence rate and the solar rotation period. \par
In the future, we want to extend our study to the full Kepler database. Such a larger sample and longer observation period will provide even better statistics than what we reported in this paper. This would enable us to split up the sample both in spectral classes and simultaneously with Rossby number, allowing for the disentangling of the effects of either on the stellar activity. Moreover, the full sample of the automated flare detection in Kepler should be cross-referenced with earlier works \citep[such as][]{davenport2016}, which would make the flare database more robust. \par
To make the statistics of the sample more robust and disentangle binarity from the flaring nature, detailed spectroscopic studies of the flare stars are required to exclude the possibility of a companion \citep{karoff2016,pedersen2017,chang2017}. In particular, the existence of flaring A-stars conflicts with our current understanding of stellar evolution, and these could use special spectroscopic scrutiny.

\acknowledgements
TVD was supported by an Odysseus grant of the FWO Vlaanderen, the IAP P7/08 CHARM (Belspo) and the GOA-2015-014 (KU~Leuven). This work was based on discussions at the ISSI and ISSI-Beijing. This project has received funding from the European Research Council (ERC) under the European Union's Horizon 2020 research and innovation programme (grant agreement No 724326).

\facilities{Kepler}

\appendix
\section{Influence of Kepler Input Catalogue}
In this appendix, we study the influence of stellar parameters taken from the Kepler Input Catalogue. To that end, we have taken the \texttt{q1\_q17\_dr25\_stellar} catalogue downloaded from \url{https://exoplanetarchive.ipac.caltech.edu/}, which contains the latest update to the stellar parameters.\par
With the new catalogue, the incidence rates of flare stars over the different spectral types are slightly altered. Most incidence rates are not too far from the earlier reported values, except for the incidence rate for A and F-stars. For the A-stars, the incidence rate seems to be nearly doubled, and the rate of the F-stars is decreased. The reason is that several flare stars which previously had an F-type have a slightly increased effective temperature in the new stellar parameter database. The new temperature crosses our (nearly arbitrarily chosen) threshold between F and A-stars of 7600K. In that sense, the numbers in Table~\ref{tab:update} give a feeling of the sensitivity of the incidence rates to the choice of temperature threshold between spectral types and perturbations in the catalogue temperatures. Probably a slightly adjusted threshold temperature would return the incidence rates to the earlier values. \par
\begin{figure}
	\includegraphics[width=\linewidth]{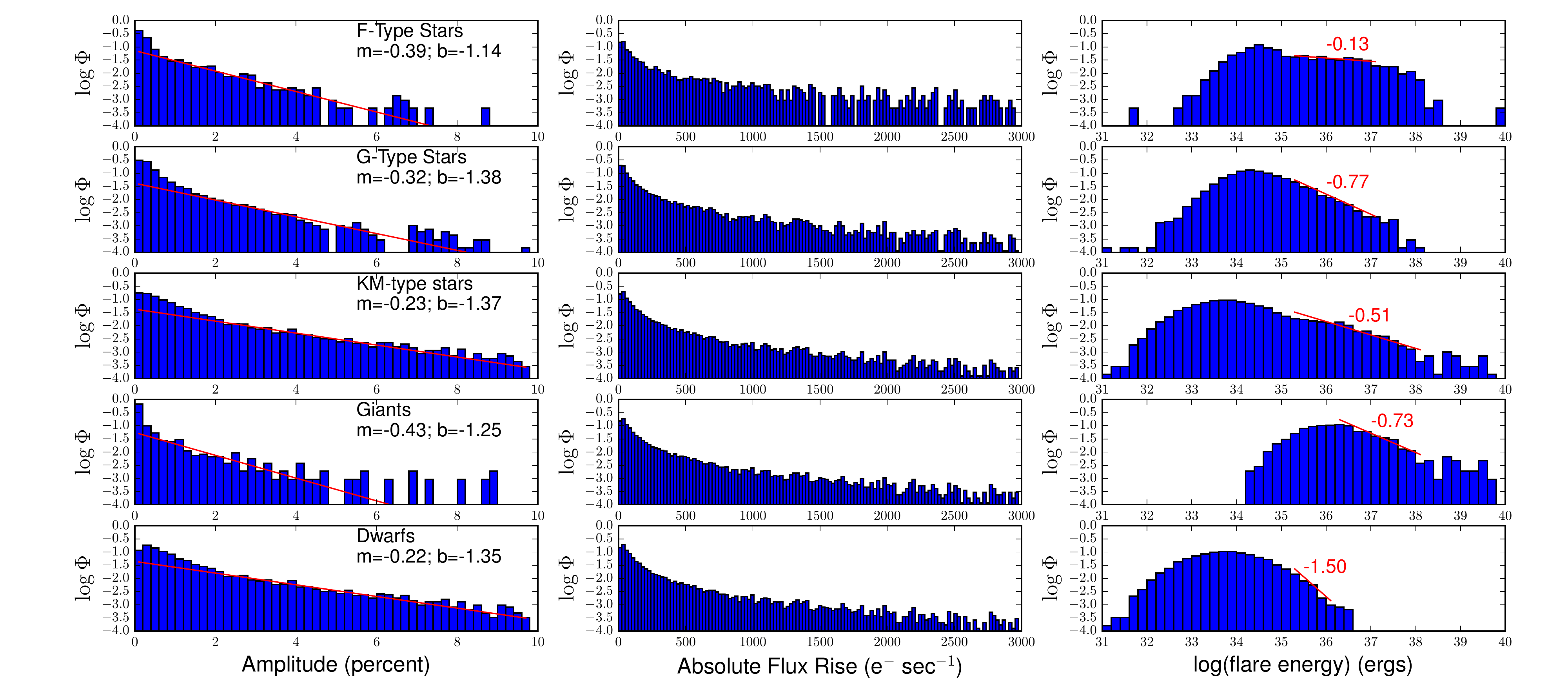}\\
	\centerline{
		\includegraphics[width=.5\linewidth]{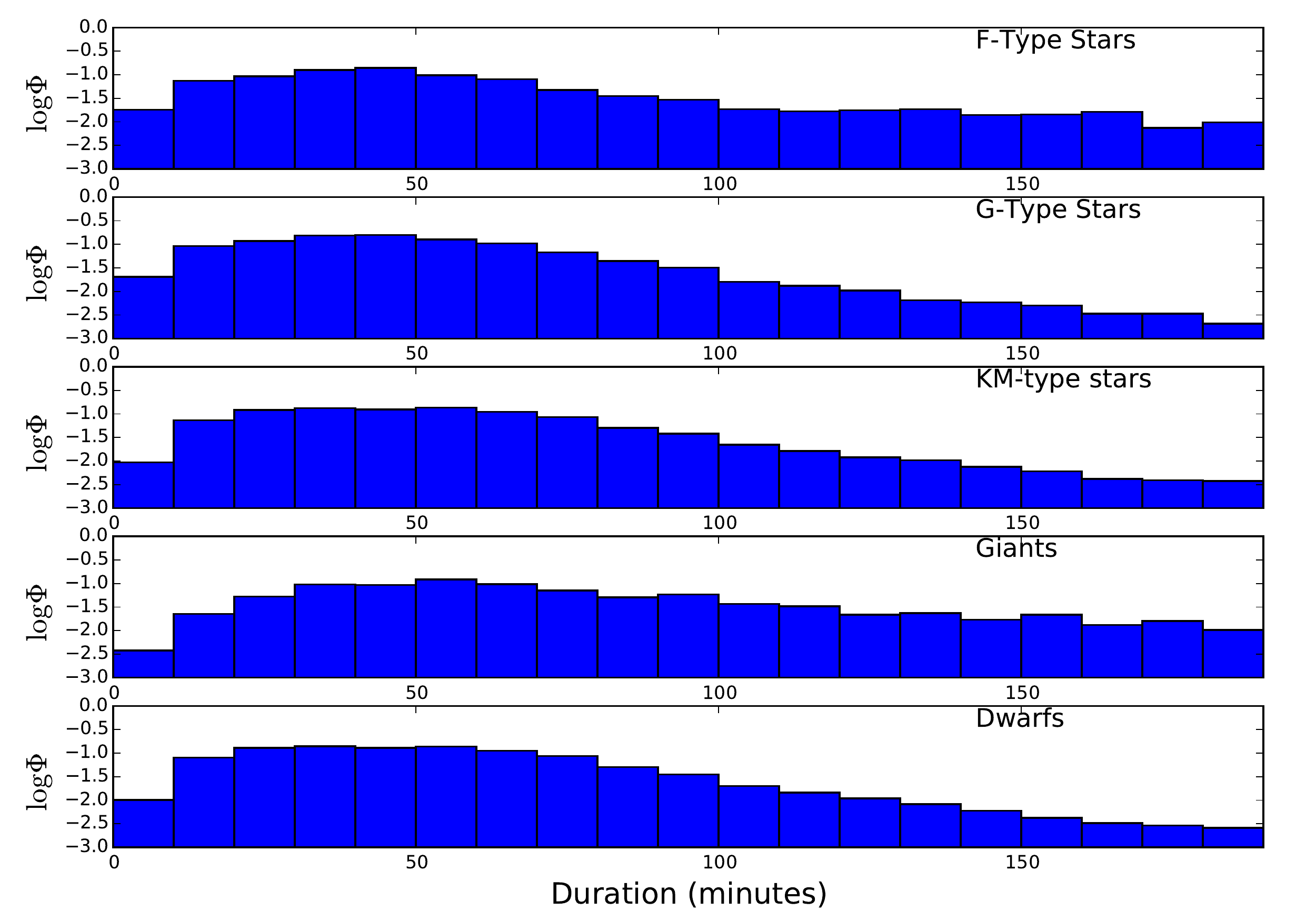}
		\includegraphics[width=.5\linewidth]{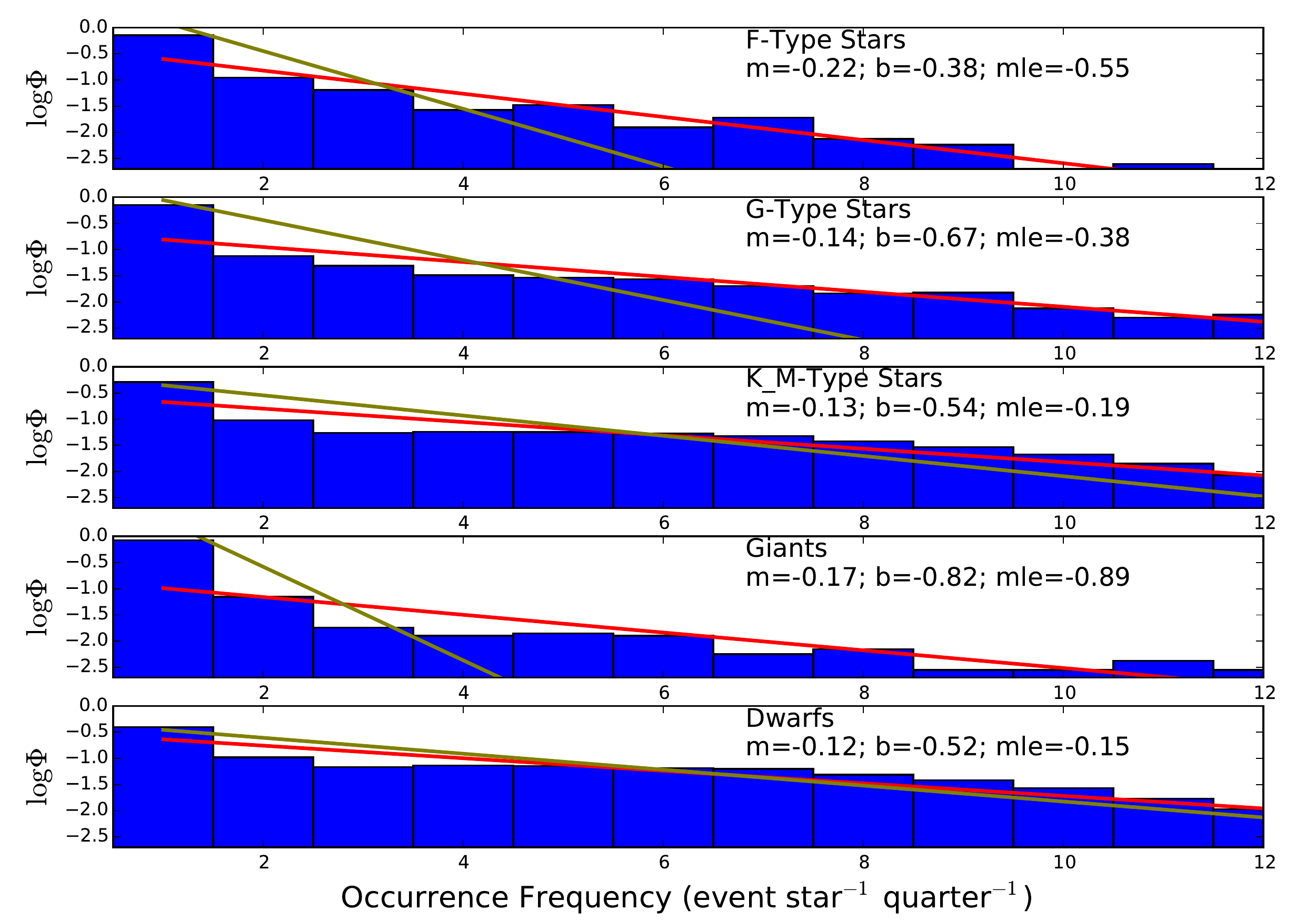}
	}
	\caption{The equivalent figures using updated stellar parameters using the \texttt{q1\_q17\_dr25\_stellar} catalogue. The top row is the same as Fig.~\ref{fig:amplitude}, bottom left is equivalent to Fig.~\ref{fig:duration} and the bottom right to Fig.~\ref{fig:occur}.}
	\label{fig:update}
\end{figure}

\begin{table}
	\caption{New incidence rates using updated stellar parameters. The table is equivalent to Table~\ref{tab:overview}.}
	\label{tab:update}
	\begin{tabular}{llll}
		\hline
		Stellar type & \# objects & \# flare stars & Incidence\\
		\hline \hline
		A+B & 2861 & 70 & 2.45\% \\
		F & 51916 & 1230 & 2.37\% \\
		G & 108463 & 3207 & 2.96\%\\
		K+M & 36583 & 2146 & 5.87\%\\
		giants & 21845 & 695 & 3.18\%\\
		\hline
	\end{tabular}
\end{table}
We have also used the updated stellar parameter catalogue to study the flare parameters. The updated versions of Figs.~\ref{fig:amplitude}-\ref{fig:occur} are displayed in Fig.~\ref{fig:update}.
From the comparison of the updated figures with Fig.~\ref{fig:amplitude}, it seems that the power law slope of the amplitude distribution is not changed, except for the giants. However, the slopes of the flare energy distribution are very sensitive to the catalogue parameters, and therefore, they are probably not so robust. The flare duration appears to be insensitive to the update, and so does the occurrence frequency.

\section{Flare light curves for A stars}
\label{sec:aflares}
In this appendix, we show the flare light curves for all detected flares on A stars, together with their pixel mask. In each of the figures, we show the intensity time series in the left panel. The time of the flare is indicated with a vertical dotted line. The middle panel shows the pixel data for the flare. The right panel shows the pixel data for the star. The latter is constructed as the average of the pixel intensities between 2 and 5 data points before the flare peak time. The data in the middle panel is the difference of the pixel intensity at the flare peak time, minus the mean pixel intensity shown in the right panel. In both the middle and right panel, the intensity is masked to only show intensity that was added to obtain the intensity time series in the left panel. \par
Some of the light curves of the A-stars show periodic modulation of the light curve (see e.g. Fig.~\ref{fig:7523115} or \ref{fig:10149211}). However, it is unclear whether these variations are due to rotational modulation due to stellar spots or periodic modulation due to stellar pulsations.
\begin{figure}
\includegraphics[width=\linewidth]{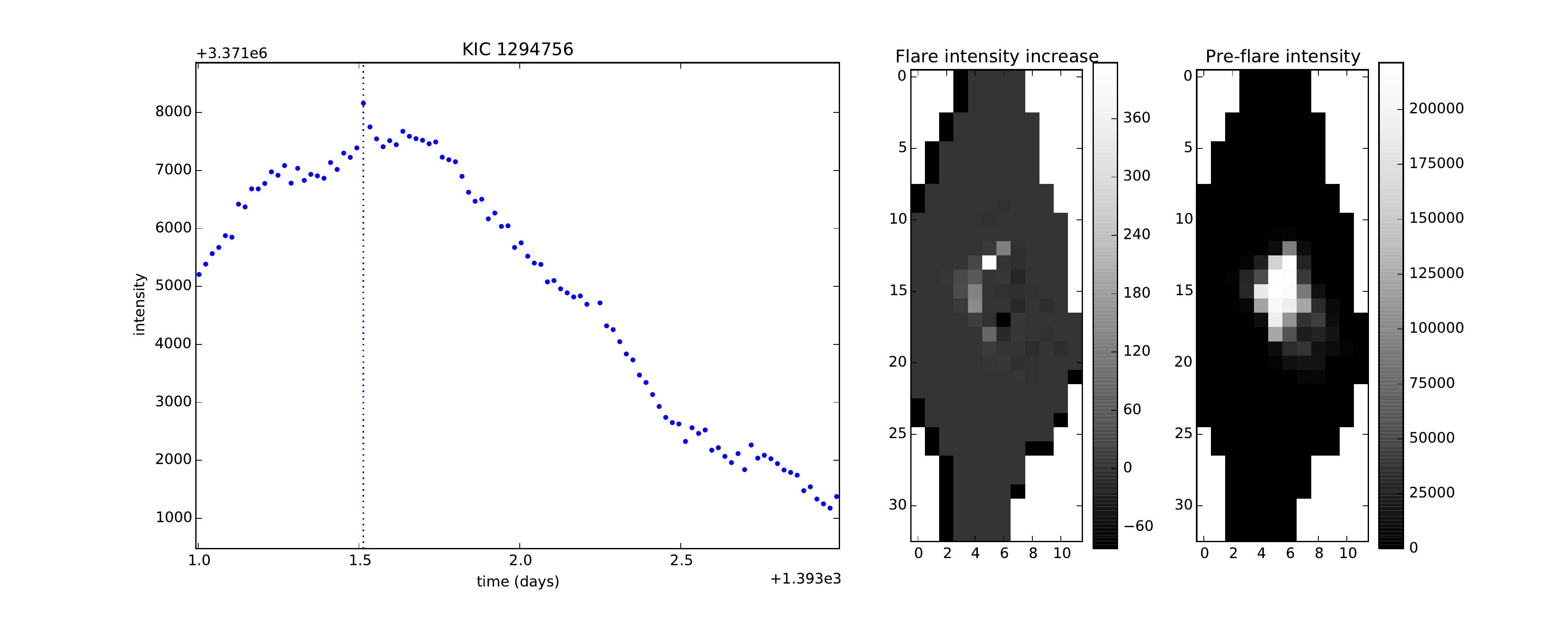}\\
\includegraphics[width=\linewidth]{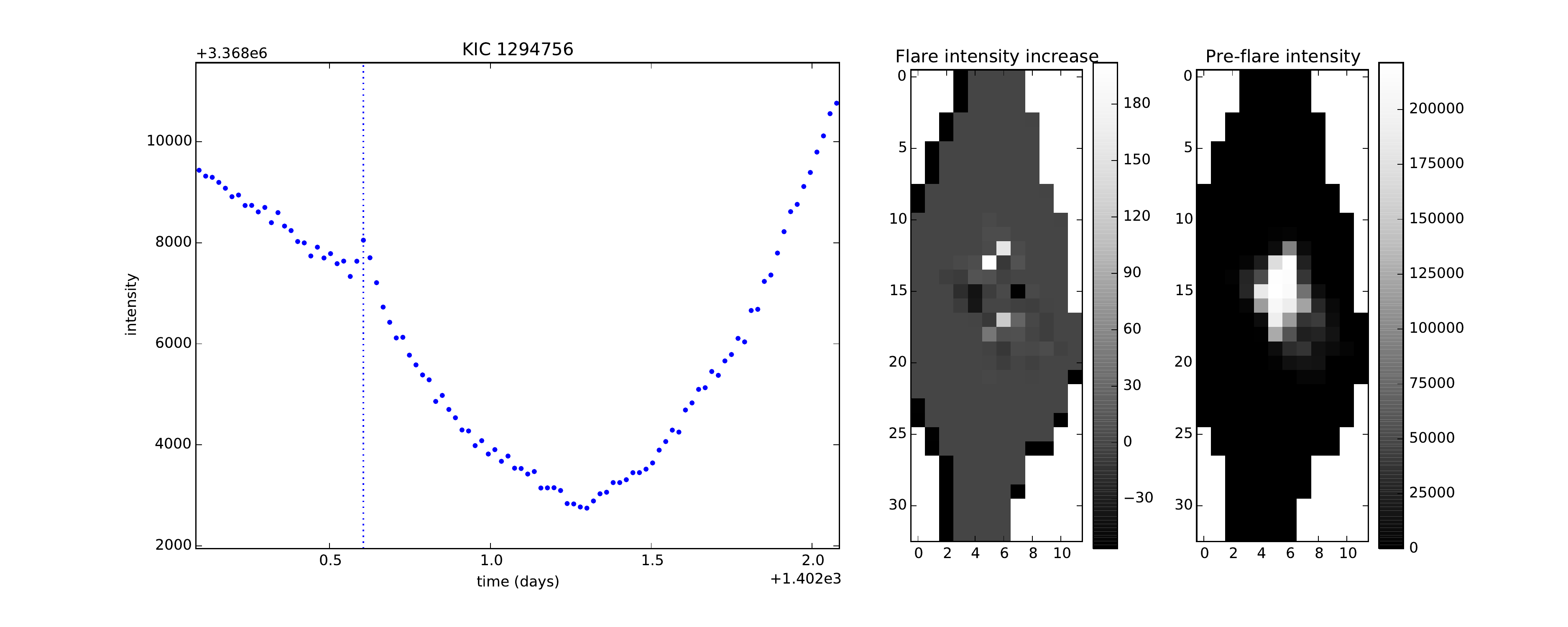}\\
\includegraphics[width=\linewidth]{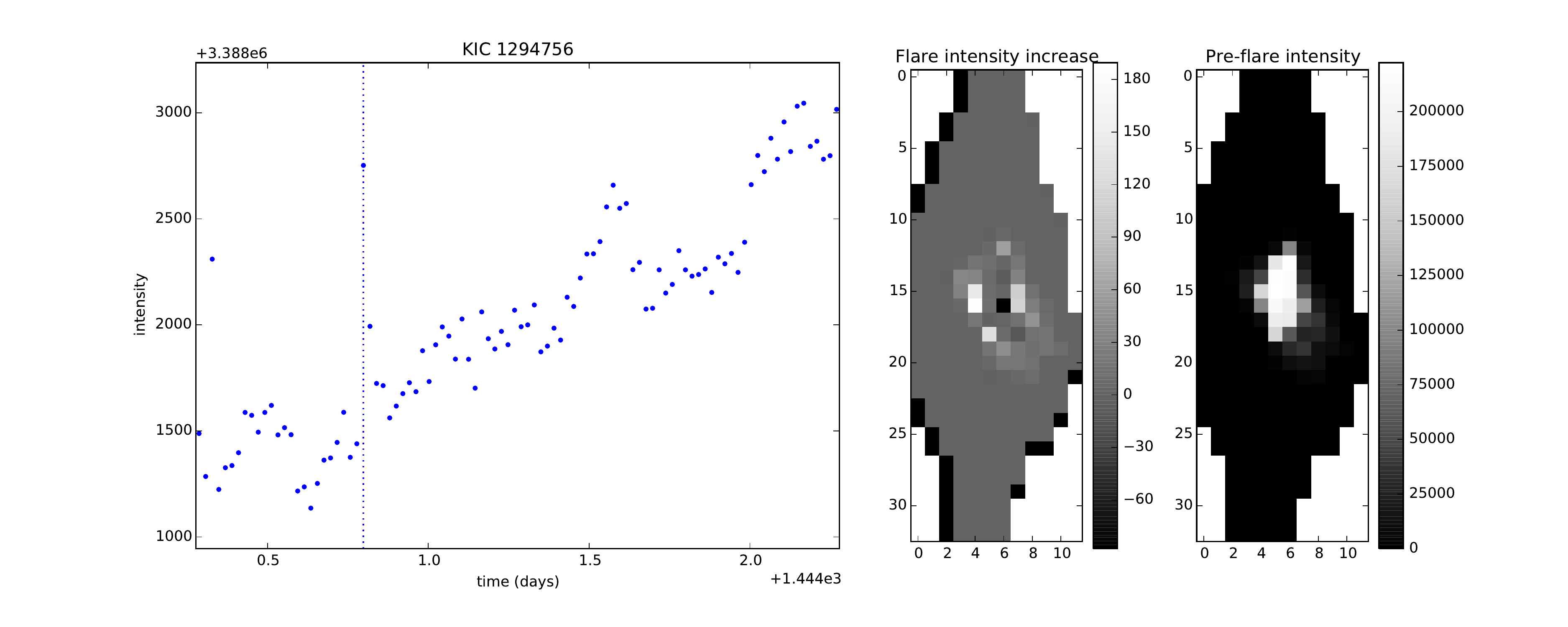}\\
\caption{Flare light curves for KIC 1294756.}
\end{figure}

\begin{figure}
\includegraphics[width=\linewidth]{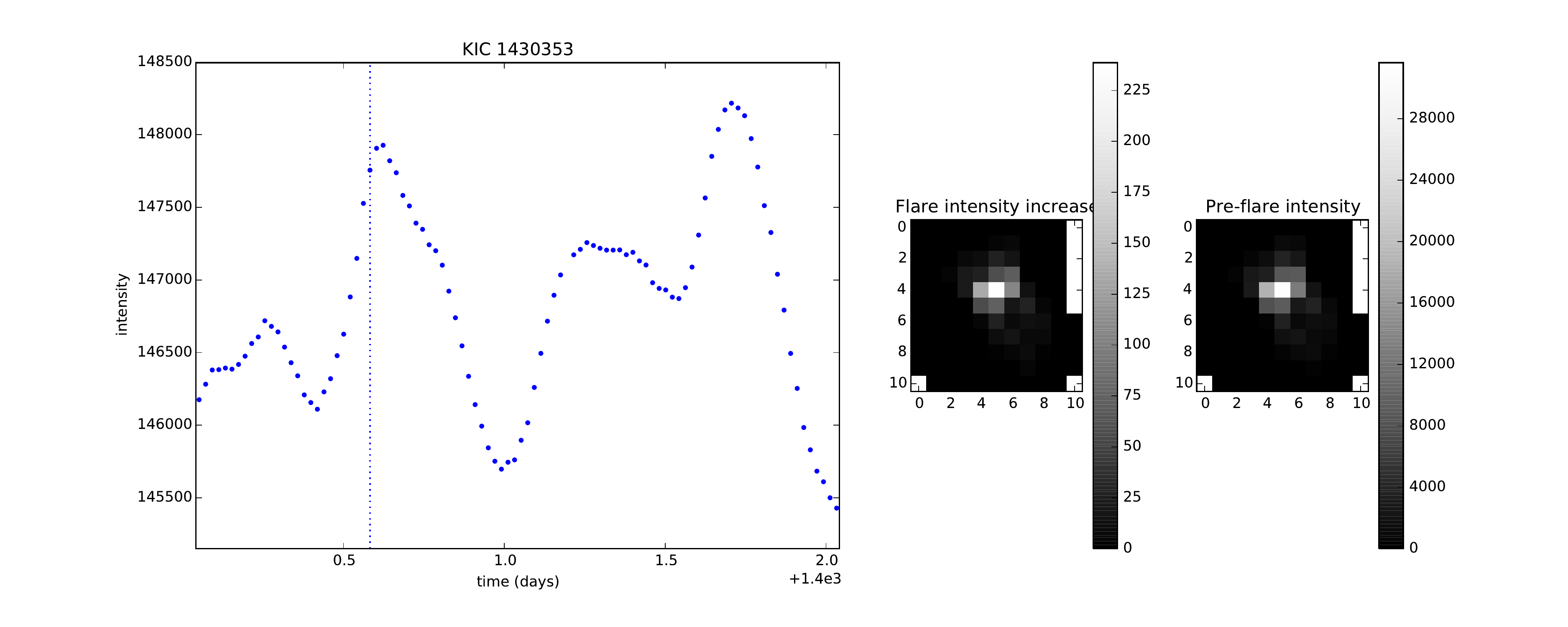}\\
\caption{Flare light curves for KIC 1430353.}
\label{fig:1430353}
\end{figure}

\begin{figure}
\includegraphics[width=\linewidth]{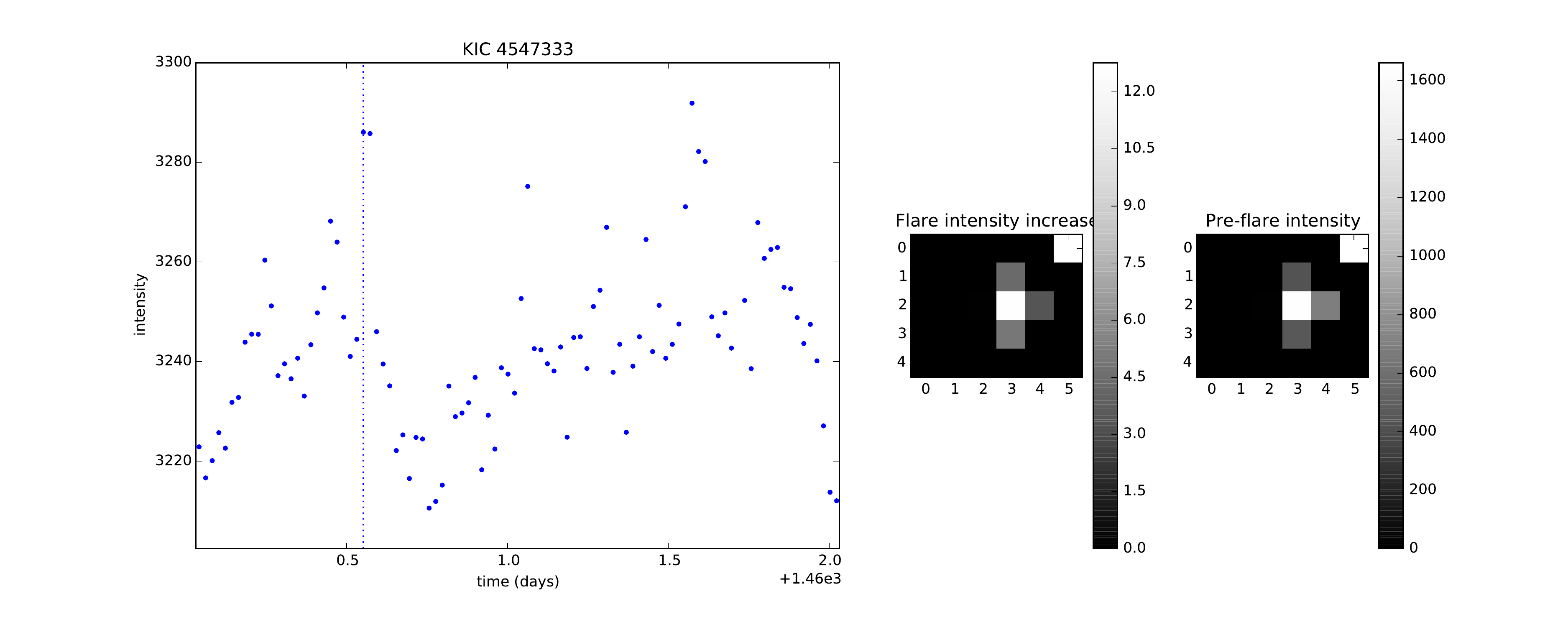}\\
\caption{Flare light curves for KIC 4547333.}
\label{fig:4547333}
\end{figure}

\begin{figure}
\includegraphics[width=\linewidth]{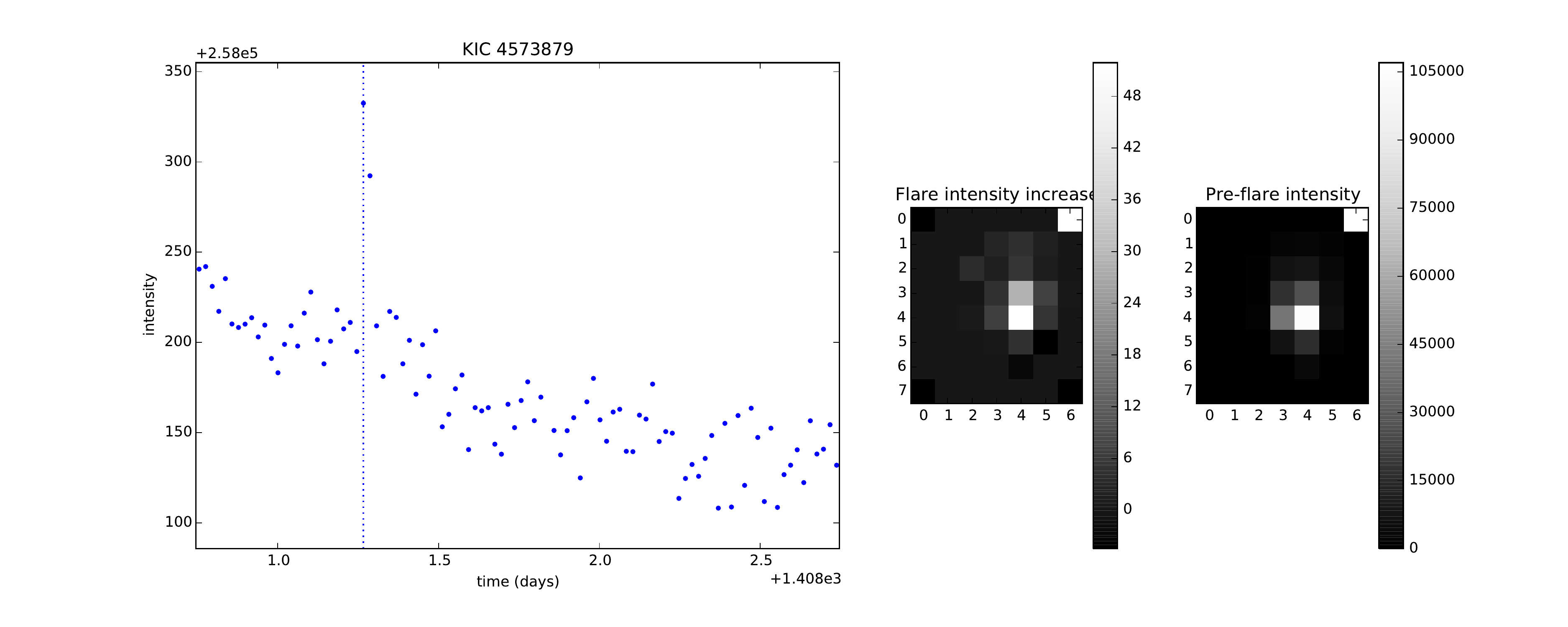}\\
\caption{Flare light curves for KIC 4573879.}
\end{figure}

\begin{figure}
\includegraphics[width=\linewidth]{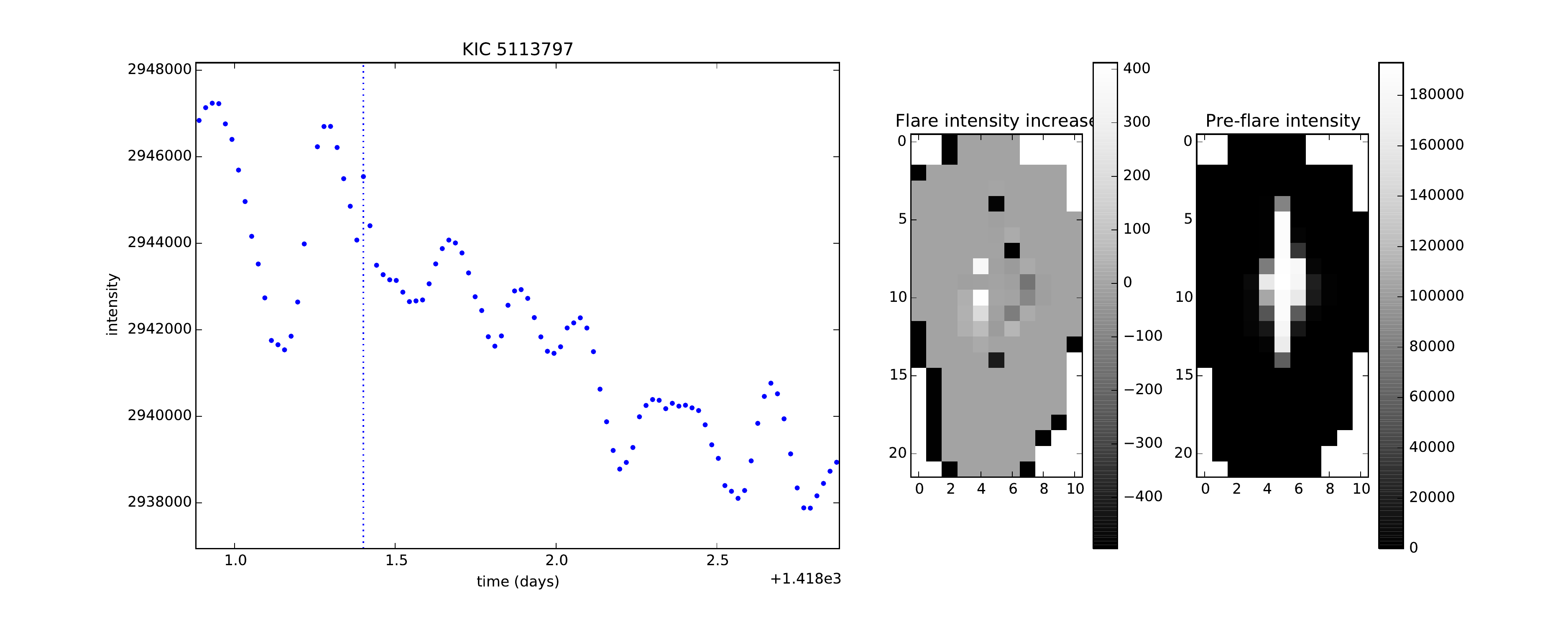}\\
\caption{Flare light curves for KIC 5113797.}
\end{figure}

\begin{figure}
\includegraphics[width=\linewidth]{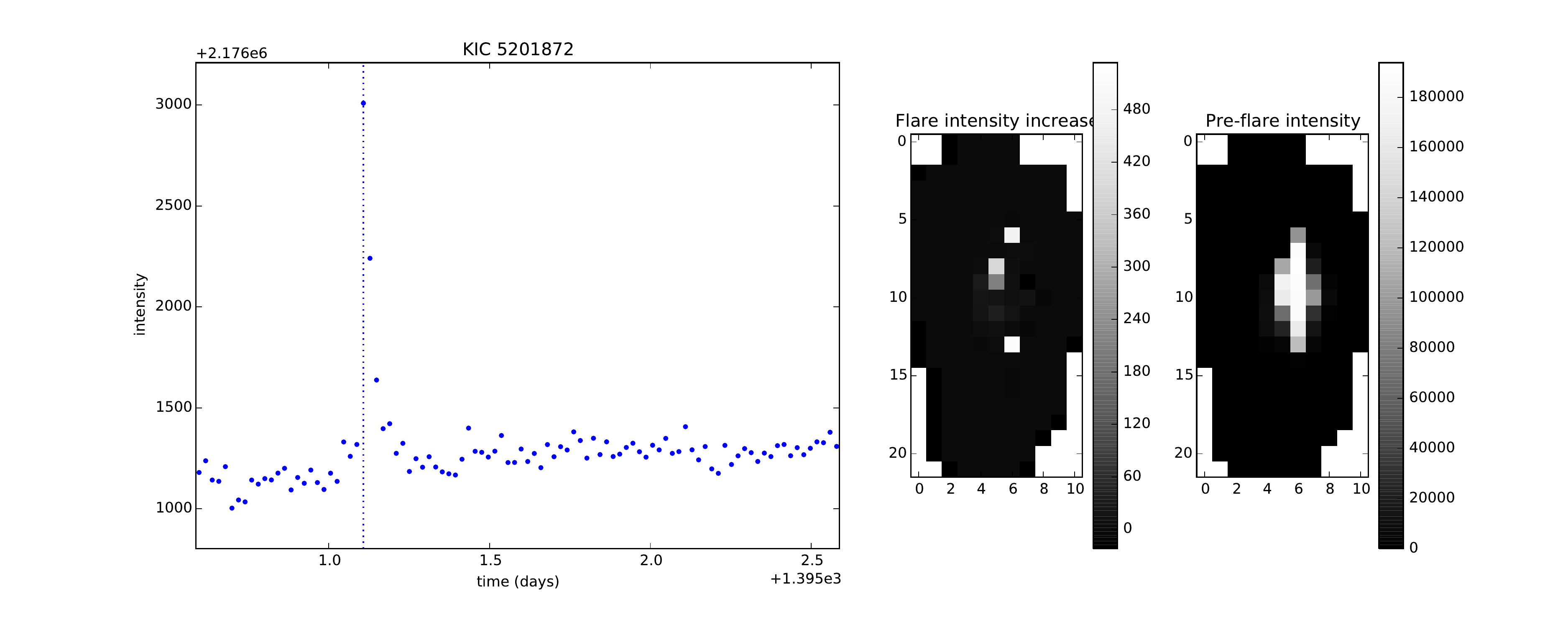}\\
\includegraphics[width=\linewidth]{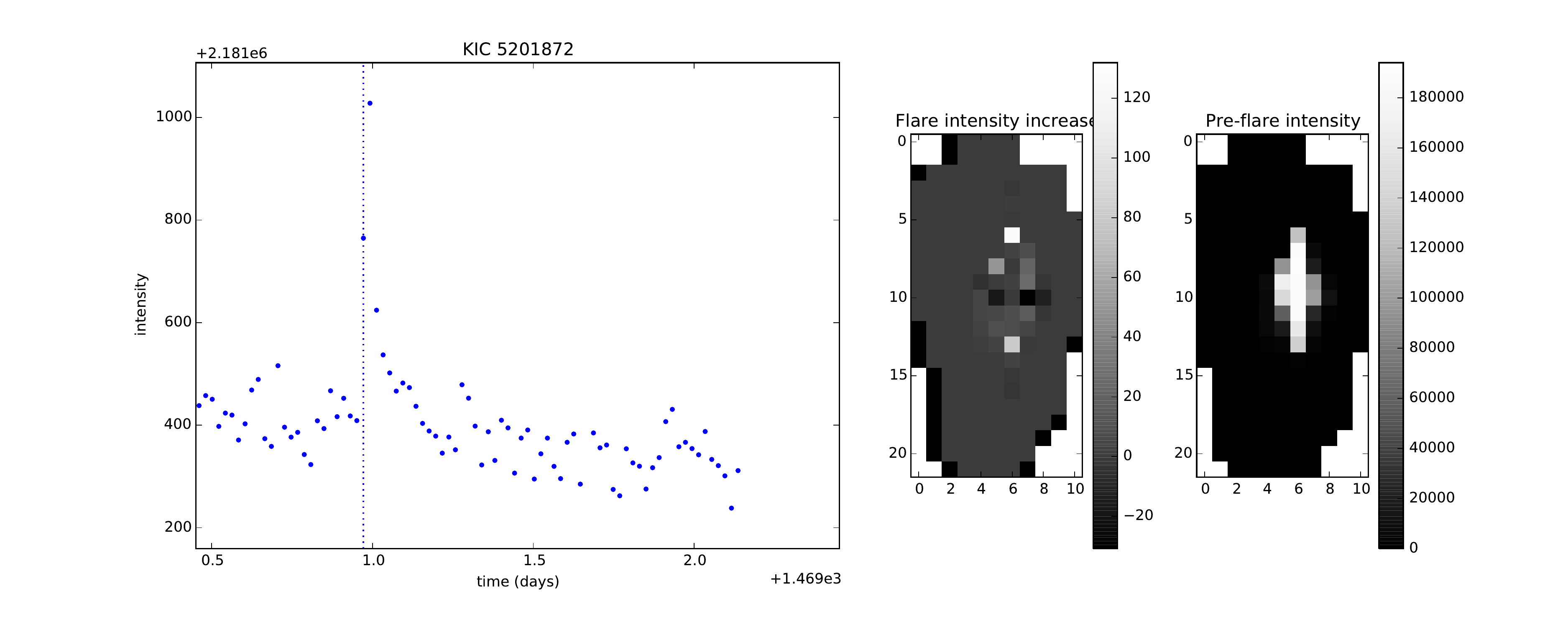}\\
\caption{Flare light curves for KIC 5201872.}
\end{figure}

\begin{figure}
\includegraphics[width=\linewidth]{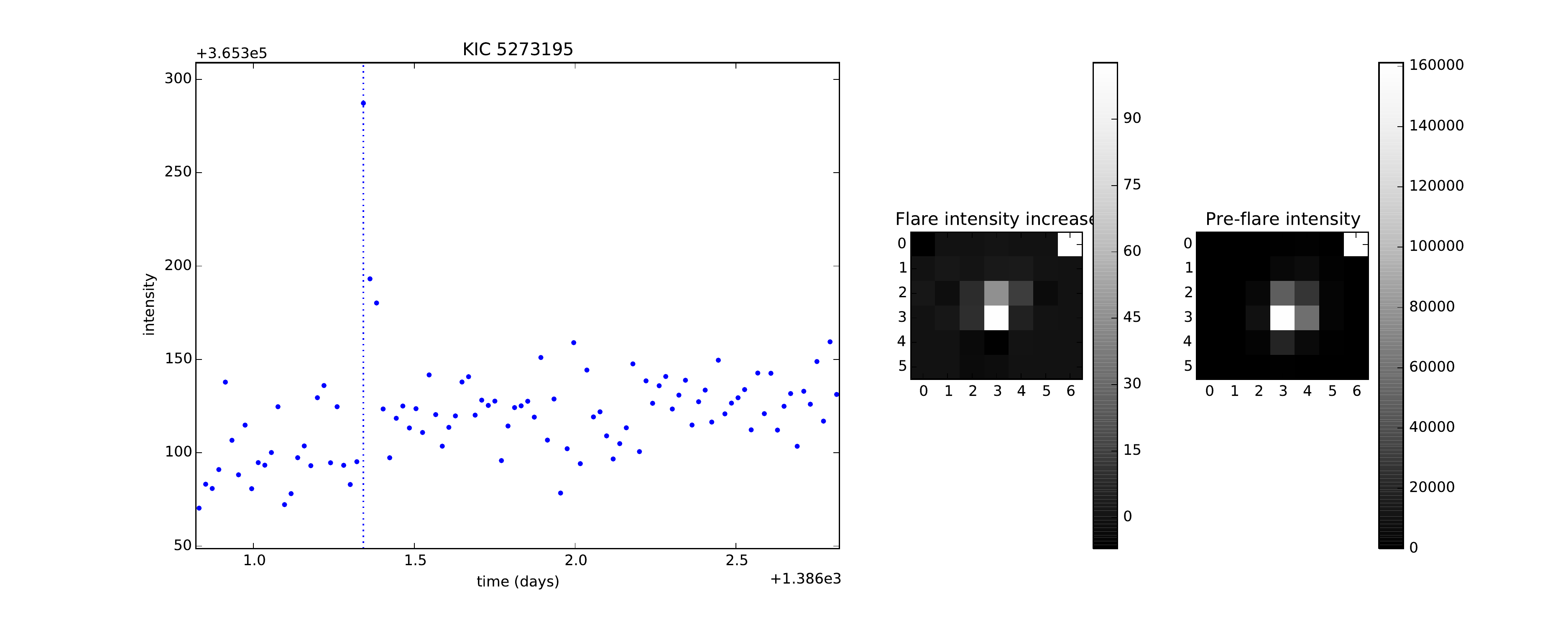}\\
\includegraphics[width=\linewidth]{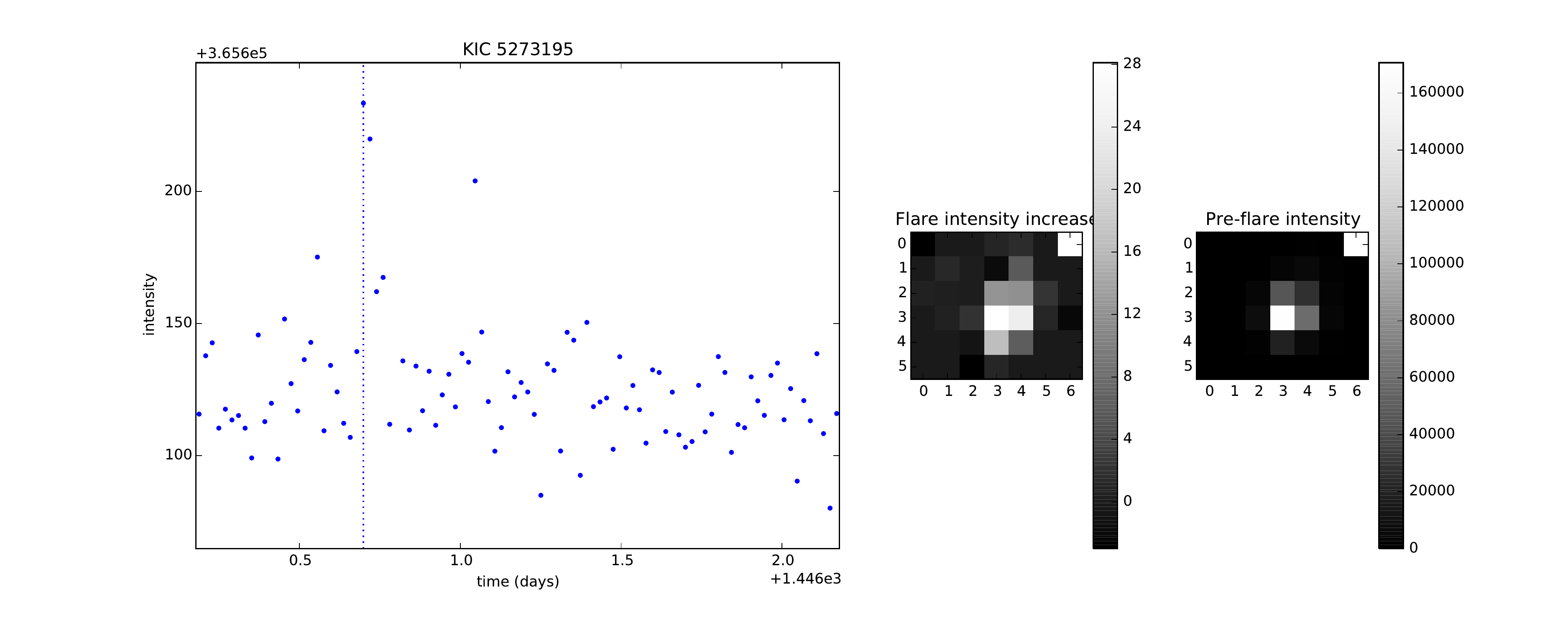}\\
\caption{Flare light curves for KIC 5273195.}
\end{figure}

\begin{figure}
\includegraphics[width=\linewidth]{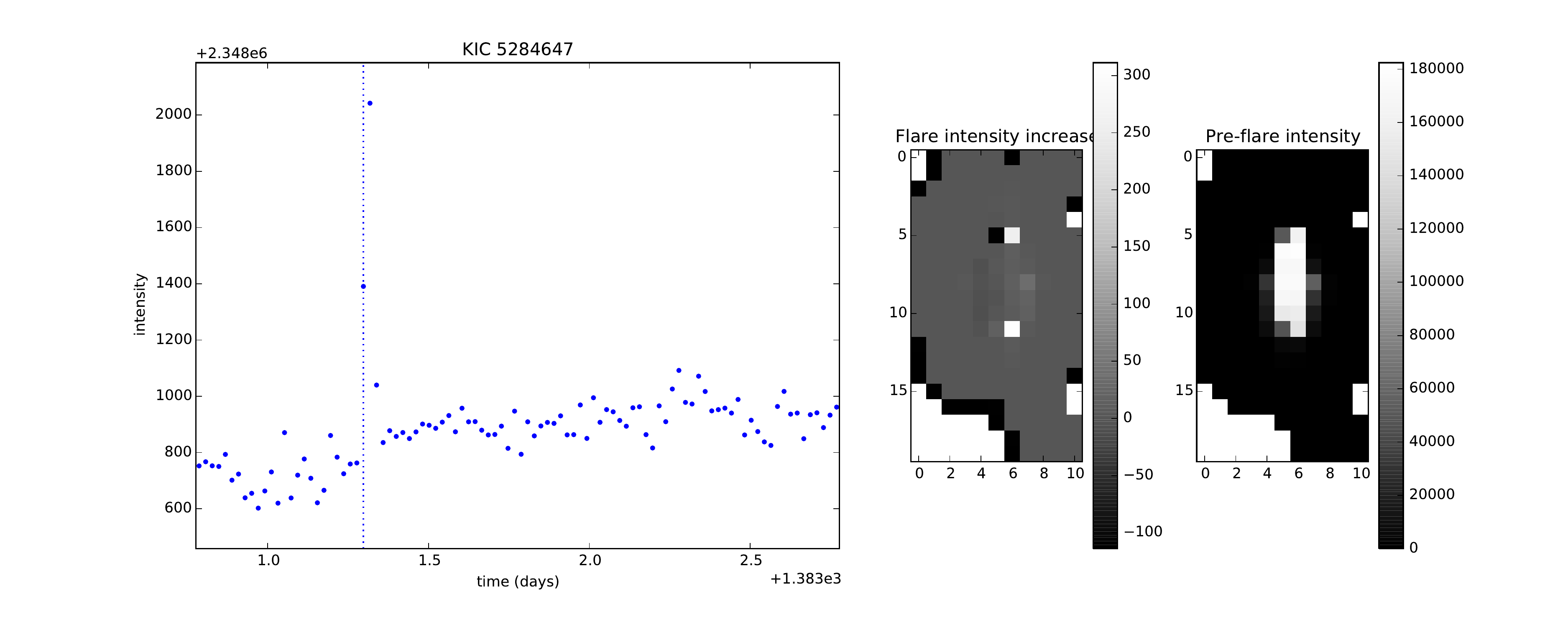}\\
\includegraphics[width=\linewidth]{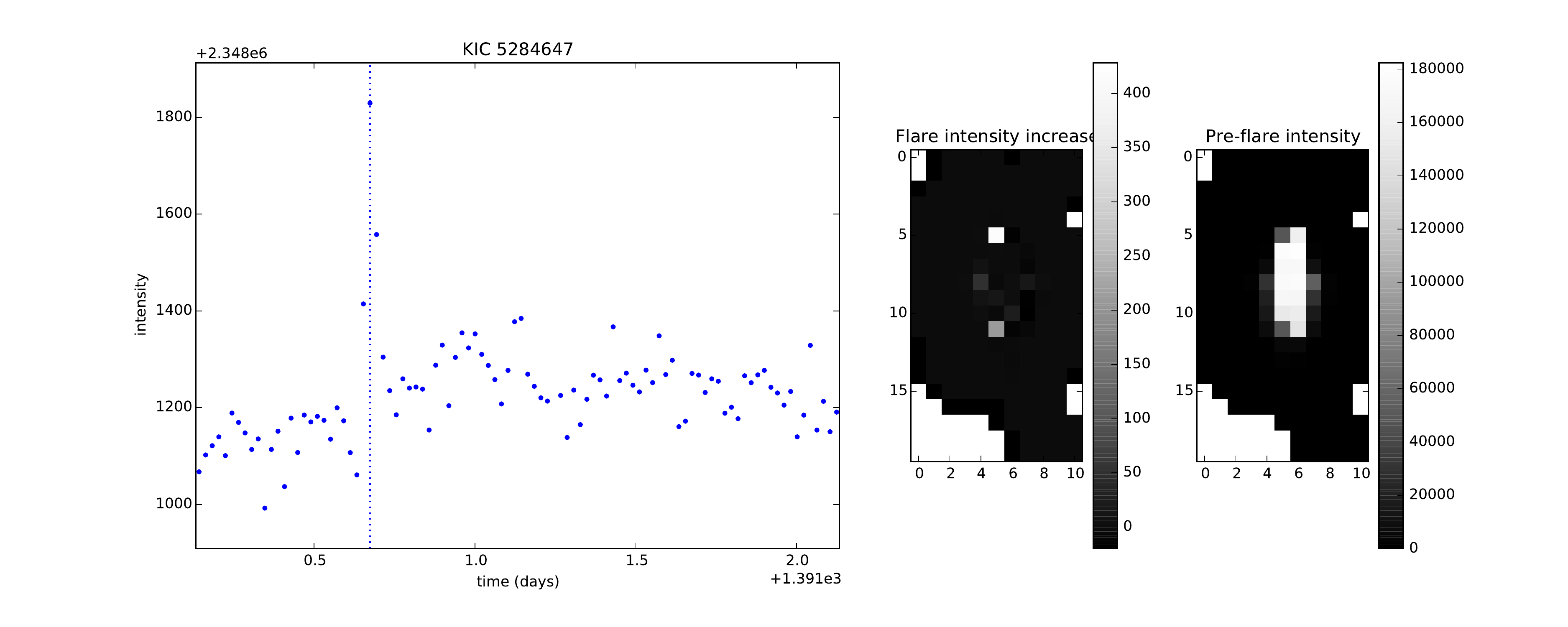}\\
\caption{Flare light curves for KIC 5284647.}
\end{figure}

\begin{figure}
\includegraphics[width=\linewidth]{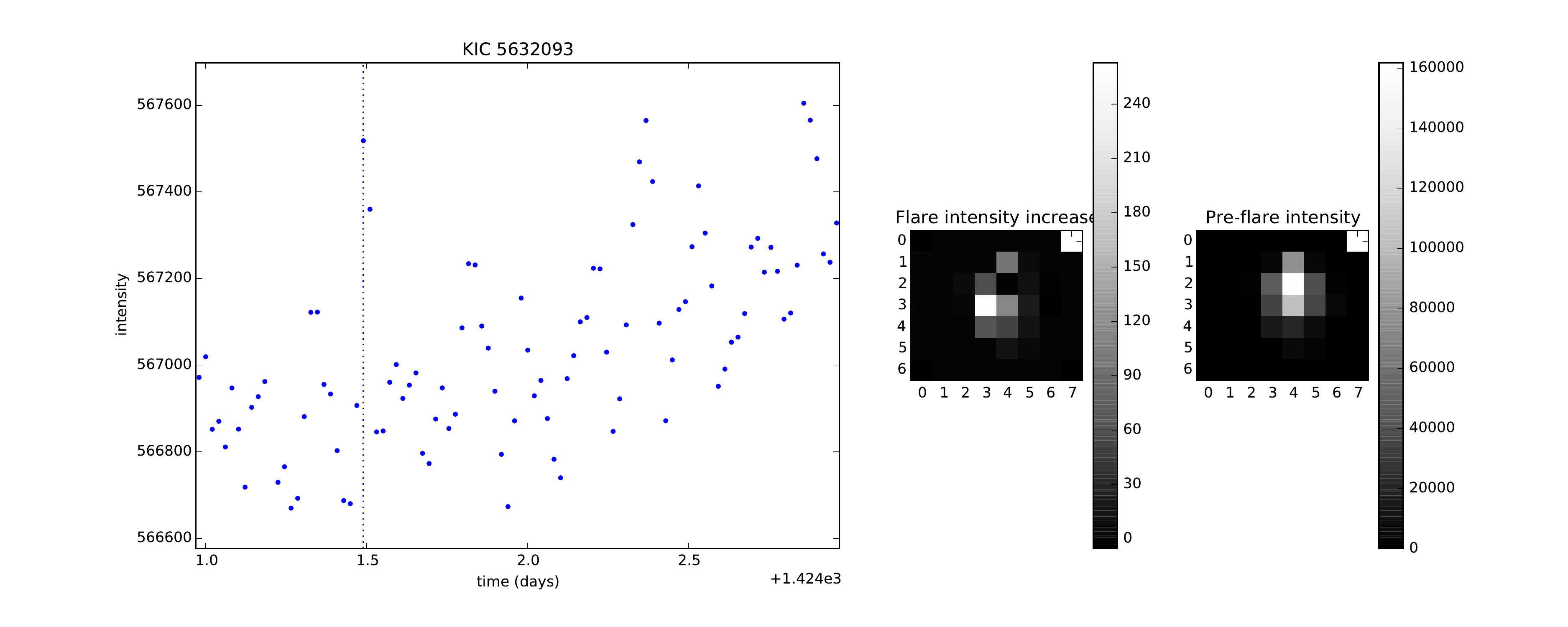}\\
\caption{Flare light curves for KIC 5632093.}
\end{figure}

\begin{figure}
\includegraphics[width=\linewidth]{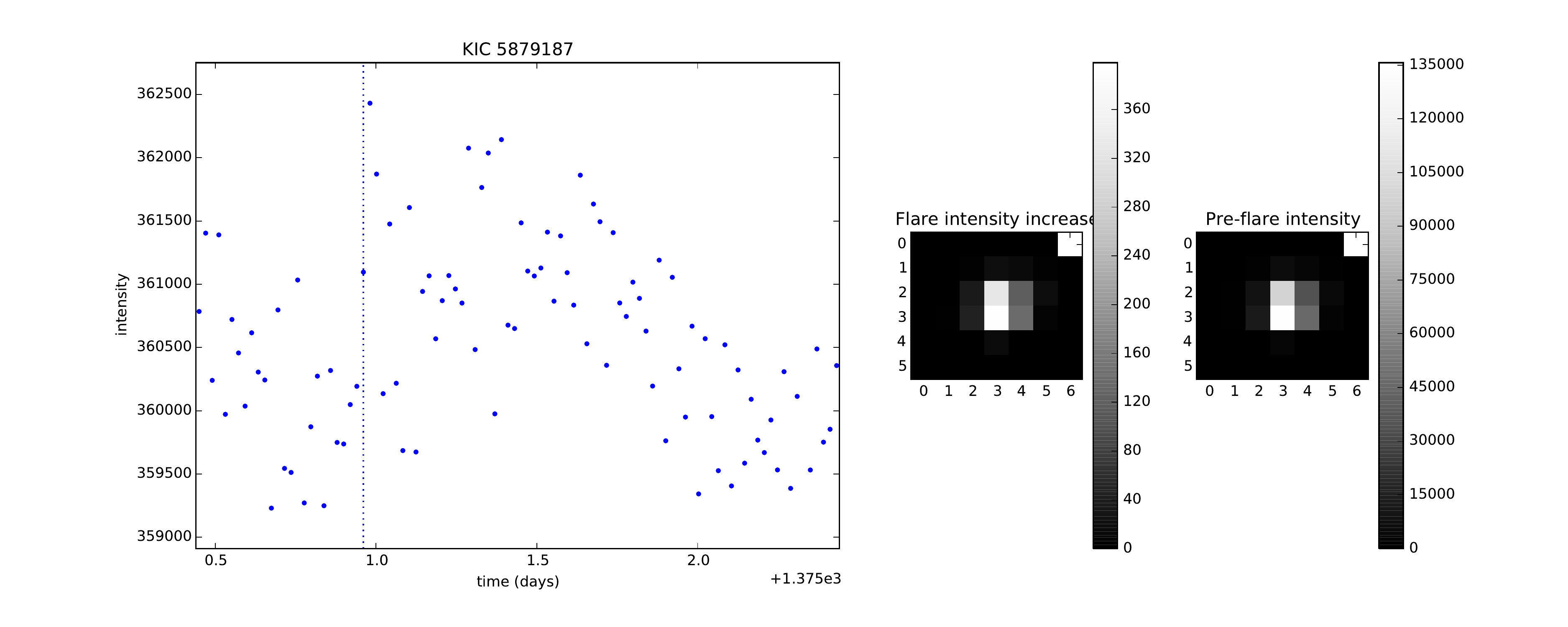}\\
\includegraphics[width=\linewidth]{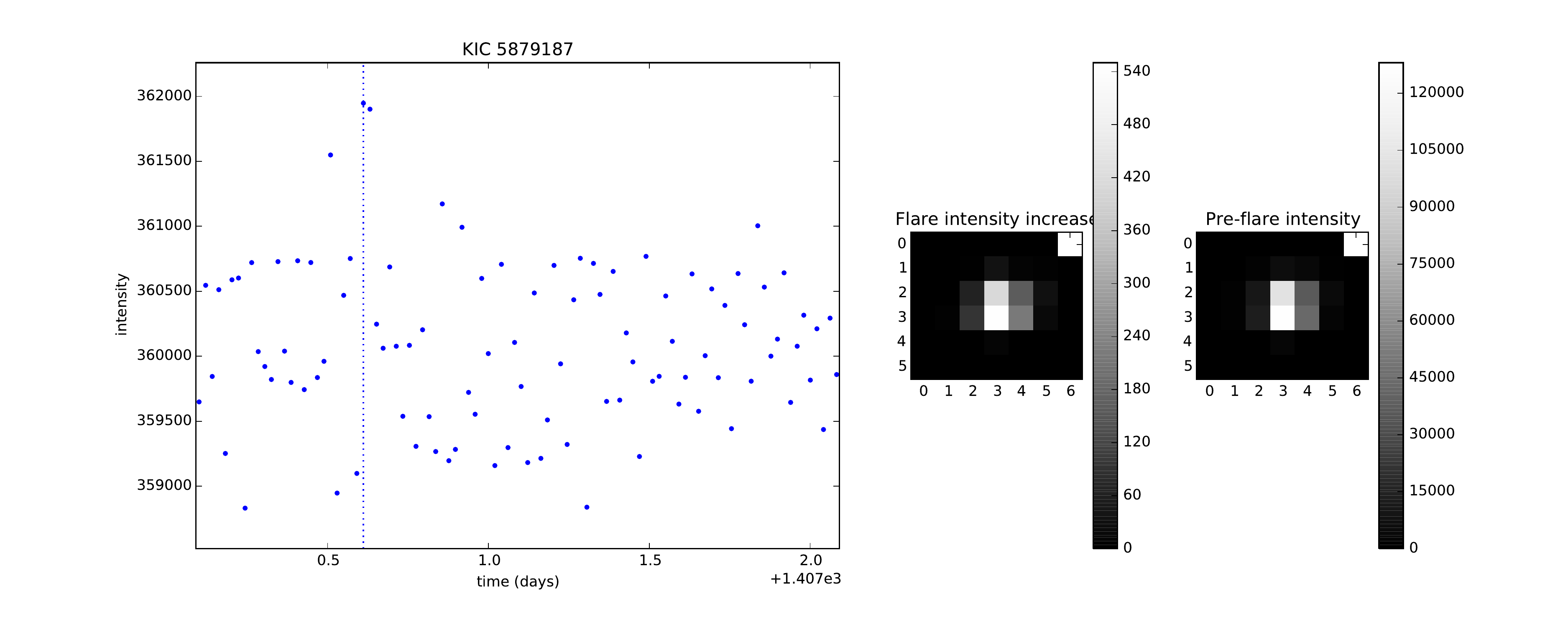}\\
\includegraphics[width=\linewidth]{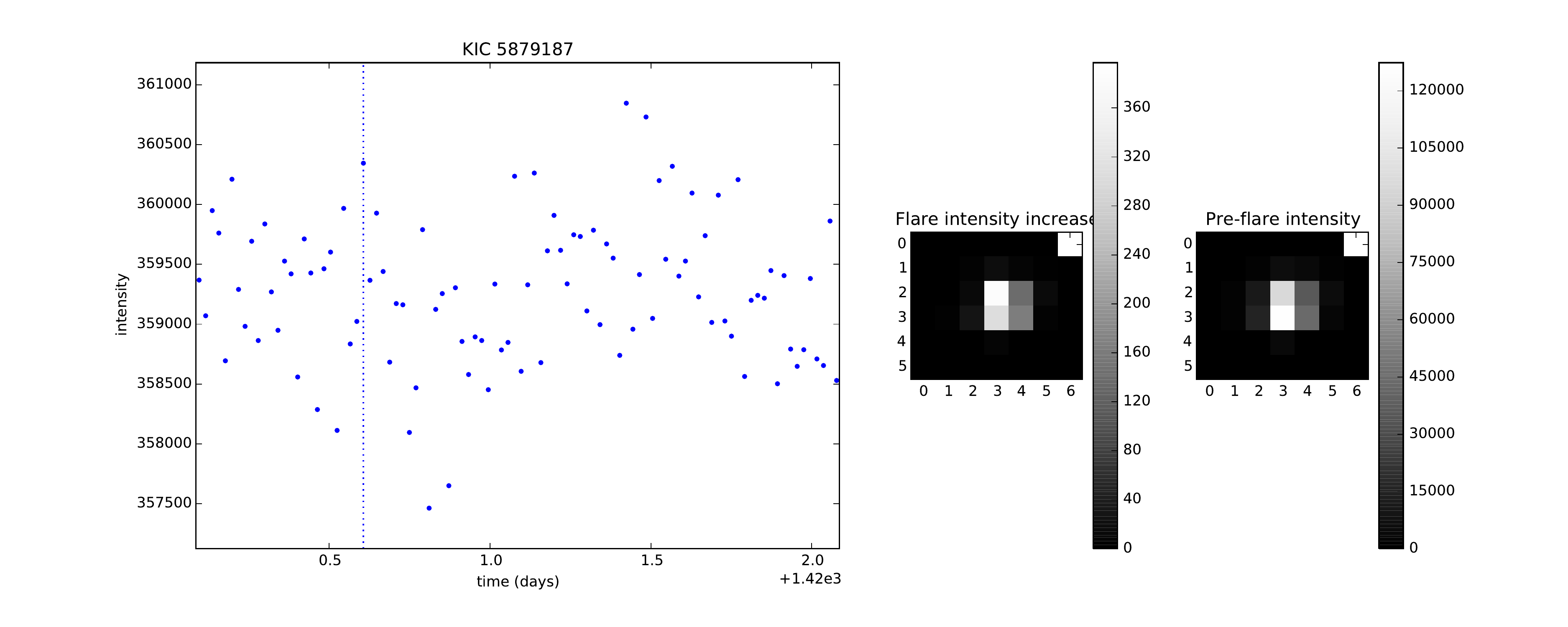}
\caption{Flare light curves for KIC 5879187.}
\label{fig:5879187a}
\end{figure}

\begin{figure}
\includegraphics[width=\linewidth]{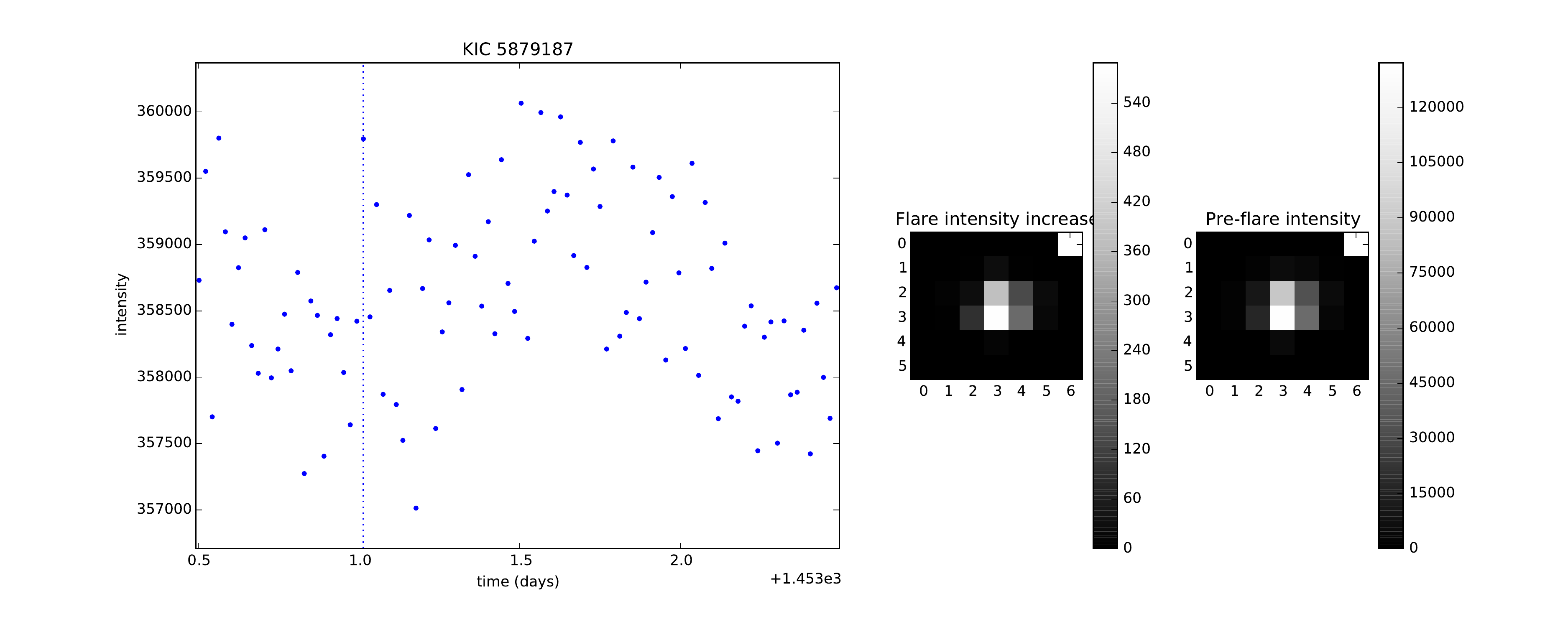}\\
\includegraphics[width=\linewidth]{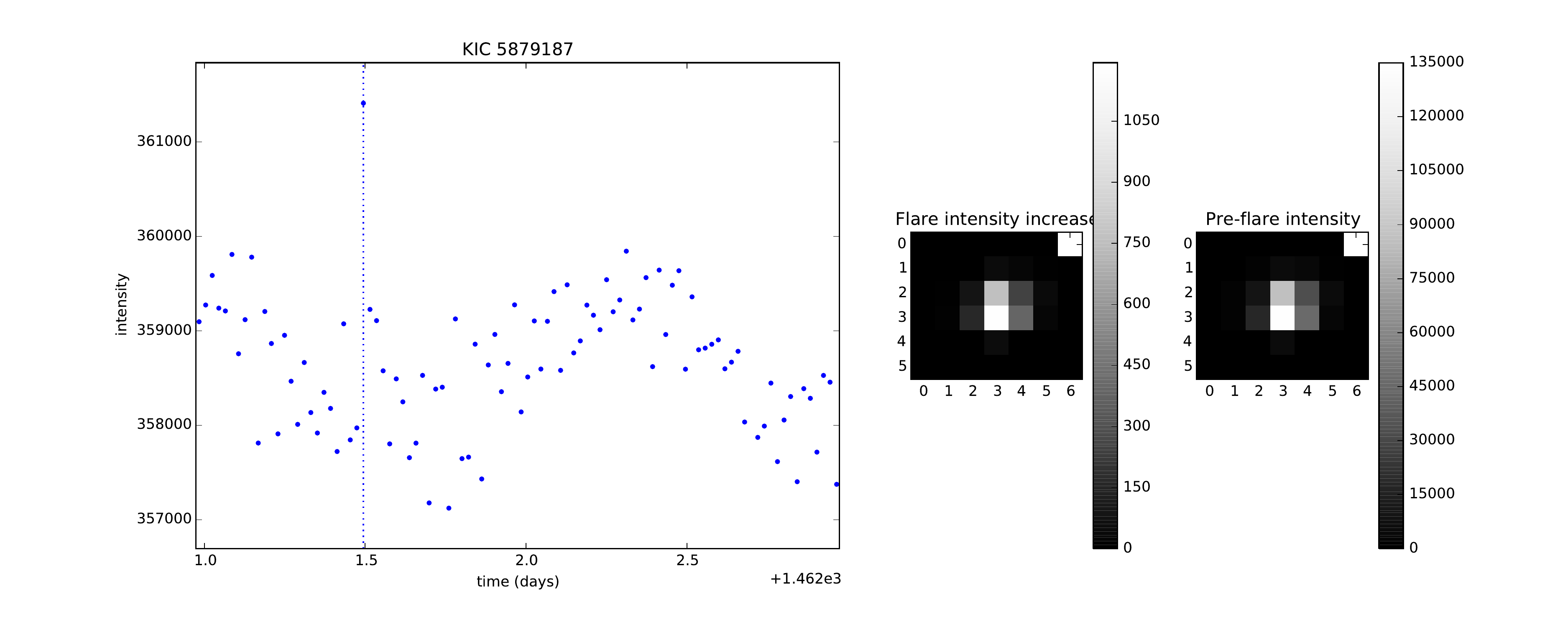}\\
\caption{Flare light curves for KIC 5879187 (continued).}
\label{fig:5879187b}
\end{figure}

\begin{figure}
\includegraphics[width=\linewidth]{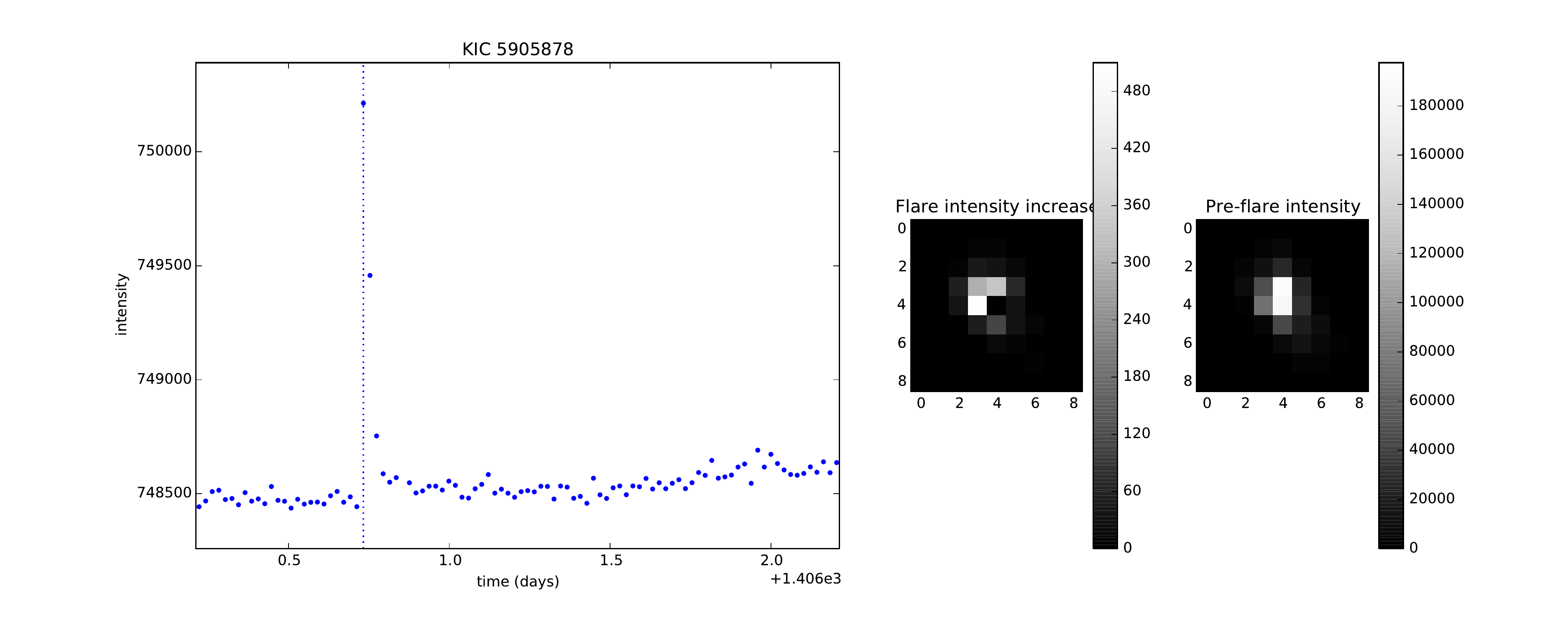}\\
\includegraphics[width=\linewidth]{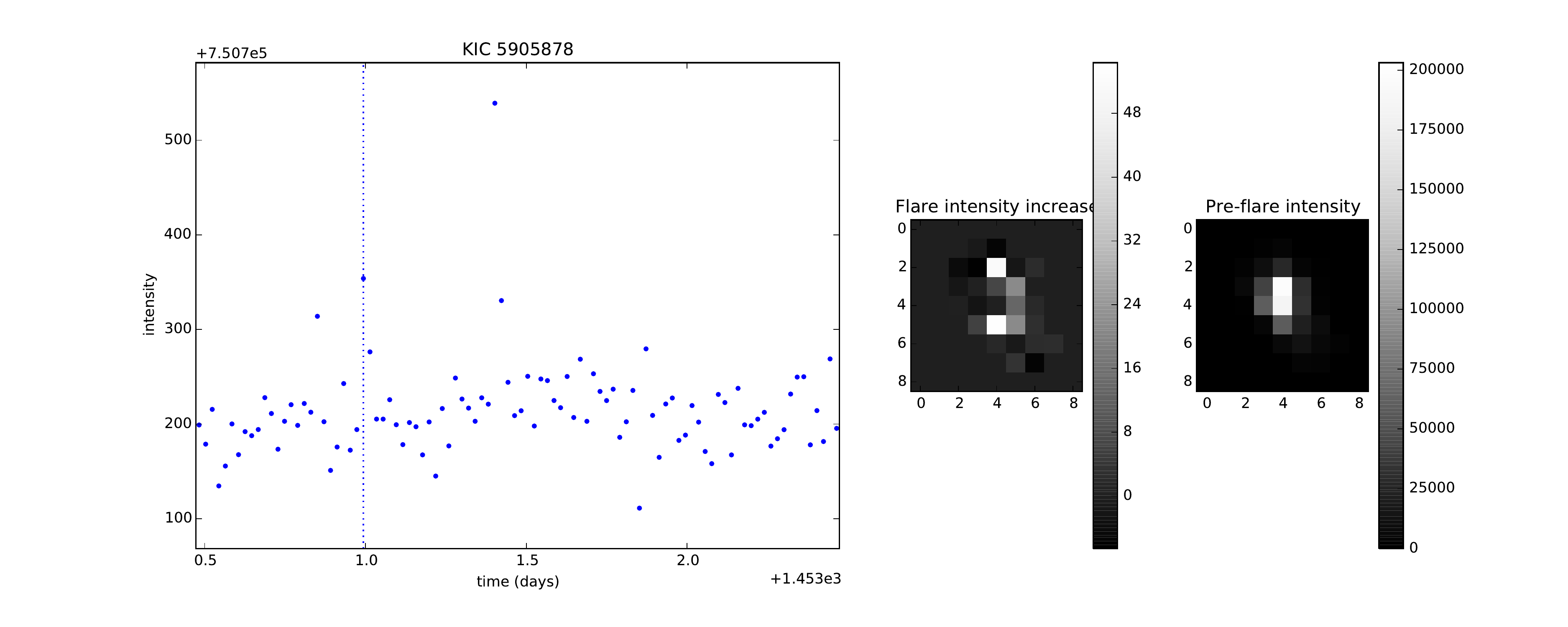}\\
\caption{Flare light curves for KIC 5905878.}
\end{figure}

\begin{figure}
\includegraphics[width=\linewidth]{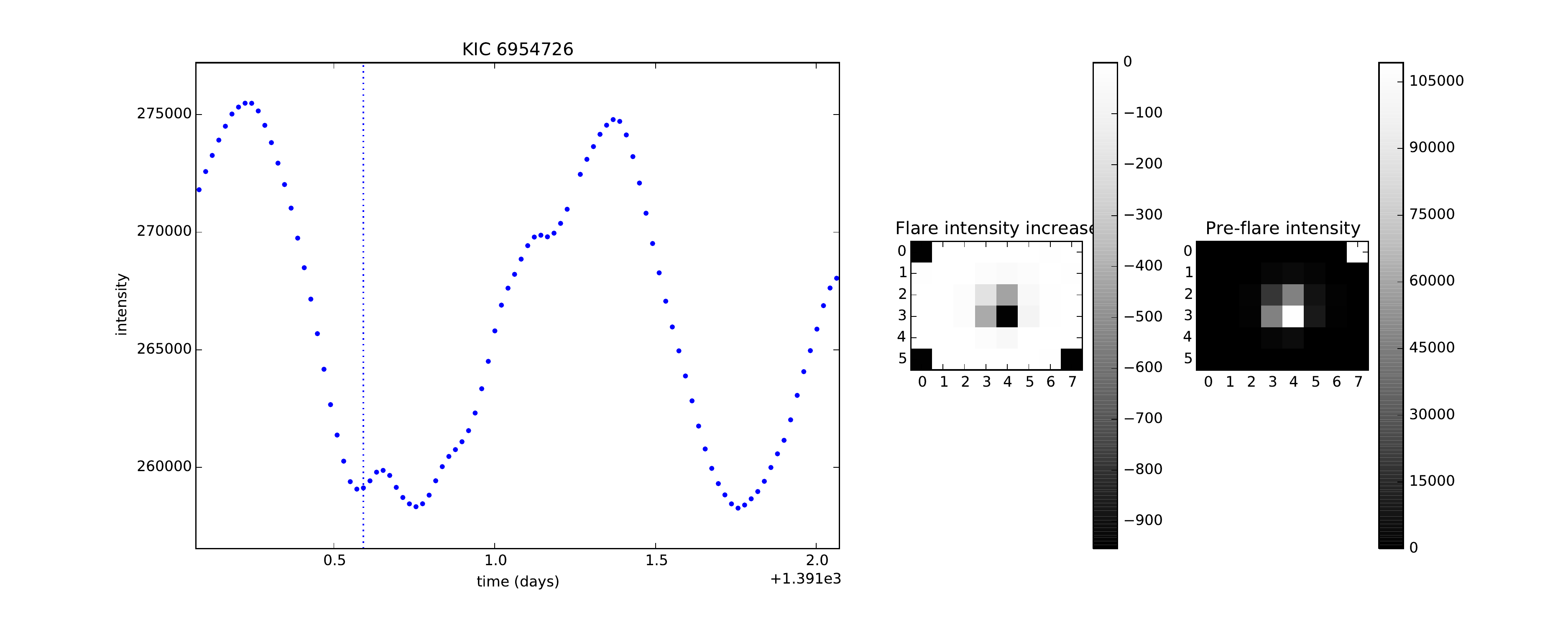}\\
\includegraphics[width=\linewidth]{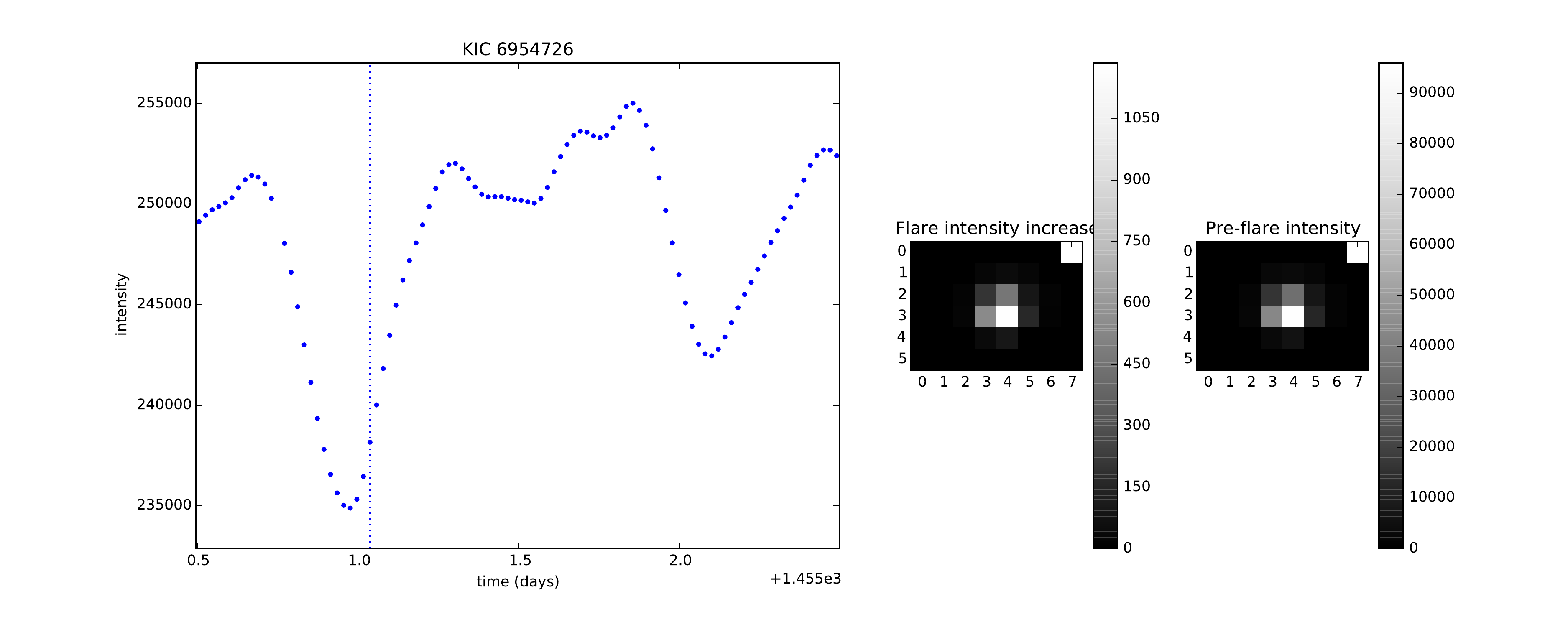}\\
\includegraphics[width=\linewidth]{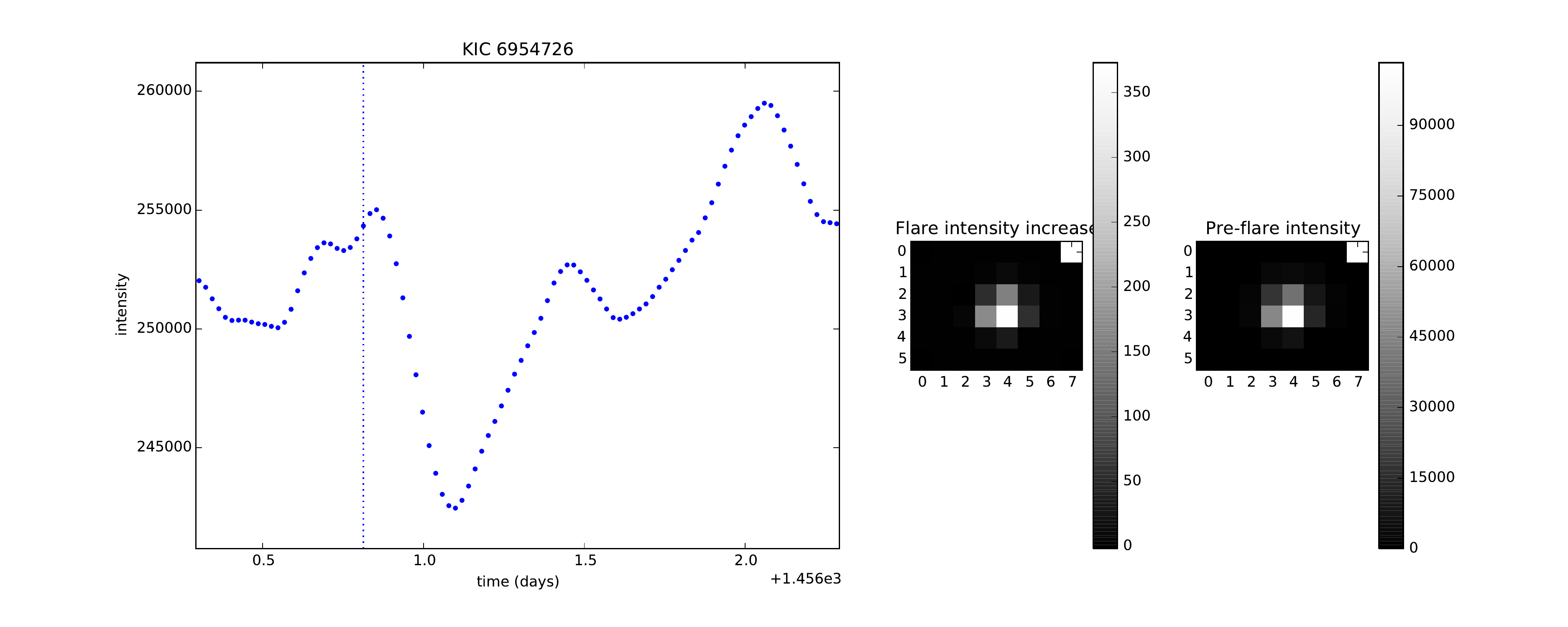}\\
\caption{Flare light curves for KIC 6954726.}
\label{fig:6954726}
\end{figure}

\begin{figure}
\includegraphics[width=\linewidth]{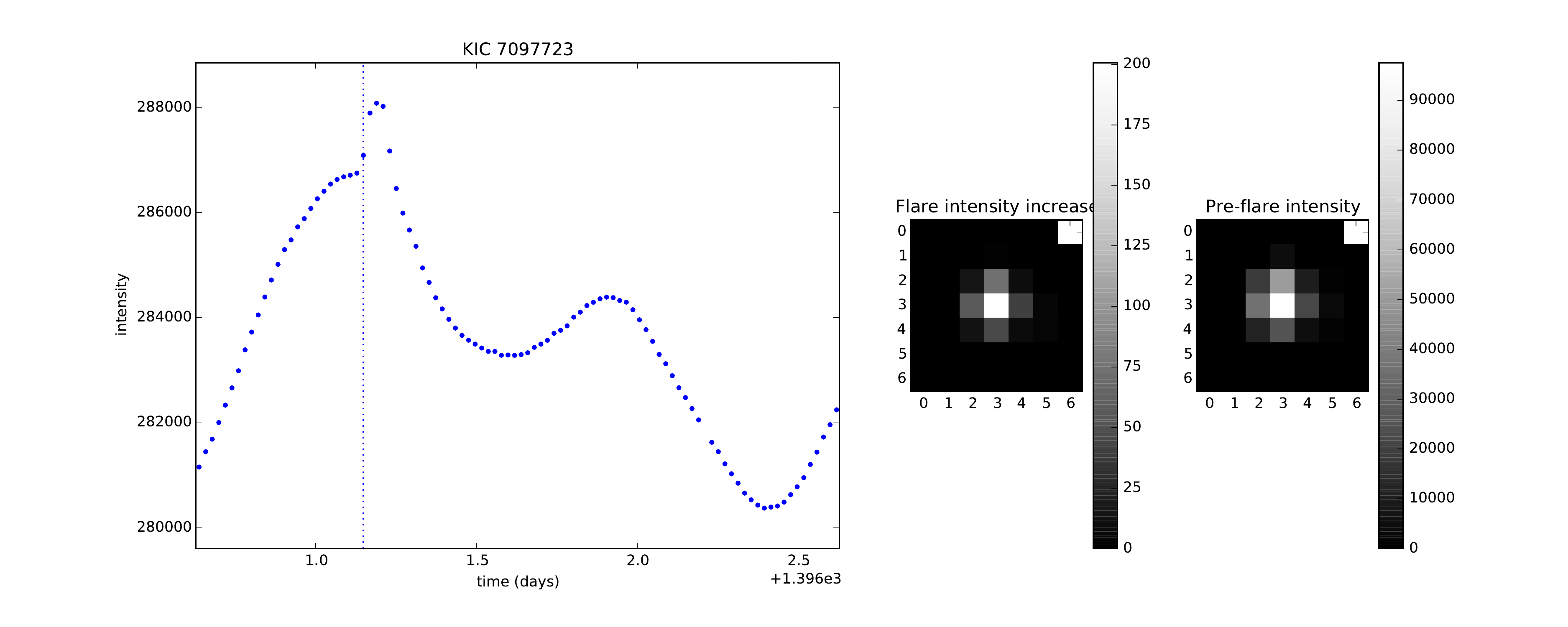}\\
\includegraphics[width=\linewidth]{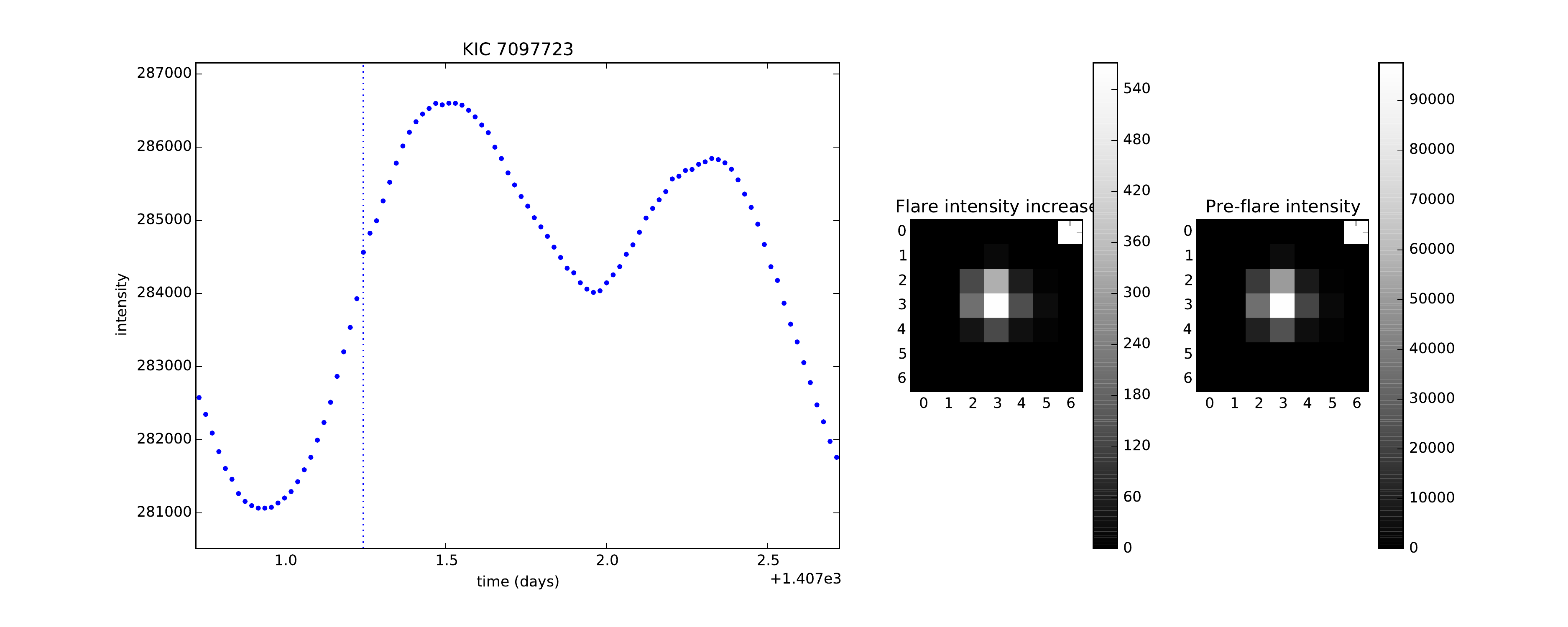}\\
\caption{Flare light curves for KIC 7097723.}
\end{figure}

\begin{figure}
\includegraphics[width=\linewidth]{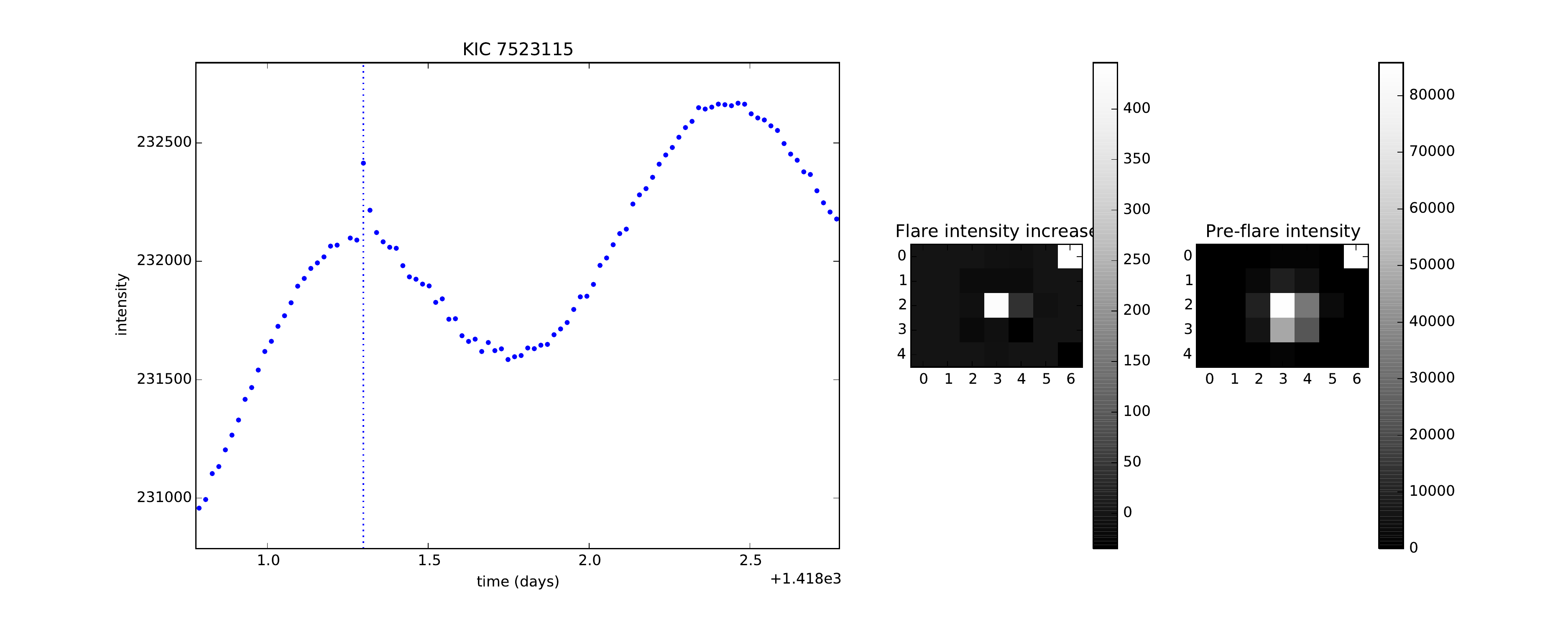}\\
\caption{Flare light curves for KIC 7523115.}
\label{fig:7523115}
\end{figure}

\begin{figure}
\includegraphics[width=\linewidth]{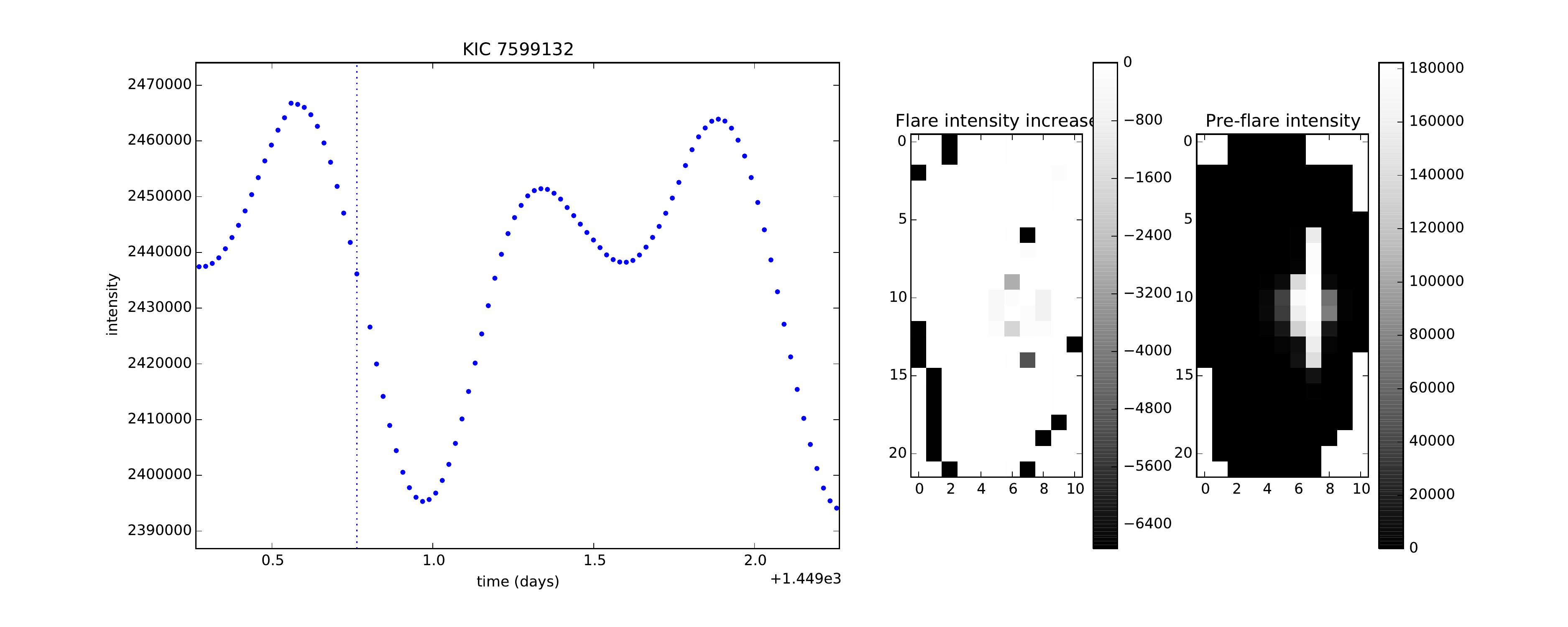}\\
\includegraphics[width=\linewidth]{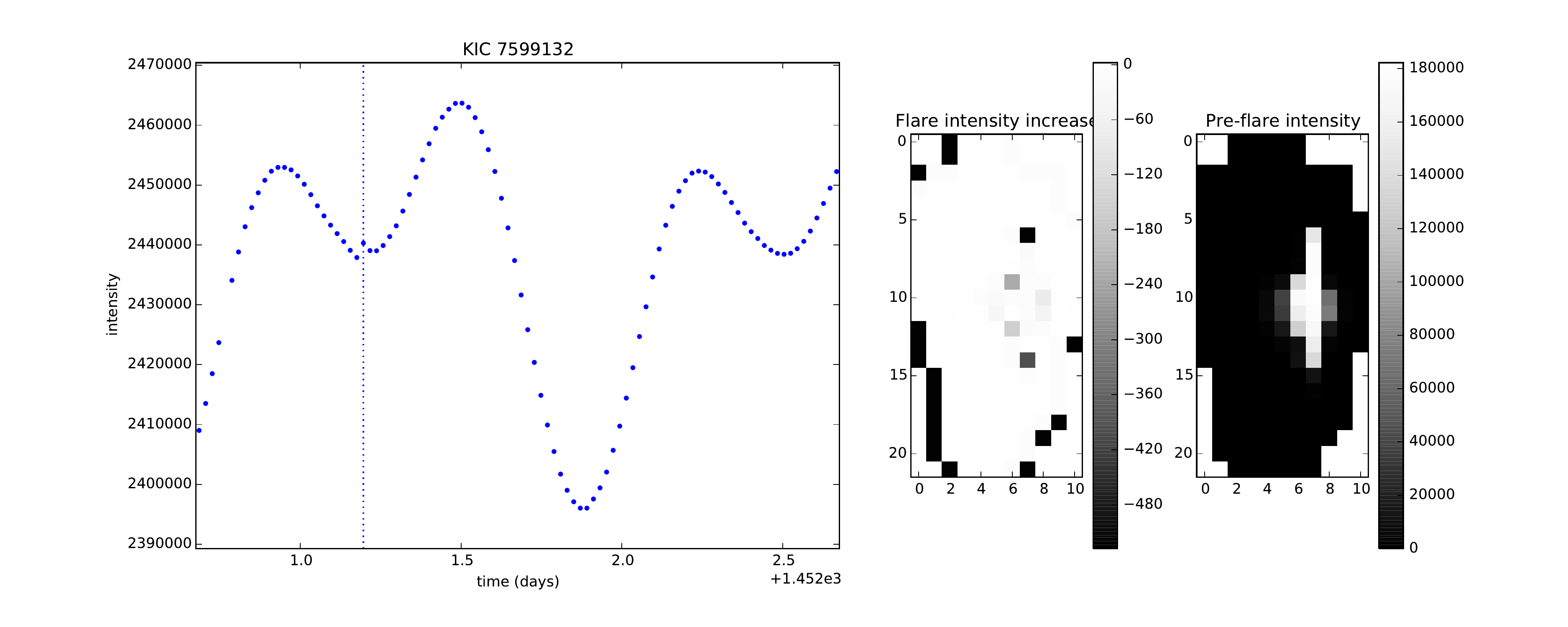}\\
\includegraphics[width=\linewidth]{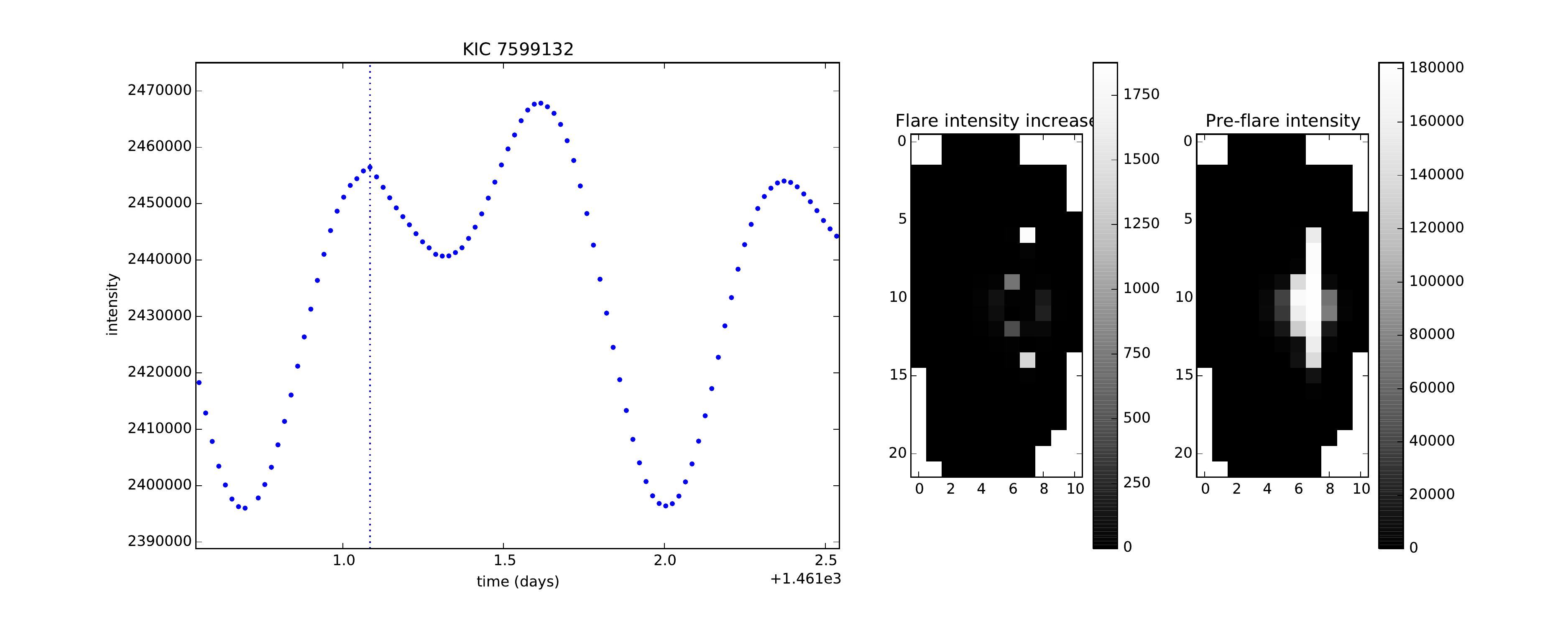}\\
\caption{Flare light curves for KIC 7599132.}
\end{figure}

\begin{figure}
\includegraphics[width=\linewidth]{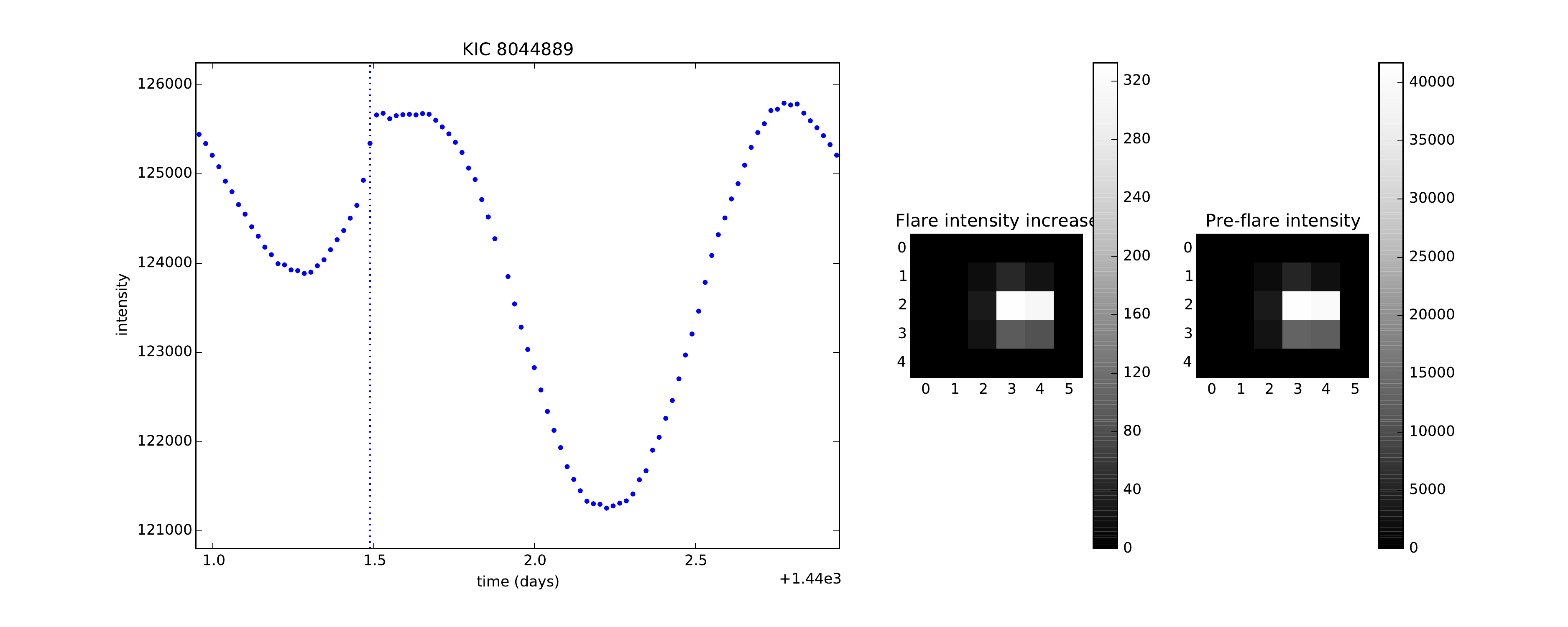}\\
\includegraphics[width=\linewidth]{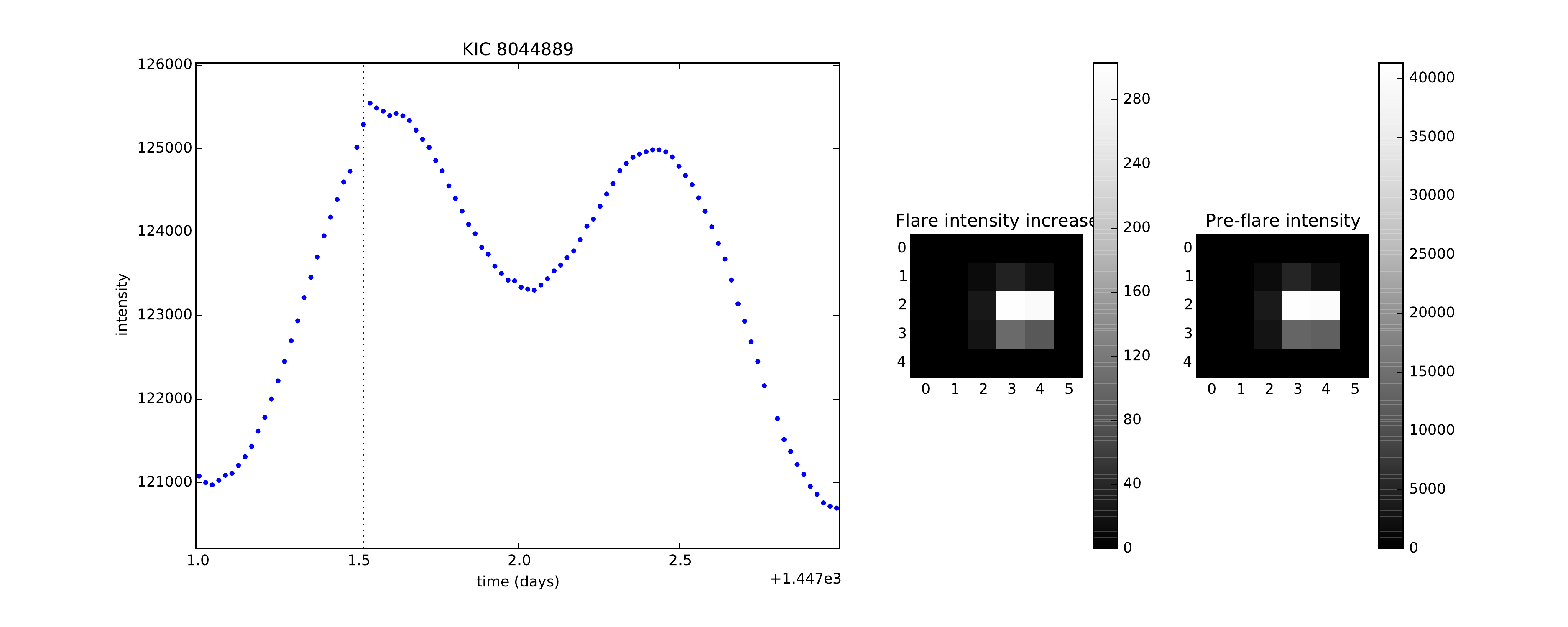}\\
\includegraphics[width=\linewidth]{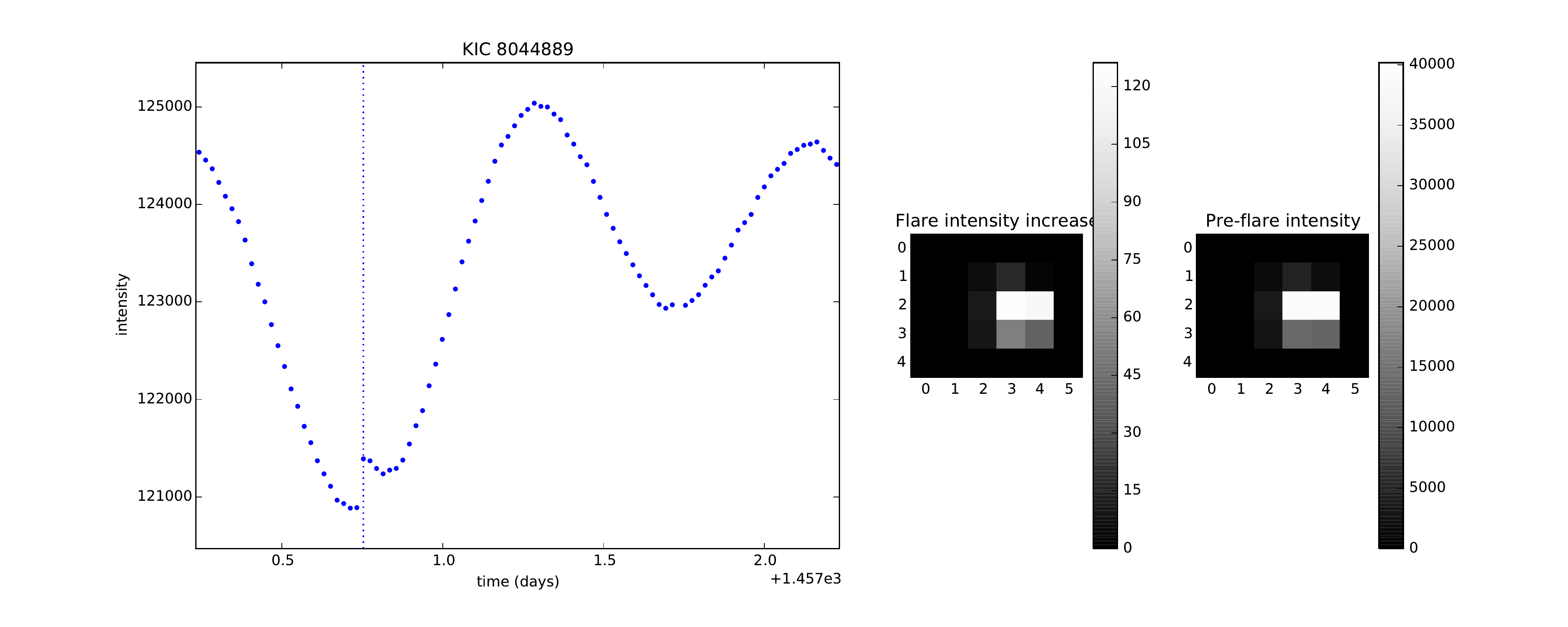}
\caption{Flare light curves for KIC 8044889.}
\end{figure}

\clearpage

\begin{figure}
\includegraphics[width=\linewidth]{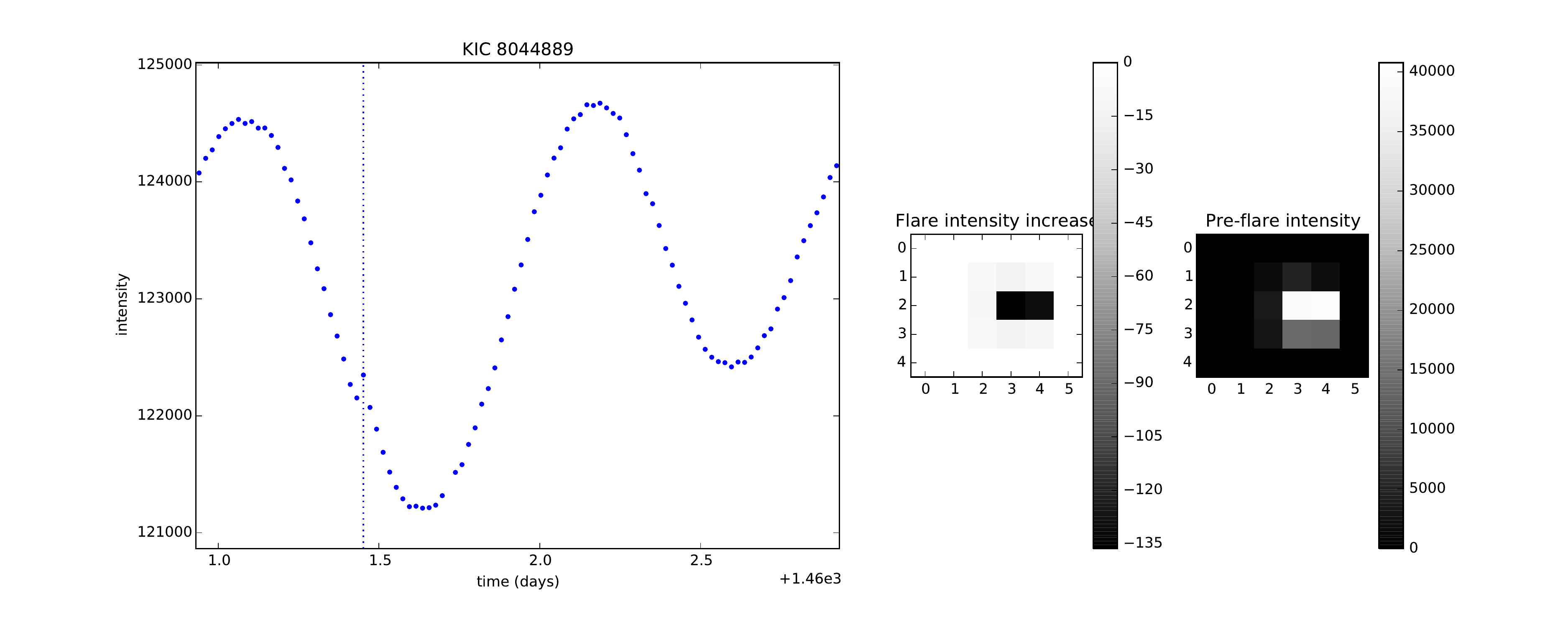}\\
\includegraphics[width=\linewidth]{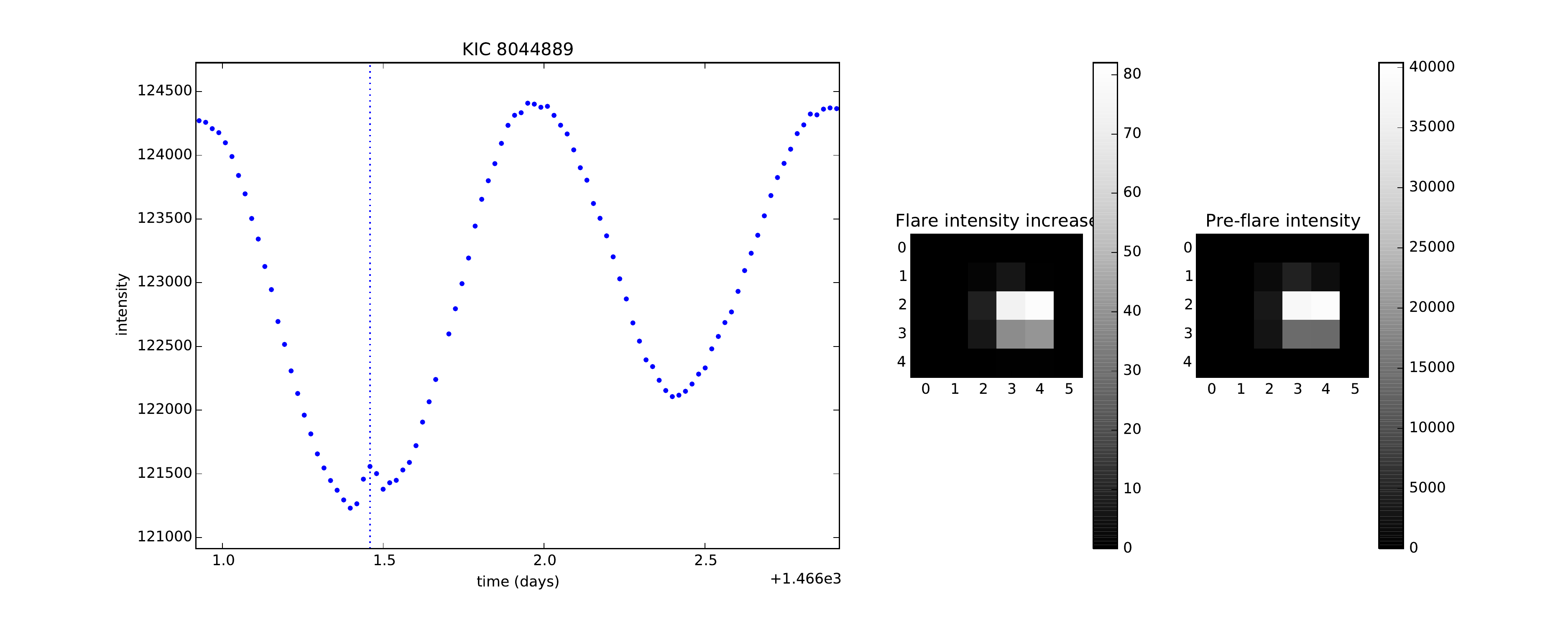}\\
\caption{Flare light curves for KIC 8044889 (continued).}
\end{figure}

\begin{figure}
\includegraphics[width=\linewidth]{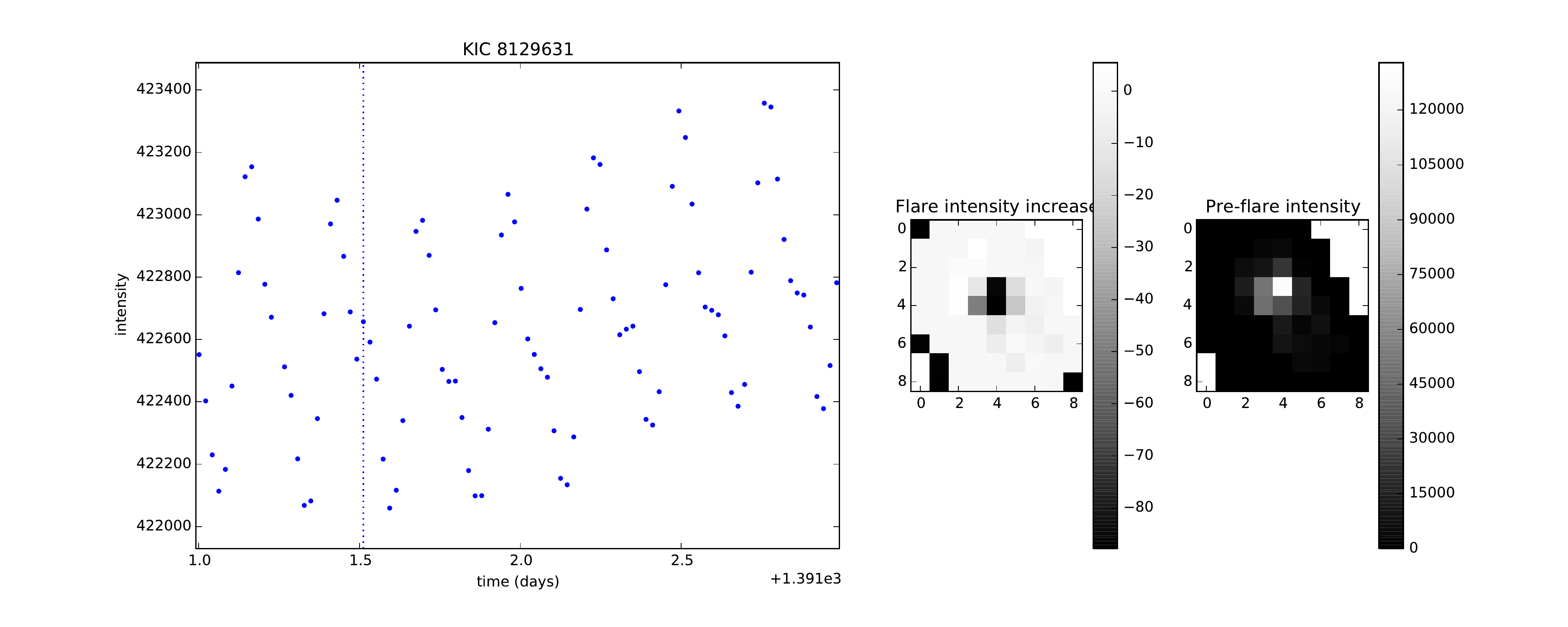}\\
\includegraphics[width=\linewidth]{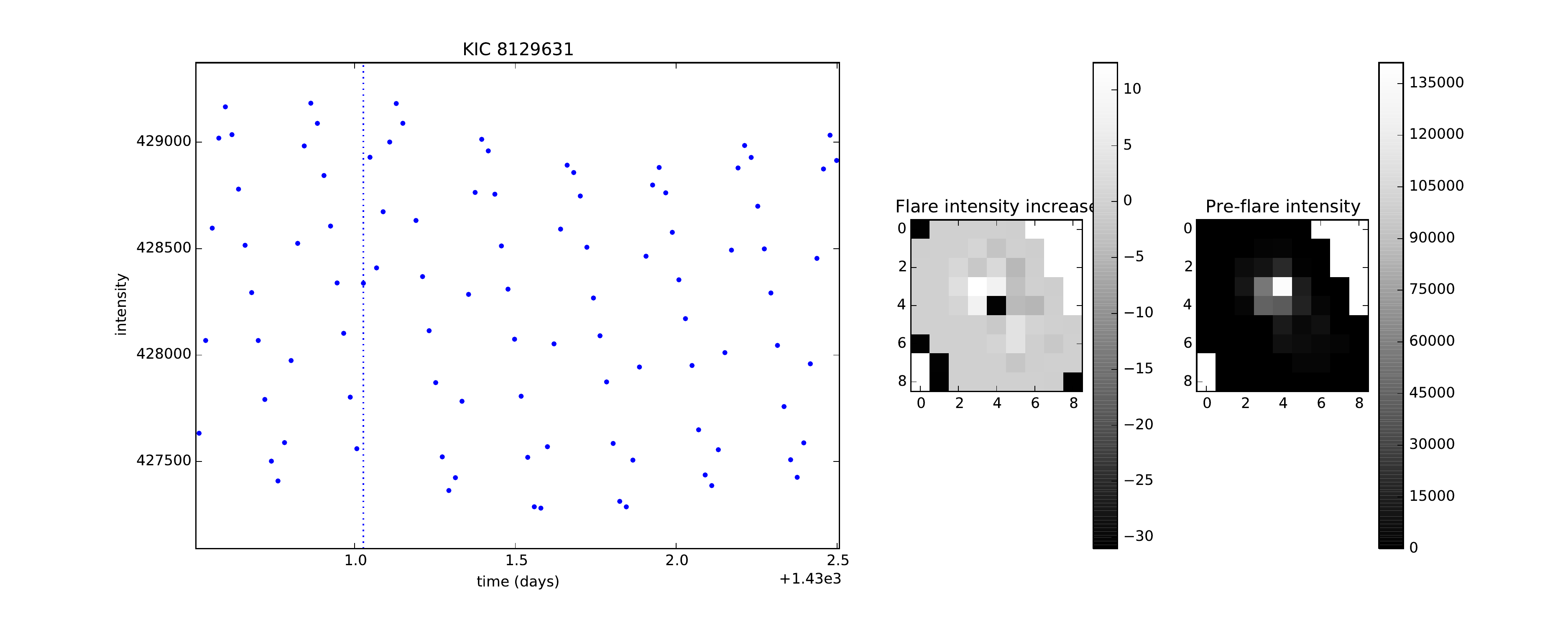}\\
\caption{Flare light curves for KIC 8129631.}
\label{fig:8129631}
\end{figure}

\begin{figure}
\includegraphics[width=\linewidth]{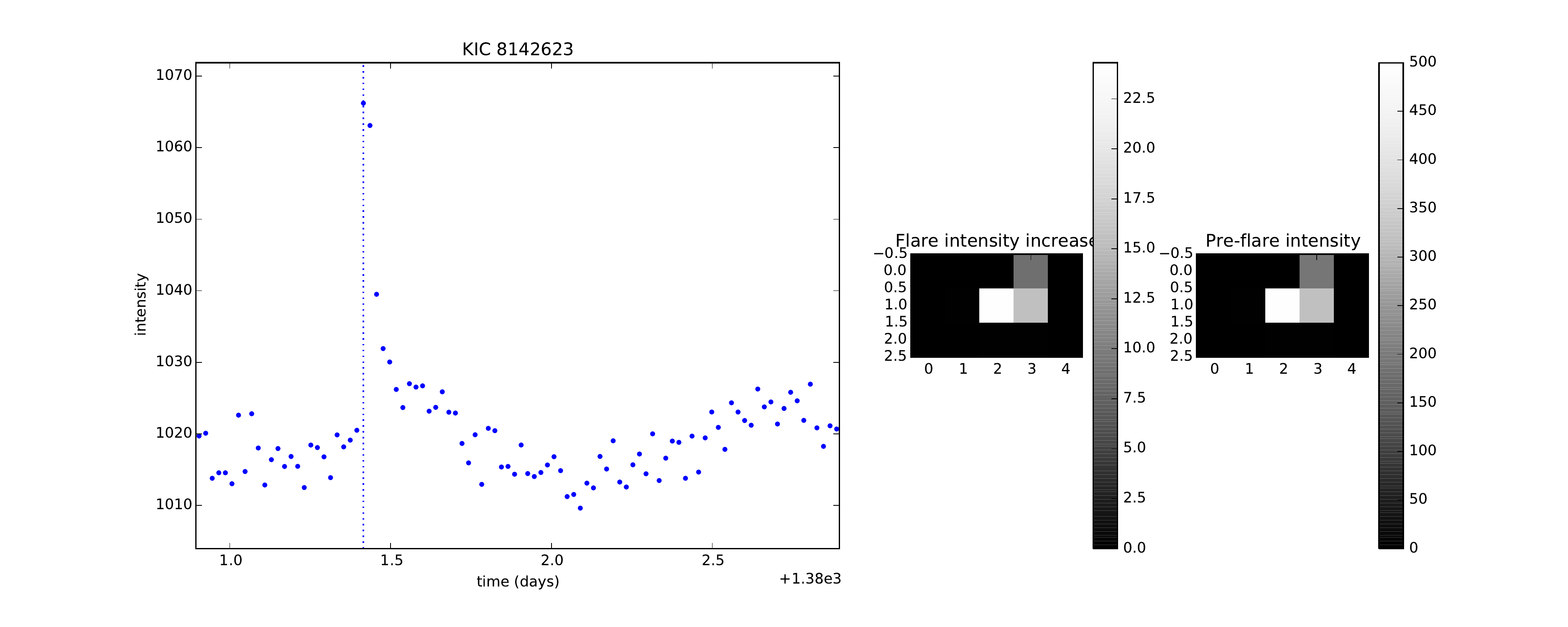}\\
\caption{Flare light curves for KIC 8142623.}
\end{figure}

\begin{figure}
\includegraphics[width=\linewidth]{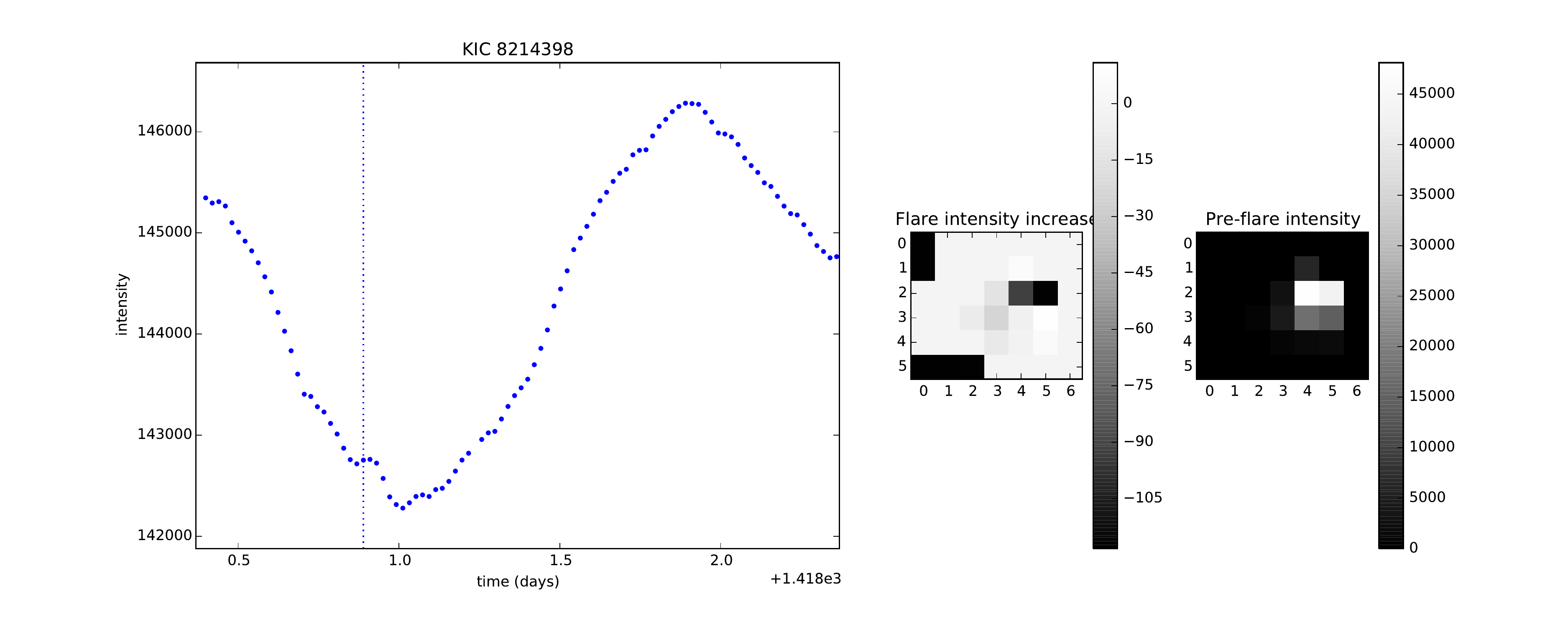}\\
\includegraphics[width=\linewidth]{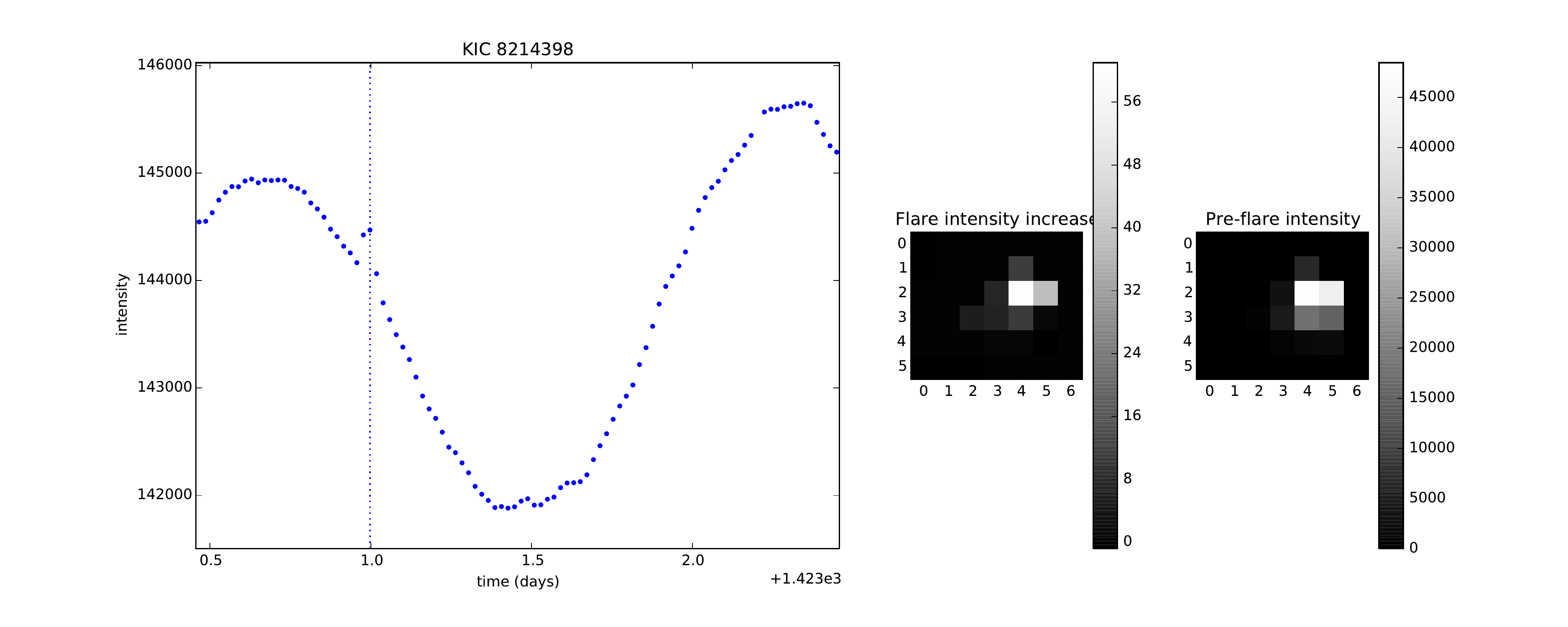}\\
\caption{Flare light curves for KIC 8214398.}
\end{figure}

\begin{figure}
\includegraphics[width=\linewidth]{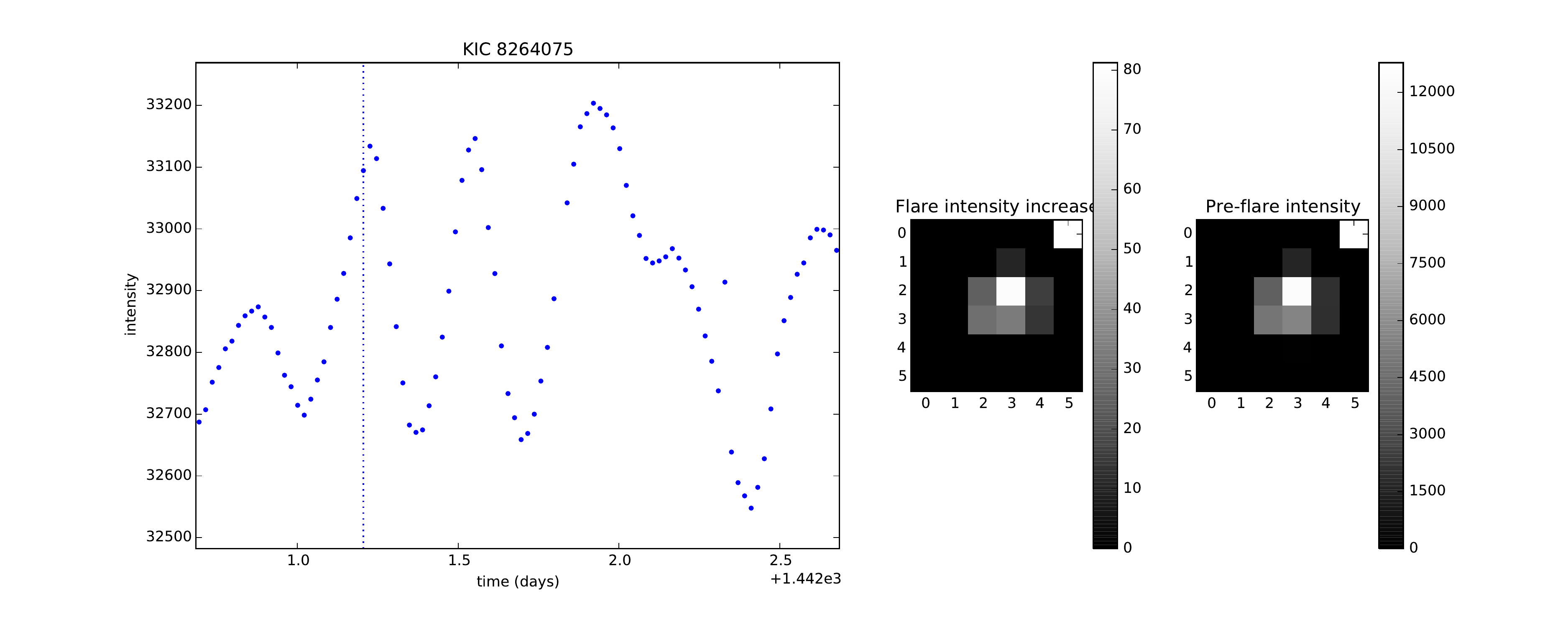}\\
\includegraphics[width=\linewidth]{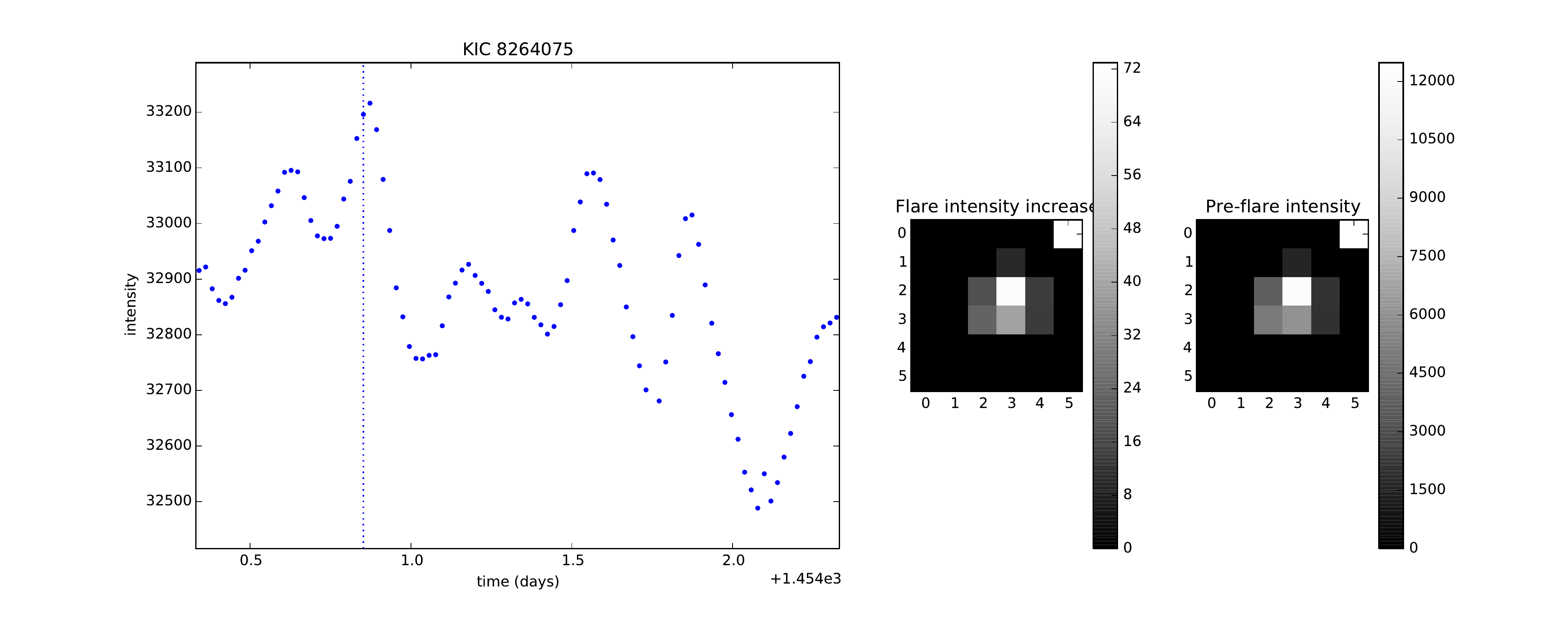}\\
\caption{Flare light curves for KIC 8264075.}
\label{fig:8264075}
\end{figure}

\begin{figure}
\includegraphics[width=\linewidth]{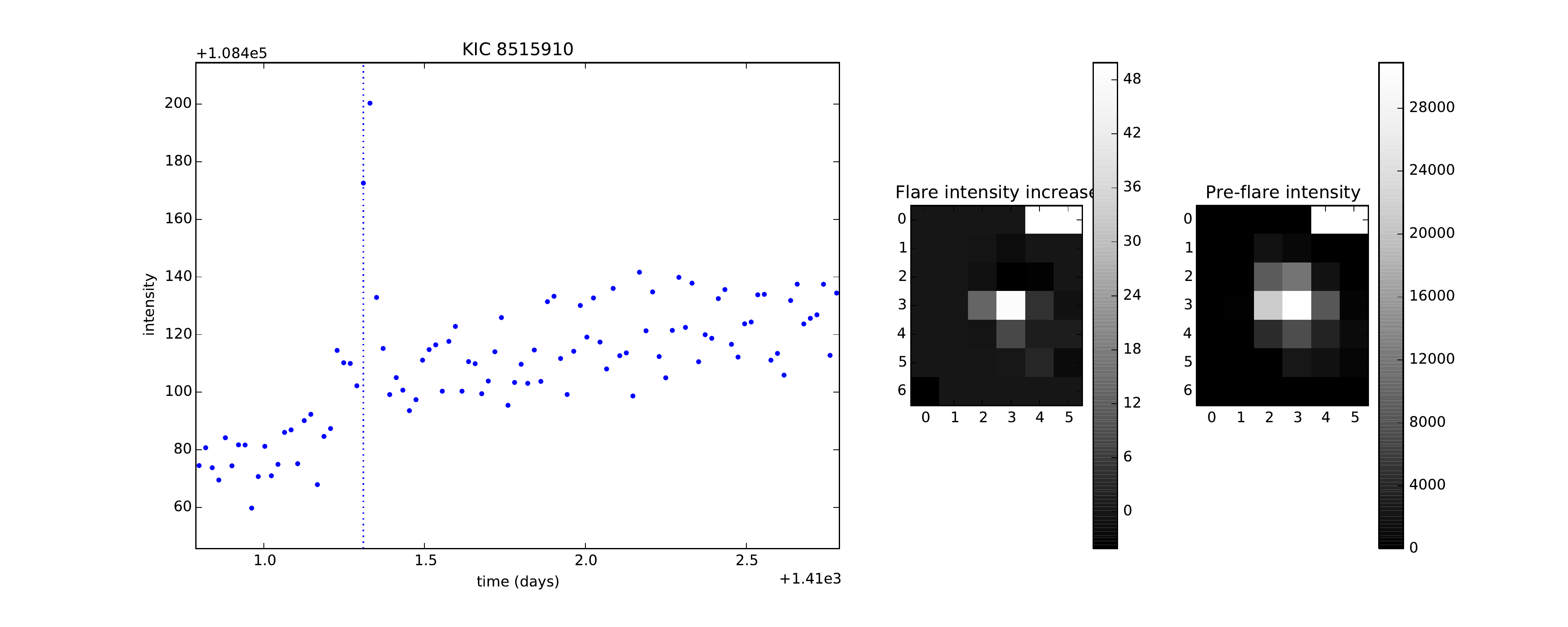}\\
\includegraphics[width=\linewidth]{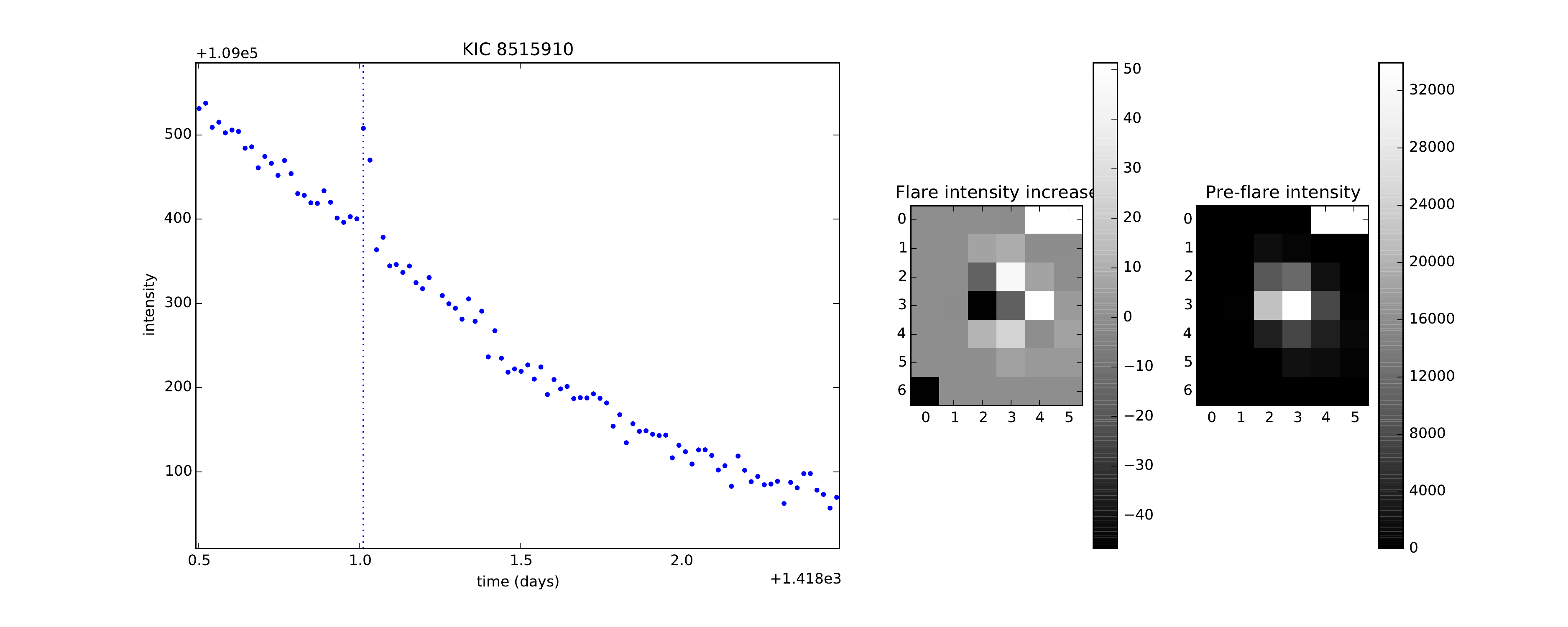}\\
\caption{Flare light curves for KIC 8515910.}
\end{figure}

\begin{figure}
\includegraphics[width=\linewidth]{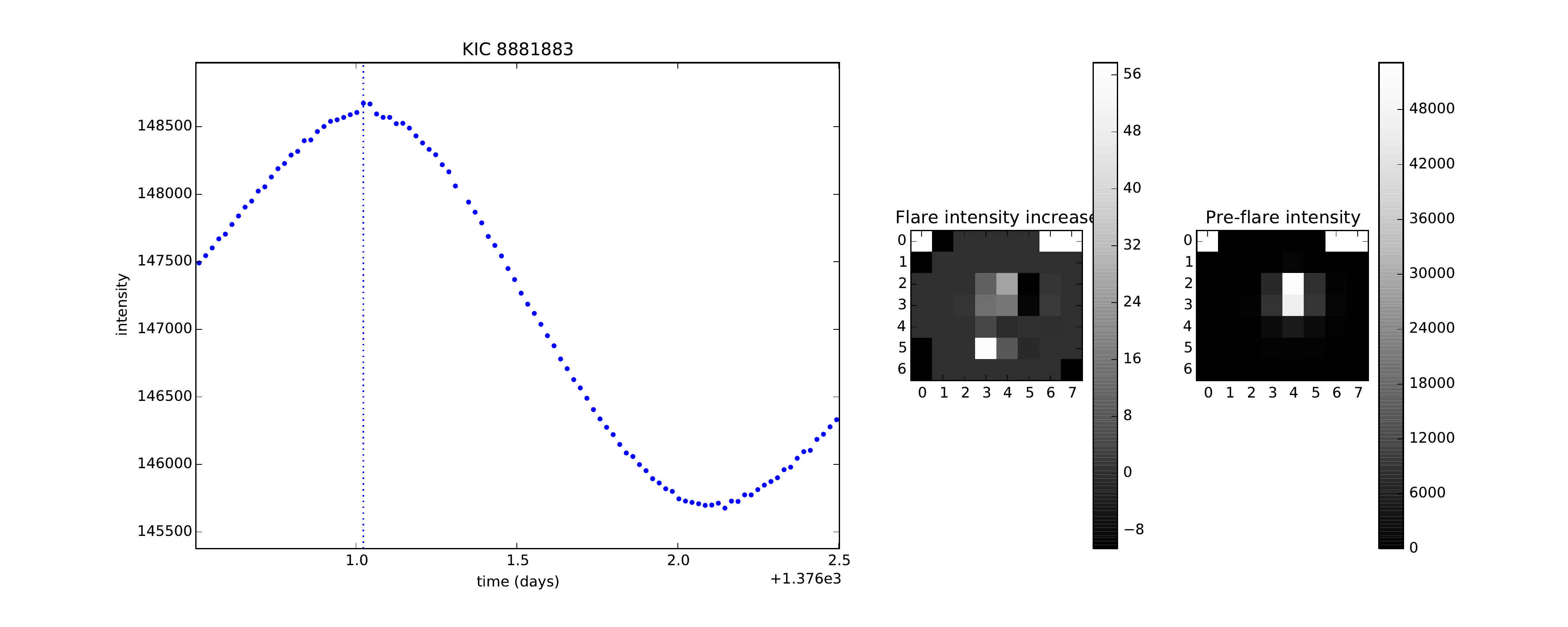}\\
\caption{Flare light curves for KIC 8881883.}
\end{figure}

\clearpage 

\begin{figure}
\includegraphics[width=\linewidth]{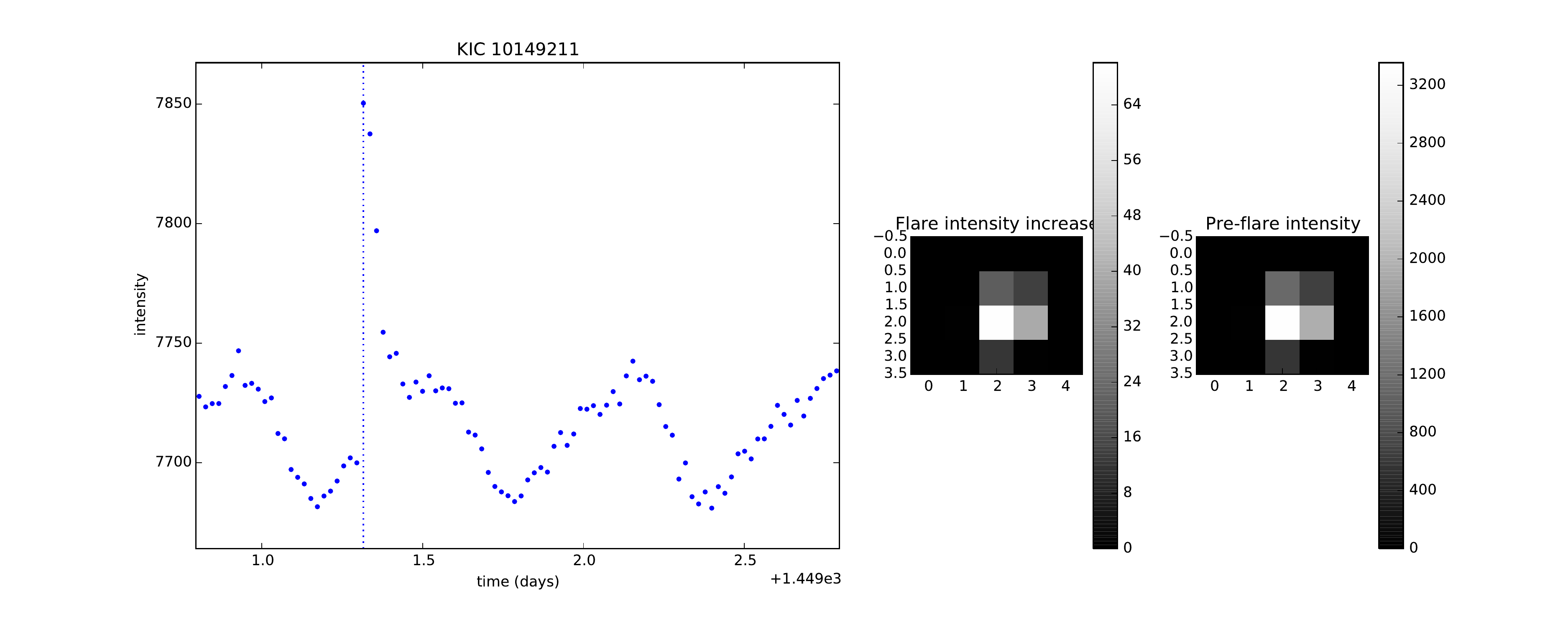}\\
\includegraphics[width=\linewidth]{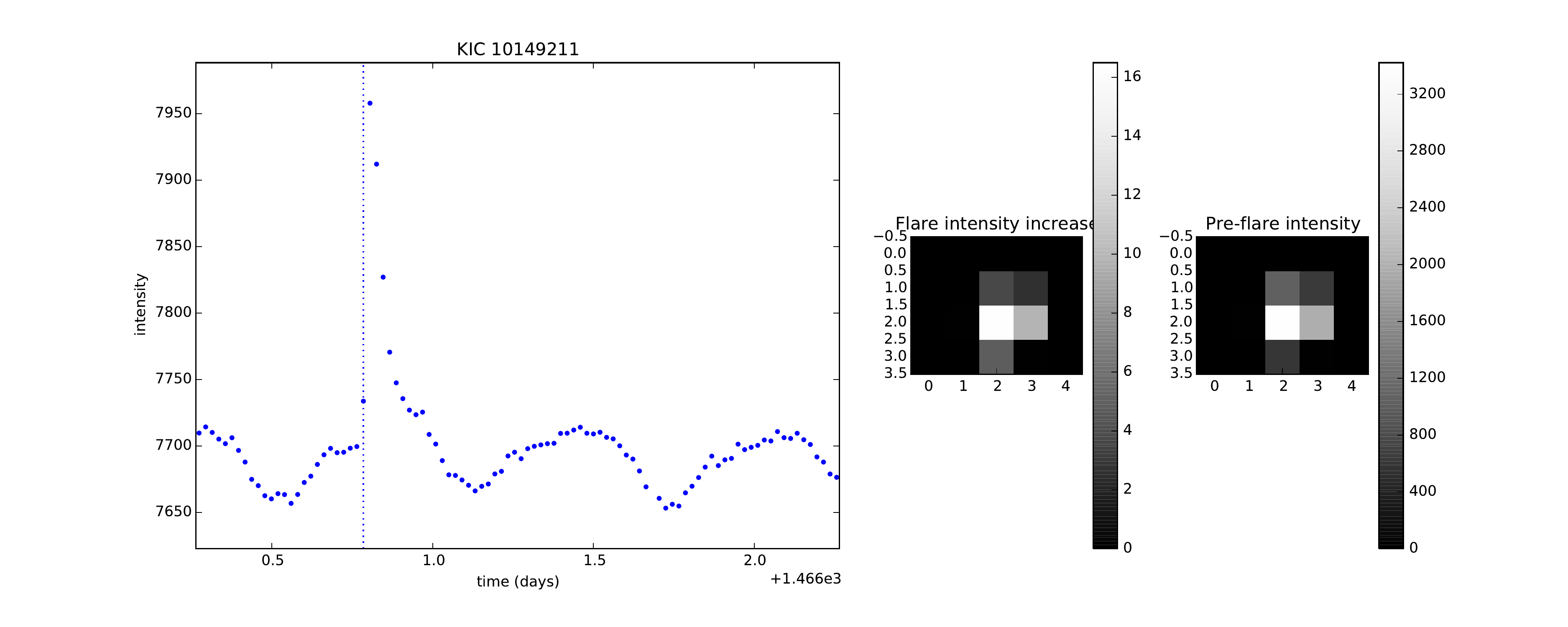}\\
\caption{Flare light curves for KIC 10149211.}
\label{fig:10149211}
\end{figure}

\begin{figure}
\includegraphics[width=\linewidth]{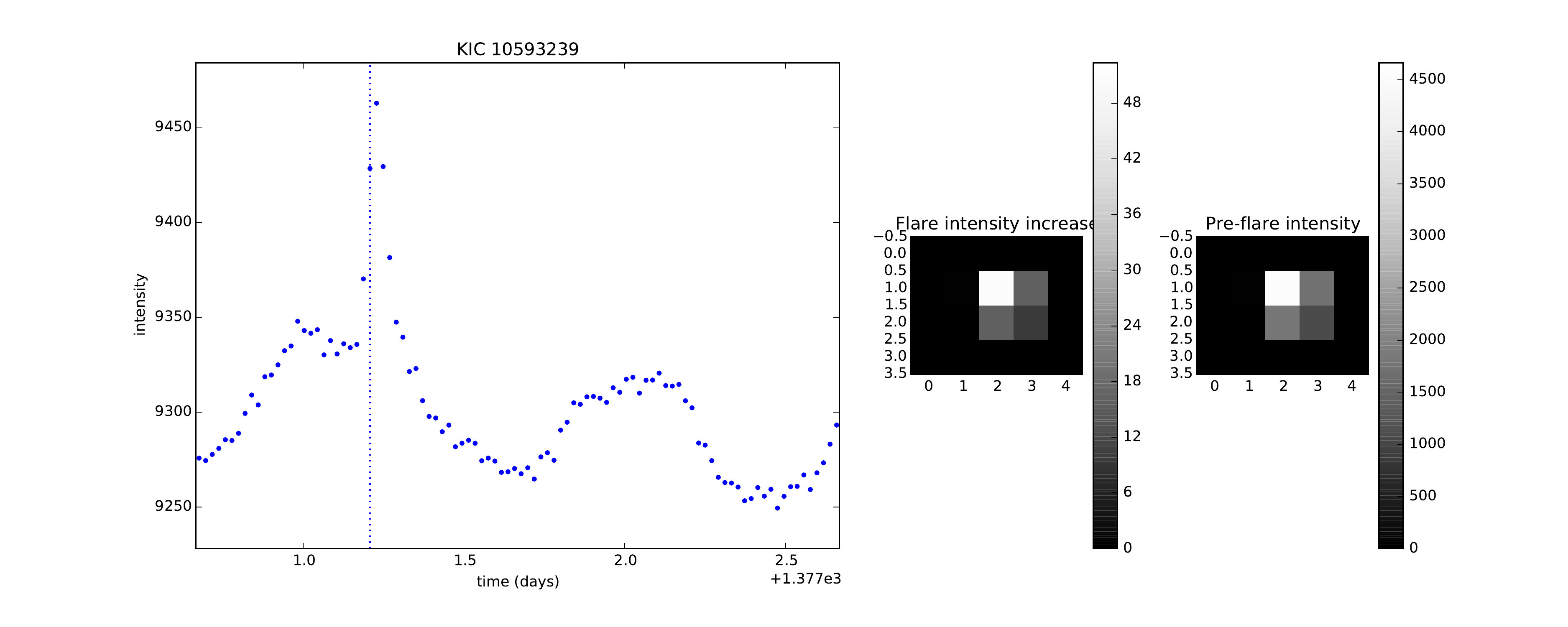}\\
\includegraphics[width=\linewidth]{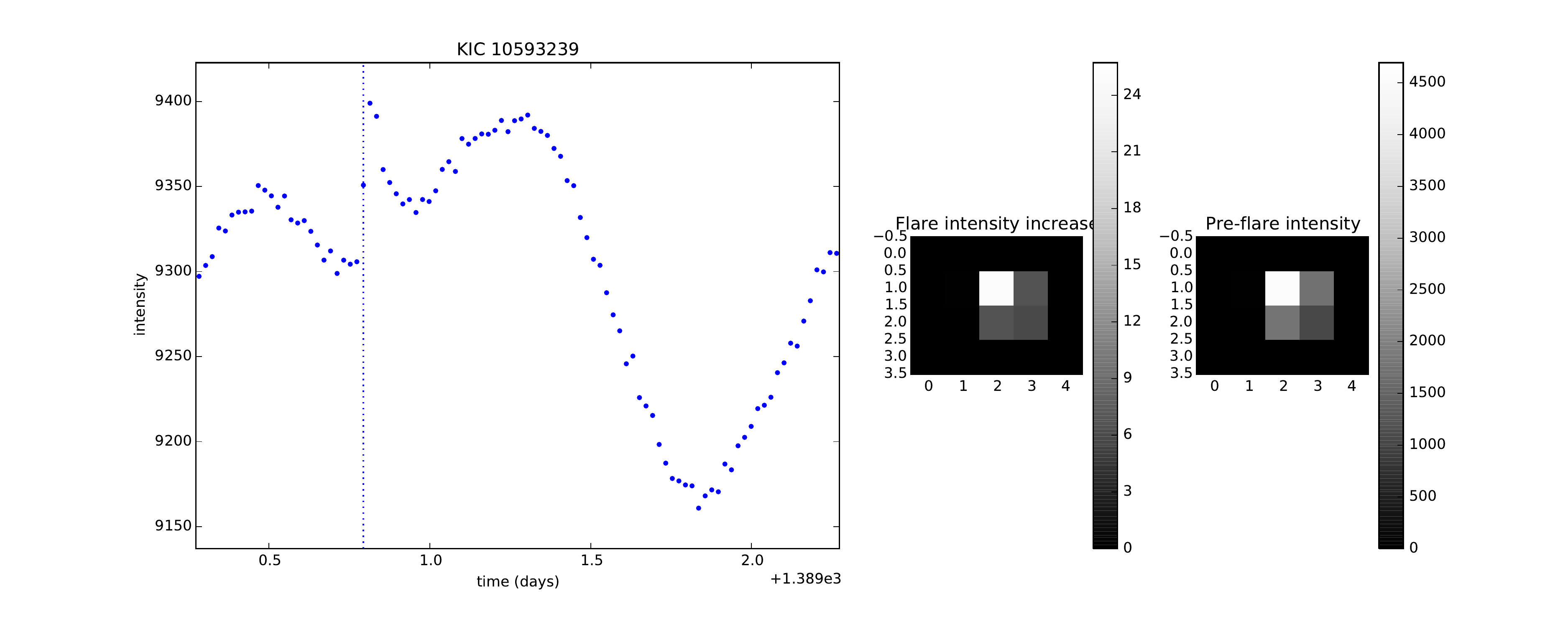}\\
\includegraphics[width=\linewidth]{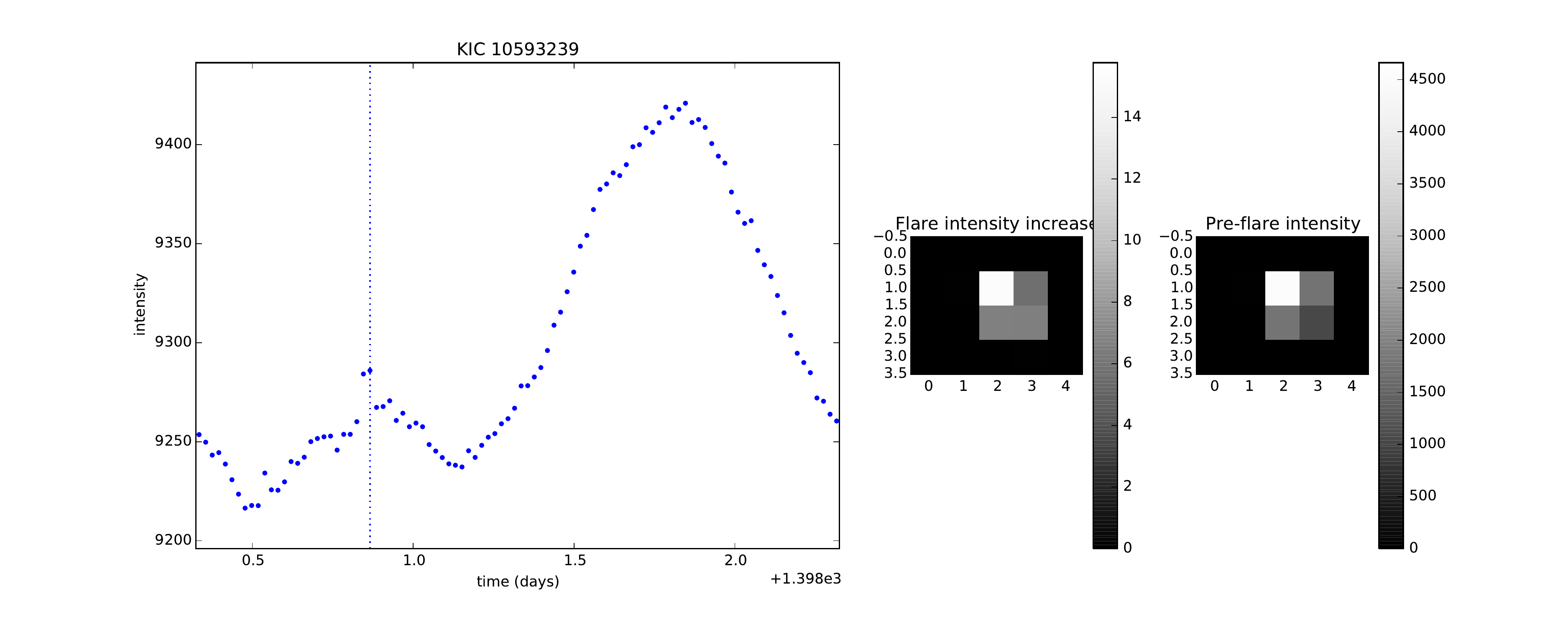}
\caption{Flare light curves for KIC 10593239.}
\end{figure}

\begin{figure}
\includegraphics[width=\linewidth]{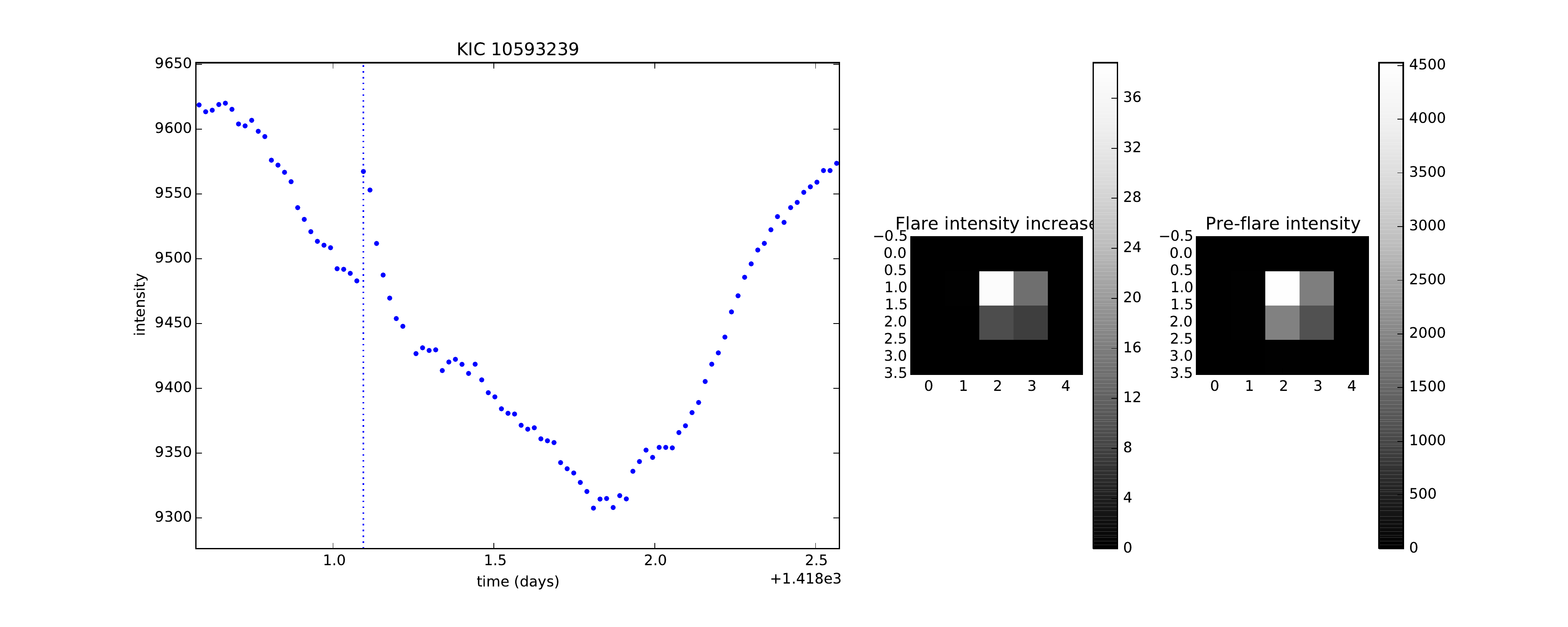}\\
\includegraphics[width=\linewidth]{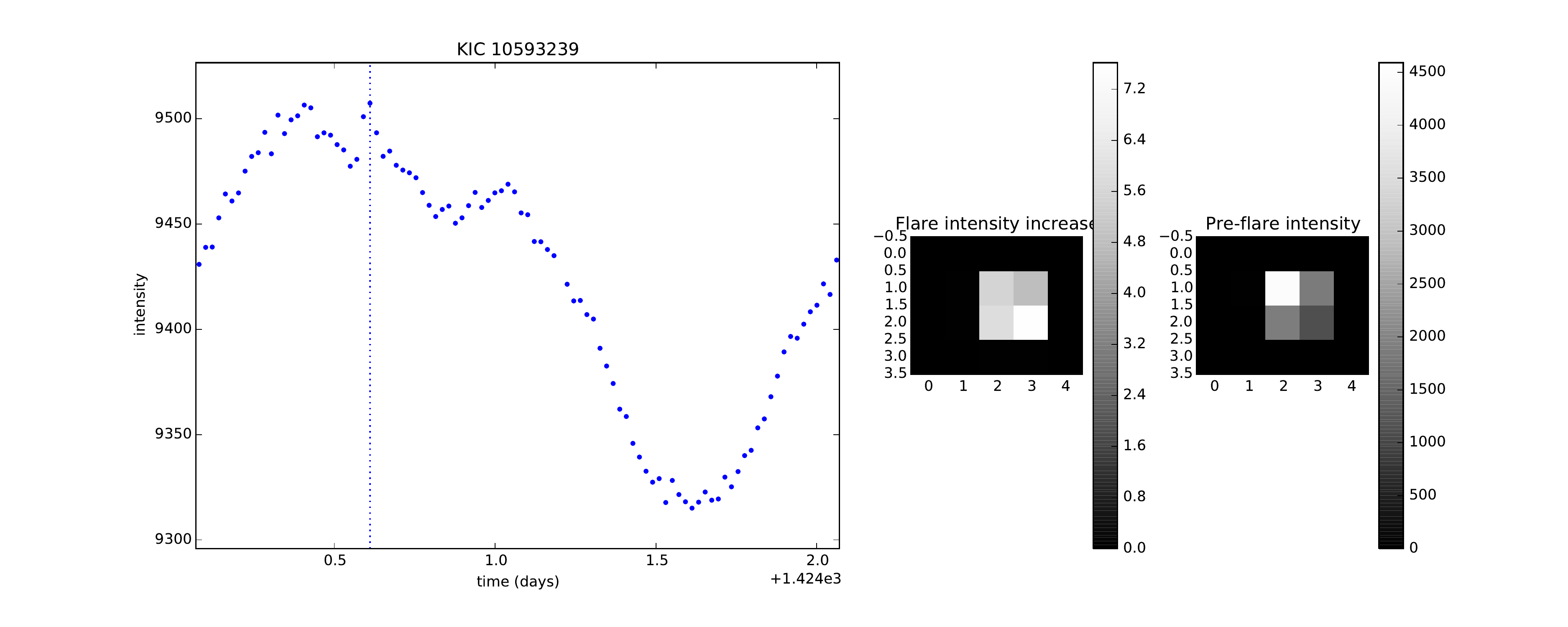}\\
\caption{Flare light curves for KIC 10593239 (continued).}
\end{figure}

\begin{figure}
\includegraphics[width=\linewidth]{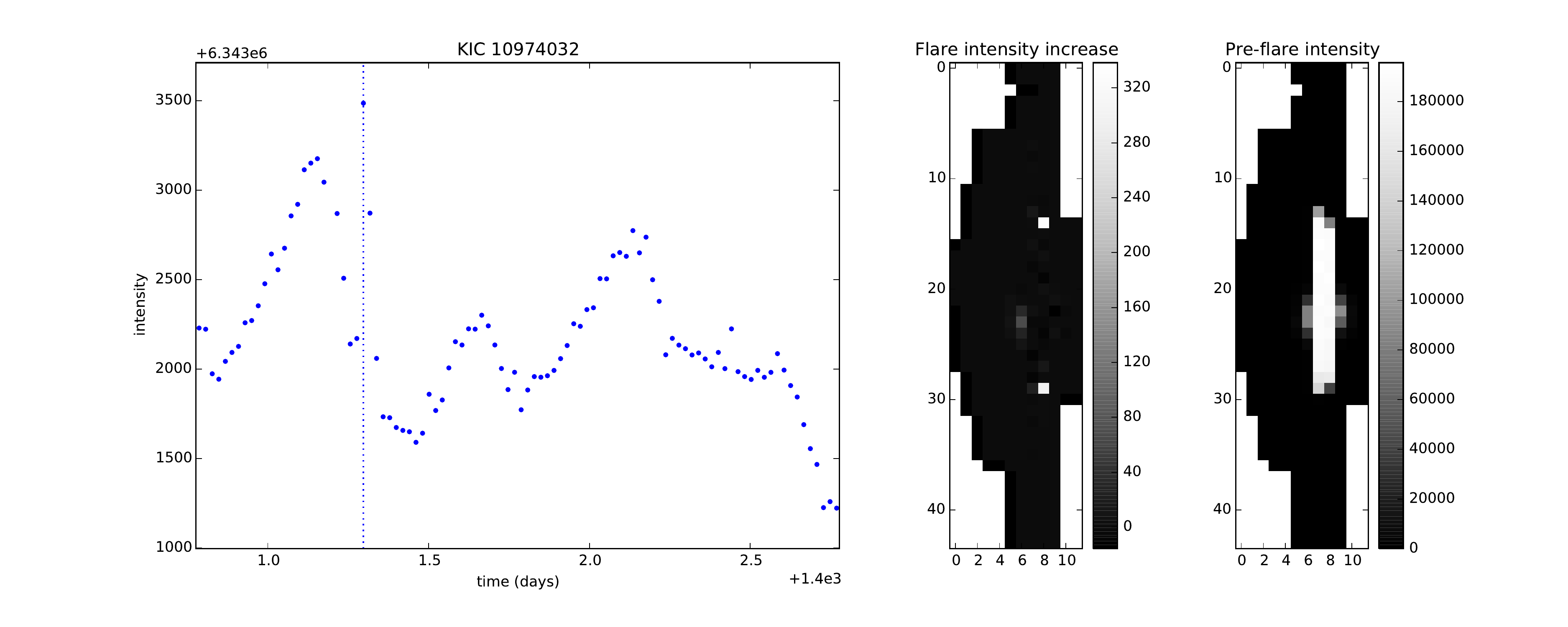}\\
\includegraphics[width=\linewidth]{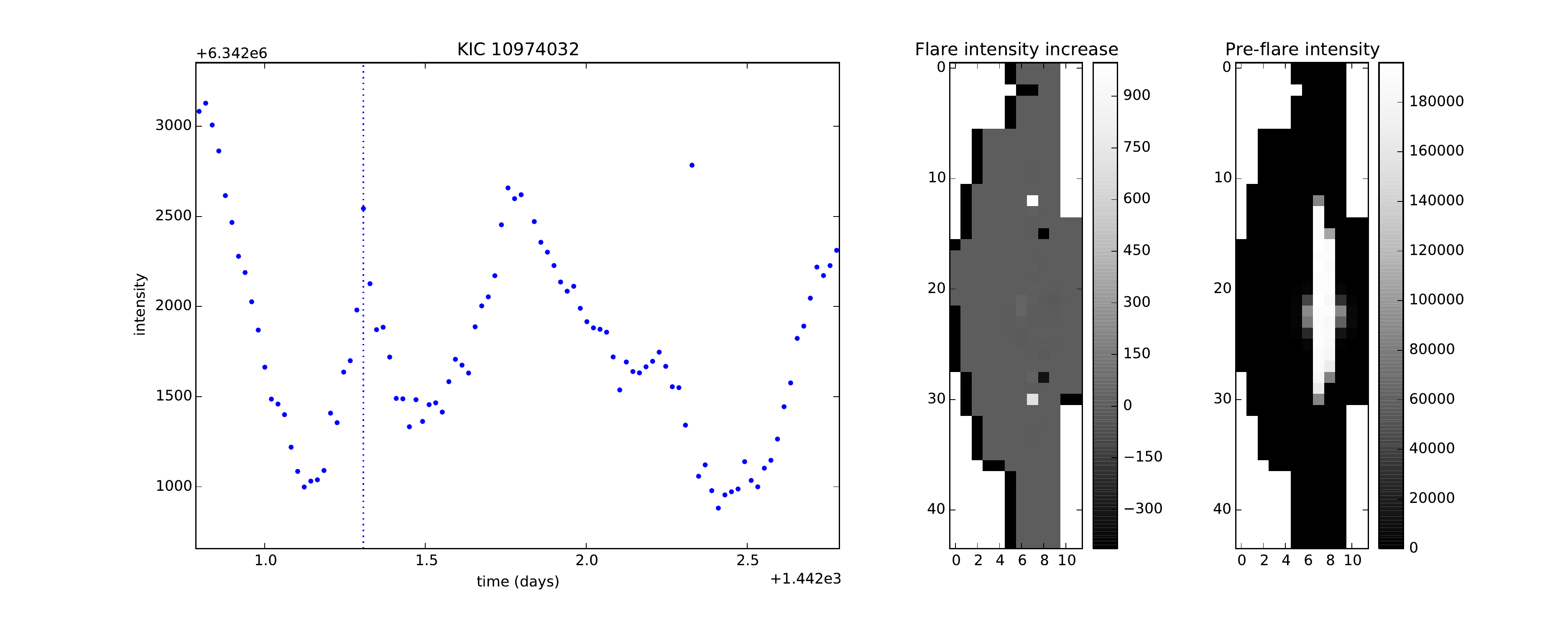}\\
\includegraphics[width=\linewidth]{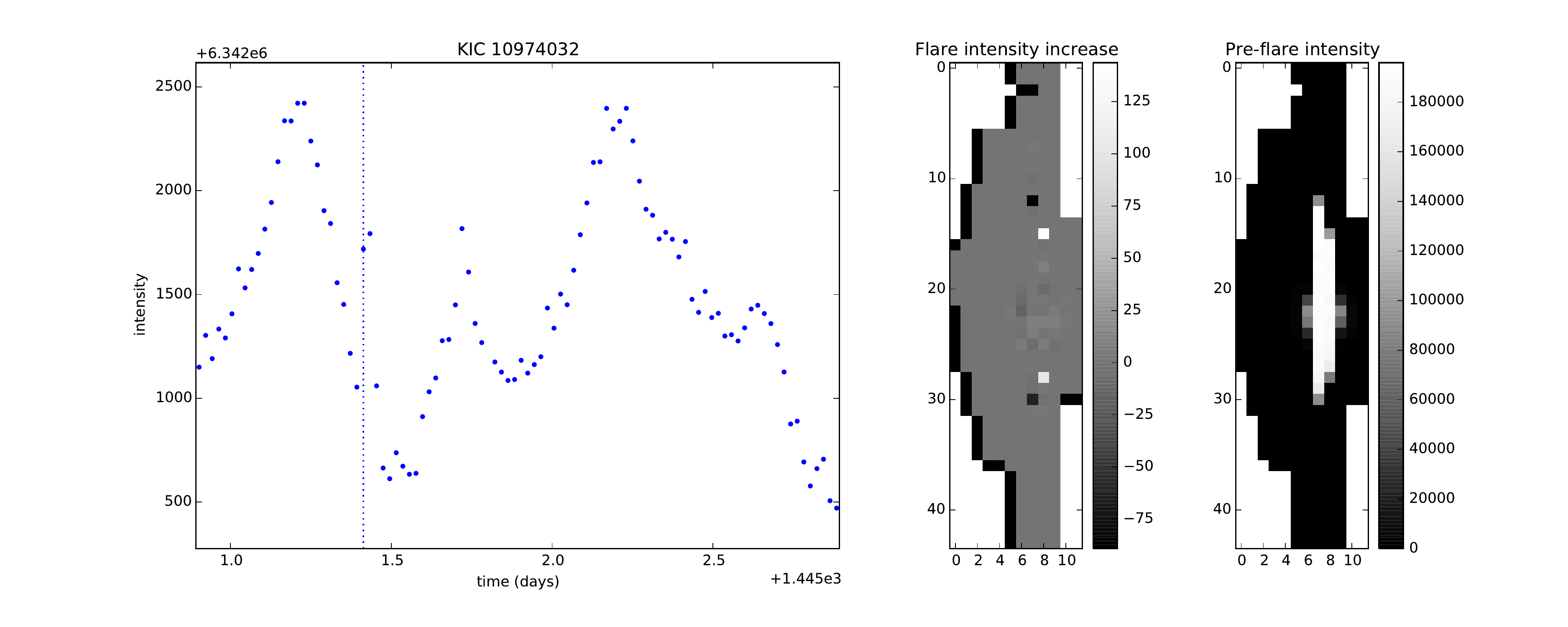}\\
\caption{Flare light curves for KIC 10974032.}
\end{figure}

\begin{figure}
\includegraphics[width=\linewidth]{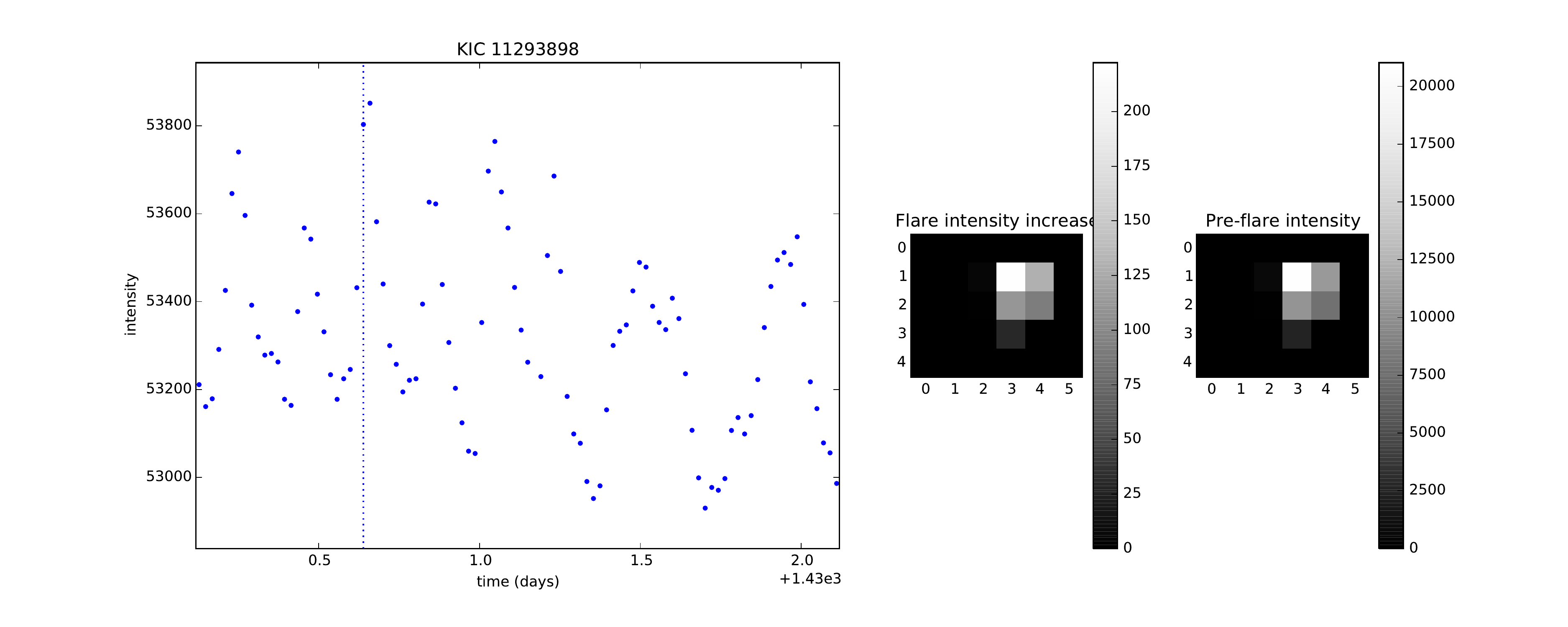}\\
\includegraphics[width=\linewidth]{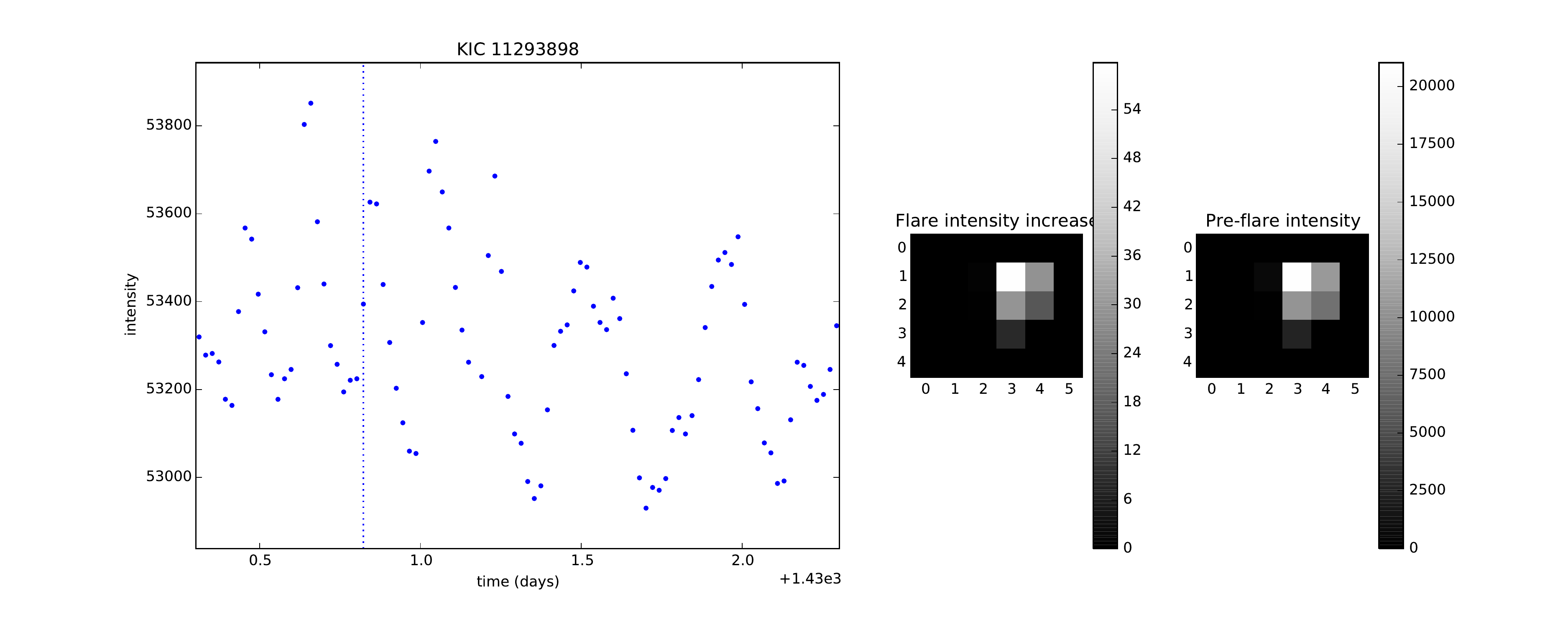}\\
\includegraphics[width=\linewidth]{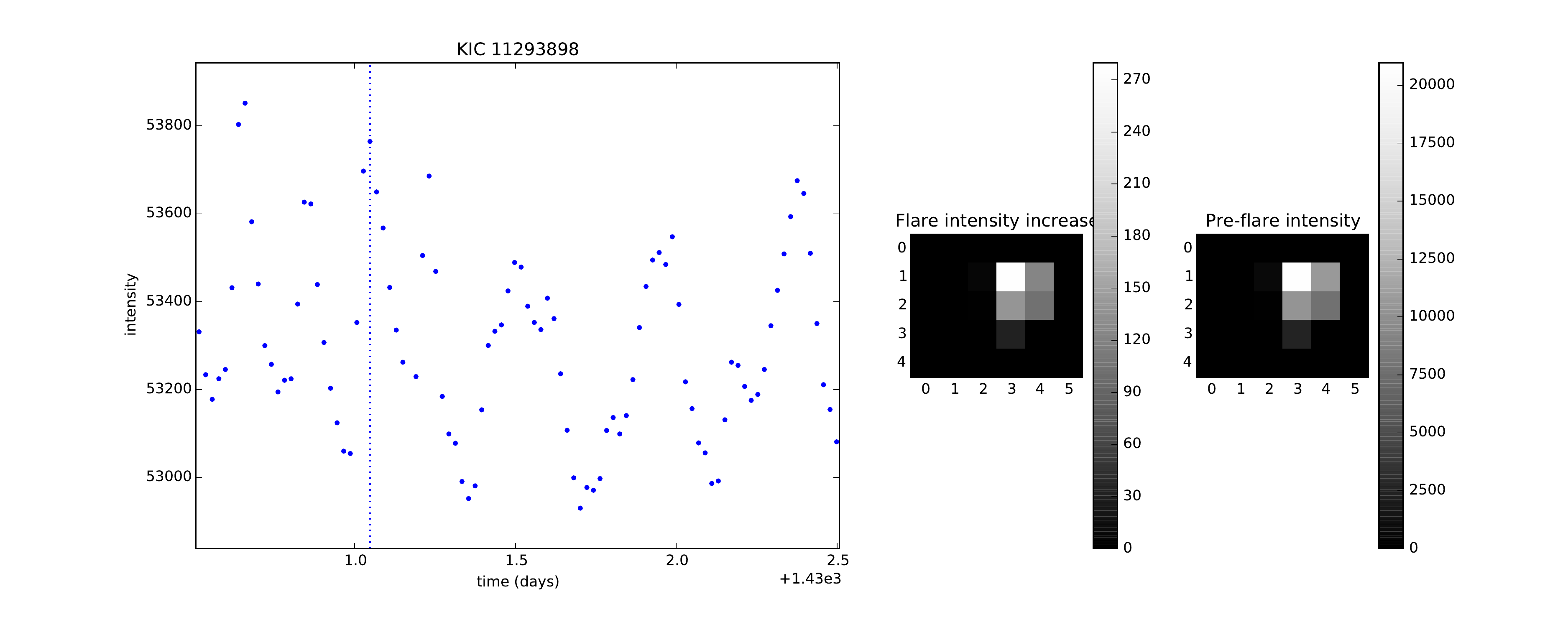}\\
\caption{Flare light curves for KIC 11293898.}
\label{fig:11293898a}
\end{figure}

\begin{figure}
\includegraphics[width=\linewidth]{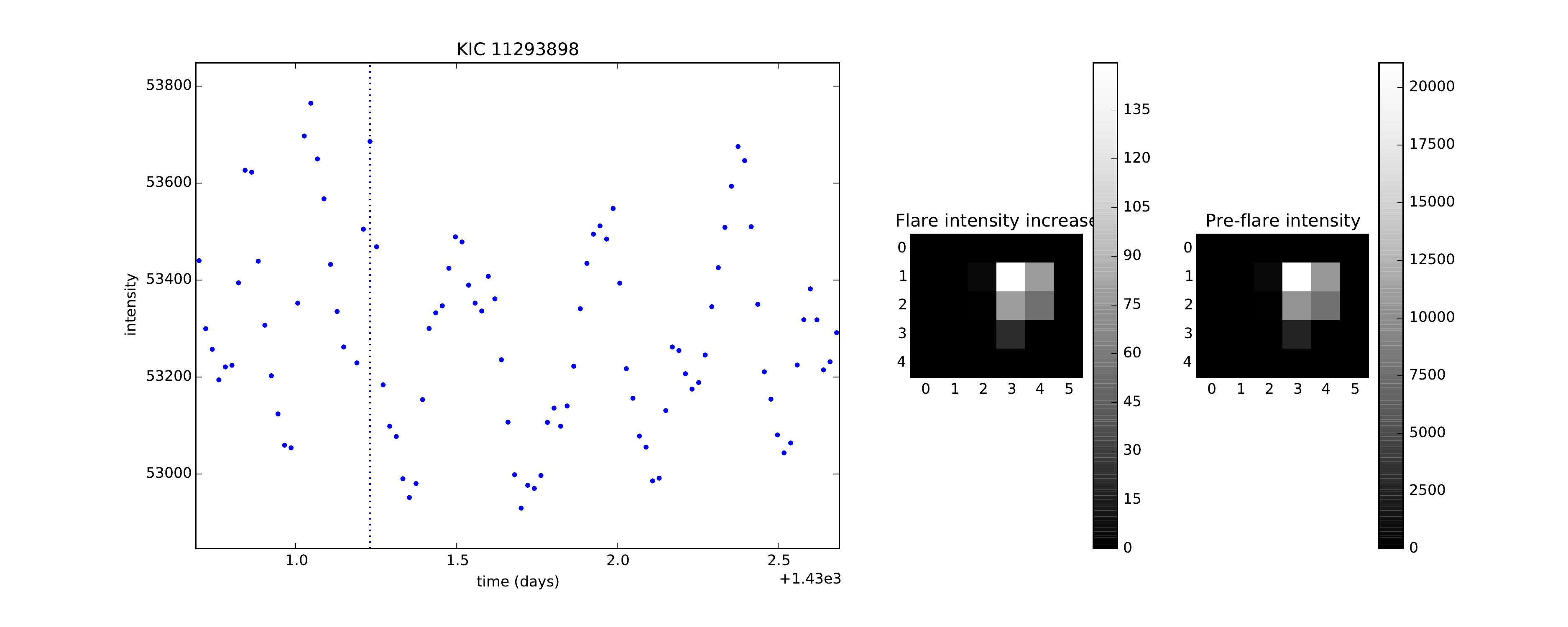}\\
\caption{Flare light curves for KIC 11293898 (continued).}
\label{fig:11293898b}
\end{figure}

\begin{figure}
\includegraphics[width=\linewidth]{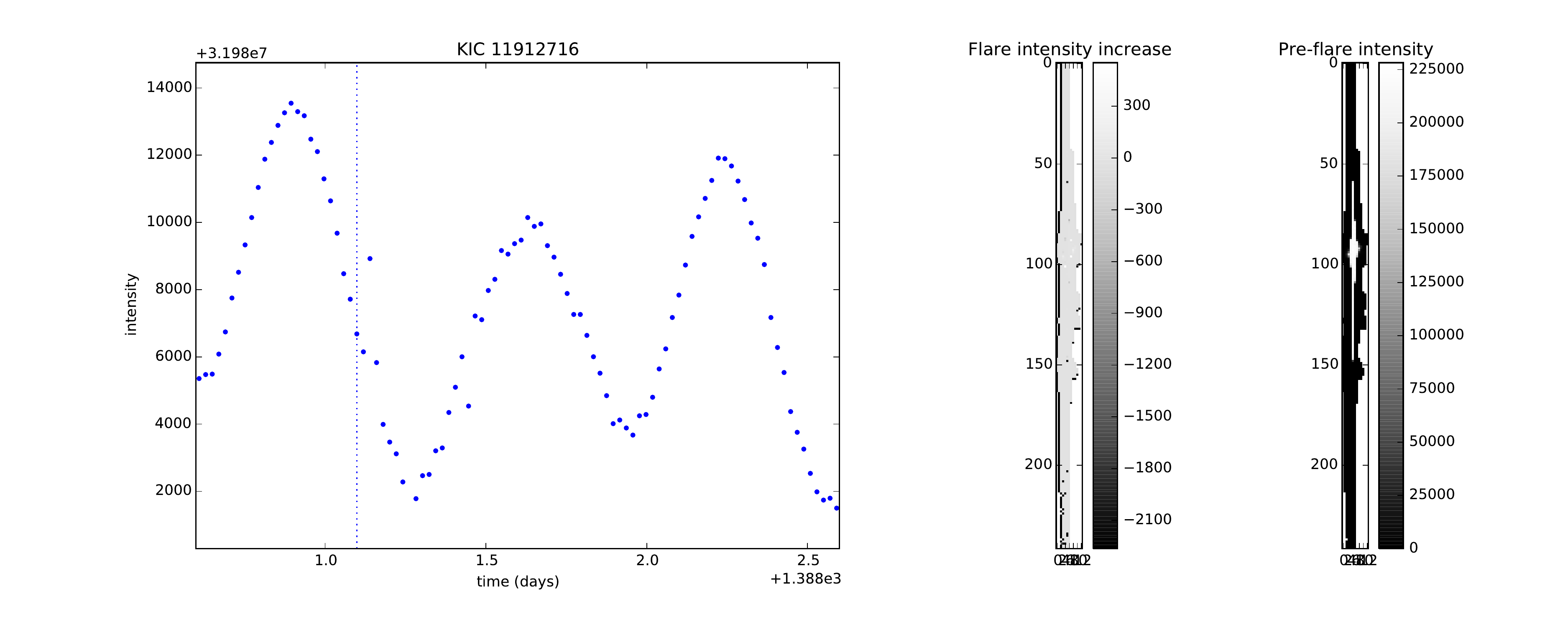}\\
\caption{Flare light curves for KIC 11912716.}
\end{figure}

\begin{figure}
\includegraphics[width=\linewidth]{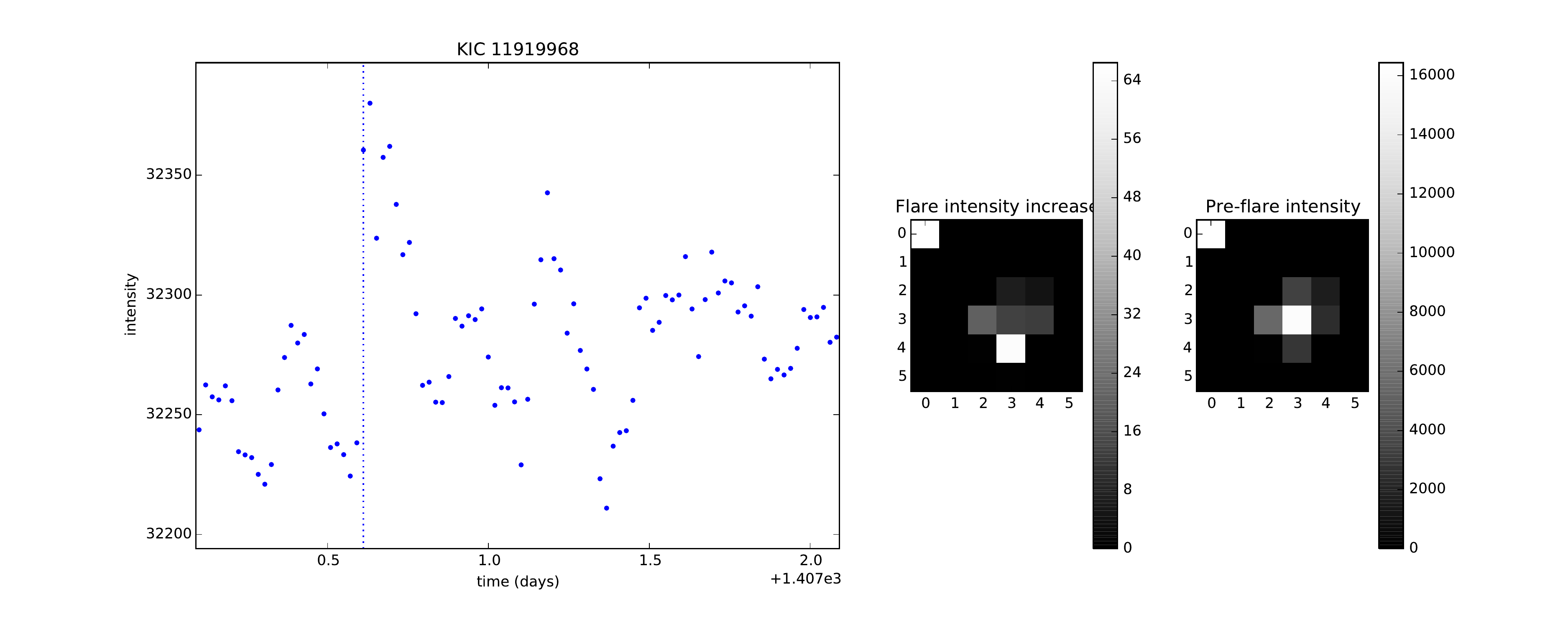}\\
\caption{Flare light curves for KIC 11919968.}
\end{figure}

\bibliographystyle{aasjournal}
\bibliography{refs}

\begin{thebibliography}{}
\expandafter\ifx\csname natexlab\endcsname\relax\def\natexlab#1{#1}\fi
\providecommand{\url}[1]{\href{#1}{#1}}

\bibitem[{{Amari} {et~al.}(2015){Amari}, {Luciani}, \& {Aly}}]{amari2015}
{Amari}, T., {Luciani}, J.-F., \& {Aly}, J.-J. 2015, \nat, 522, 188

\bibitem[{{Anfinogentov} {et~al.}(2013){Anfinogentov}, {Nakariakov},
  {Mathioudakis}, {Van Doorsselaere}, \& {Kowalski}}]{anfinogentov2013}
{Anfinogentov}, S., {Nakariakov}, V.~M., {Mathioudakis}, M., {Van
  Doorsselaere}, T., \& {Kowalski}, A.~F. 2013, \apj, 773, 156

\bibitem[{{Balona}(2012)}]{balona2012}
{Balona}, L.~A. 2012, \mnras, 423, 3420

\bibitem[{{Balona}(2013)}]{balona2013}
---. 2013, \mnras, 431, 2240

\bibitem[{{Balona}(2015)}]{balona2015b}
---. 2015, \mnras, 447, 2714

\bibitem[{{Balona} {et~al.}(2015){Balona}, {Broomhall}, {Kosovichev},
  {Nakariakov}, {Pugh}, \& {Van Doorsselaere}}]{balona2015}
{Balona}, L.~A., {Broomhall}, A.-M., {Kosovichev}, A., {et~al.} 2015, \mnras,
  450, 956

\bibitem[{{Candelaresi} {et~al.}(2014){Candelaresi}, {Hillier}, {Maehara},
  {Brandenburg}, \& {Shibata}}]{candelaresi2014}
{Candelaresi}, S., {Hillier}, A., {Maehara}, H., {Brandenburg}, A., \&
  {Shibata}, K. 2014, \apj, 792, 67

\bibitem[{{Chang} {et~al.}(2017){Chang}, {Song}, {Luo}, {Huang}, {Ip}, {Fu},
  {Zhang}, {Hou}, {Cao}, \& {Wang}}]{chang2017}
{Chang}, H.-Y., {Song}, Y.-H., {Luo}, A.-L., {et~al.} 2017, \apj, 834, 92

\bibitem[{{Cho} {et~al.}(2016){Cho}, {Cho}, {Nakariakov}, {Kim}, \&
  {Kumar}}]{cho2016}
{Cho}, I.-H., {Cho}, K.-S., {Nakariakov}, V.~M., {Kim}, S., \& {Kumar}, P.
  2016, \apj, 830, 110

\bibitem[{{Davenport}(2016)}]{davenport2016}
{Davenport}, J.~R.~A. 2016, \apj, 829, 23

\bibitem[{{Degroote} {et~al.}(2009){Degroote}, {Briquet}, {Catala},
  {Uytterhoeven}, {Lefever}, {Morel}, {Aerts}, {Carrier}, {Auvergne}, {Baglin},
  \& {Michel}}]{degroote2009}
{Degroote}, P., {Briquet}, M., {Catala}, C., {et~al.} 2009, \aap, 506, 111

\bibitem[{{Gaulme} {et~al.}(2014){Gaulme}, {Jackiewicz}, {Appourchaux}, \&
  {Mosser}}]{gaulme2014}
{Gaulme}, P., {Jackiewicz}, J., {Appourchaux}, T., \& {Mosser}, B. 2014, \apj,
  785, 5

\bibitem[{{Harper} {et~al.}(2013){Harper}, {O'Riain}, \& {Ayres}}]{harper2013}
{Harper}, G.~M., {O'Riain}, N., \& {Ayres}, T.~R. 2013, \mnras, 428, 2064

\bibitem[{{Hawley} {et~al.}(2014){Hawley}, {Davenport}, {Kowalski},
  {Wisniewski}, {Hebb}, {Deitrick}, \& {Hilton}}]{hawley2014}
{Hawley}, S.~L., {Davenport}, J.~R.~A., {Kowalski}, A.~F., {et~al.} 2014, \apj,
  797, 121

\bibitem[{Johnson \& Kotz(1970)}]{johnsonkotz}
Johnson, N.~L., \& Kotz, S. 1970, Continuous univariate distributions, Wiley
  series in probability and mathematical statistics (New York (N.Y.): Wiley)

\bibitem[{{Karoff} {et~al.}(2016){Karoff}, {Knudsen}, {De Cat}, {Bonanno},
  {Fogtmann-Schulz}, {Fu}, {Frasca}, {Inceoglu}, {Olsen}, {Zhang}, {Hou},
  {Wang}, {Shi}, \& {Zhang}}]{karoff2016}
{Karoff}, C., {Knudsen}, M.~F., {De Cat}, P., {et~al.} 2016, Nature
  Communications, 7, 11058

\bibitem[{{Karovska} {et~al.}(2005){Karovska}, {Schlegel}, {Hack}, {Raymond},
  \& {Wood}}]{karovska2005}
{Karovska}, M., {Schlegel}, E., {Hack}, W., {Raymond}, J.~C., \& {Wood}, B.~E.
  2005, \apjl, 623, L137

\bibitem[{{Konstantinova-Antova} {et~al.}(2008){Konstantinova-Antova},
  {Auri{\`e}re}, {Iliev}, {Cabanac}, {Donati}, {Mouillet}, \&
  {Petit}}]{konstantinova2008}
{Konstantinova-Antova}, R., {Auri{\`e}re}, M., {Iliev}, I.~K., {et~al.} 2008,
  \aap, 480, 475

\bibitem[{{Kowalski} {et~al.}(2010){Kowalski}, {Hawley}, {Holtzman},
  {Wisniewski}, \& {Hilton}}]{kowalski2010}
{Kowalski}, A.~F., {Hawley}, S.~L., {Holtzman}, J.~A., {Wisniewski}, J.~P., \&
  {Hilton}, E.~J. 2010, \apjl, 714, L98

\bibitem[{{Kowalski} {et~al.}(2013){Kowalski}, {Hawley}, {Wisniewski}, {Osten},
  {Hilton}, {Holtzman}, {Schmidt}, \& {Davenport}}]{kowalski2013}
{Kowalski}, A.~F., {Hawley}, S.~L., {Wisniewski}, J.~P., {et~al.} 2013, \apjs,
  207, 15

\bibitem[{{Lurie} {et~al.}(2015){Lurie}, {Davenport}, {Hawley}, {Wilkinson},
  {Wisniewski}, {Kowalski}, \& {Hebb}}]{lurie2015}
{Lurie}, J.~C., {Davenport}, J.~R.~A., {Hawley}, S.~L., {et~al.} 2015, \apj,
  800, 95

\bibitem[{{Maehara} {et~al.}(2015){Maehara}, {Shibayama}, {Notsu}, {Notsu},
  {Honda}, {Nogami}, \& {Shibata}}]{maehara2015}
{Maehara}, H., {Shibayama}, T., {Notsu}, Y., {et~al.} 2015, Earth, Planets, and
  Space, 67, 59

\bibitem[{{Maehara} {et~al.}(2012){Maehara}, {Shibayama}, {Notsu}, {Notsu},
  {Nagao}, {Kusaba}, {Satoshi}, {Satoshi}, {Nogami}, \&
  {Shibata}}]{maehara2012}
{Maehara}, M., {Shibayama}, T., {Notsu}, S., {et~al.} 2012, \nat

\bibitem[{{Mathioudakis} {et~al.}(2003){Mathioudakis}, {Seiradakis},
  {Williams}, {Avgoloupis}, {Bloomfield}, \& {McAteer}}]{mathioudakis2003}
{Mathioudakis}, M., {Seiradakis}, J.~H., {Williams}, D.~R., {et~al.} 2003,
  \aap, 403, 1101

\bibitem[{{McQuillan} {et~al.}(2014{\natexlab{a}}){McQuillan}, {Mazeh}, \&
  {Aigrain}}]{mcquillan2014}
{McQuillan}, A., {Mazeh}, T., \& {Aigrain}, S. 2014{\natexlab{a}}, \apjs, 211,
  24

\bibitem[{{McQuillan} {et~al.}(2014{\natexlab{b}}){McQuillan}, {Mazeh}, \&
  {Aigrain}}]{mcquillan2014catalog}
---. 2014{\natexlab{b}}, VizieR Online Data Catalog, 221

\bibitem[{{Mitra-Kraev} {et~al.}(2005){Mitra-Kraev}, {Harra}, {Williams}, \&
  {Kraev}}]{mitra-kraev2005}
{Mitra-Kraev}, U., {Harra}, L.~K., {Williams}, D.~R., \& {Kraev}, E. 2005,
  \aap, 436, 1041

\bibitem[{{Notsu} {et~al.}(2015){Notsu}, {Honda}, {Maehara}, {Notsu},
  {Shibayama}, {Nogami}, \& {Shibata}}]{notsu2015}
{Notsu}, Y., {Honda}, S., {Maehara}, H., {et~al.} 2015, \pasj, 67, 33

\bibitem[{{Pedersen} {et~al.}(2017){Pedersen}, {Antoci}, {Korhonen}, {White},
  {Jessen-Hansen}, {Lehtinen}, {Nikbakhsh}, \& {Viuho}}]{pedersen2017}
{Pedersen}, M.~G., {Antoci}, V., {Korhonen}, H., {et~al.} 2017, \mnras, 466,
  3060

\bibitem[{{Pitkin} {et~al.}(2014){Pitkin}, {Williams}, {Fletcher}, \&
  {Grant}}]{pitkin2014}
{Pitkin}, M., {Williams}, D., {Fletcher}, L., \& {Grant}, S.~D.~T. 2014,
  \mnras, 445, 2268

\bibitem[{{Pugh} {et~al.}(2016){Pugh}, {Armstrong}, {Nakariakov}, \&
  {Broomhall}}]{pugh2016}
{Pugh}, C.~E., {Armstrong}, D.~J., {Nakariakov}, V.~M., \& {Broomhall}, A.-M.
  2016, \mnras, 459, 3659

\bibitem[{{Pugh} {et~al.}(2015){Pugh}, {Nakariakov}, \& {Broomhall}}]{pugh2015}
{Pugh}, C.~E., {Nakariakov}, V.~M., \& {Broomhall}, A.-M. 2015, \apjl, 813, L5

\bibitem[{{Ramsay} \& {Doyle}(2014)}]{ramsay2014}
{Ramsay}, G., \& {Doyle}, J.~G. 2014, \mnras, 442, 2926

\bibitem[{{Ramsay} {et~al.}(2013){Ramsay}, {Doyle}, {Hakala}, {Garcia-Alvarez},
  {Brooks}, {Barclay}, \& {Still}}]{ramsay2013}
{Ramsay}, G., {Doyle}, J.~G., {Hakala}, P., {et~al.} 2013, \mnras, 434, 2451

\bibitem[{{Shakhovskaia}(1989)}]{shakhovskaia1989}
{Shakhovskaia}, N.~I. 1989, \solphys, 121, 375

\bibitem[{{Shibata} {et~al.}(2013){Shibata}, {Isobe}, {Hillier}, {Choudhuri},
  {Maehara}, {Ishii}, {Shibayama}, {Notsu}, {Notsu}, {Nagao}, {Honda}, \&
  {Nogami}}]{shibata2013}
{Shibata}, K., {Isobe}, H., {Hillier}, A., {et~al.} 2013, \pasj, 65,
  arXiv:1212.1361

\bibitem[{{Shibayama} {et~al.}(2013){Shibayama}, {Maehara}, {Notsu}, {Notsu},
  {Nagao}, {Honda}, {Ishii}, {Nogami}, \& {Shibata}}]{shibayama2013}
{Shibayama}, T., {Maehara}, H., {Notsu}, S., {et~al.} 2013, \apj S., 209, 5

\bibitem[{{Simon} \& {Drake}(1989)}]{simon1989}
{Simon}, T., \& {Drake}, S.~A. 1989, \apj, 346, 303

\bibitem[{{Srivastava} {et~al.}(2013){Srivastava}, {Lalitha}, \&
  {Pandey}}]{srivastava2013}
{Srivastava}, A.~K., {Lalitha}, S., \& {Pandey}, J.~C. 2013, \apjl, 778, L28

\bibitem[{{Sun} {et~al.}(2015){Sun}, {Cheng}, {Ding}, {Guo}, {Priest},
  {Parnell}, {Edwards}, {Zhang}, {Chen}, \& {Fang}}]{sun2015}
{Sun}, J.~Q., {Cheng}, X., {Ding}, M.~D., {et~al.} 2015, Nature Communications,
  6, 7598

\bibitem[{{Van Doorsselaere} {et~al.}(2016){Van Doorsselaere}, {Kupriyanova},
  \& {Yuan}}]{vd2016}
{Van Doorsselaere}, T., {Kupriyanova}, E.~G., \& {Yuan}, D. 2016, \solphys,
  291, 3143

\bibitem[{{Vlemmings} {et~al.}(2015){Vlemmings}, {Ramstedt}, {O'Gorman},
  {Humphreys}, {Wittkowski}, {Baudry}, \& {Karovska}}]{vlemmings2015}
{Vlemmings}, W.~H.~T., {Ramstedt}, S., {O'Gorman}, E., {et~al.} 2015, \aap,
  577, L4

\bibitem[{{Walkowicz} {et~al.}(2011){Walkowicz}, {Basri}, {Batalha},
  {Gilliland}, {Jenkins}, {Borucki}, {Koch}, {Caldwell}, {Dupree}, {Latham},
  {Meibom}, {Howell}, {Brown}, \& {Bryson}}]{walkowicz2011}
{Walkowicz}, L.~M., {Basri}, G., {Batalha}, N., {et~al.} 2011, \aj, 141, 50

\bibitem[{{Welsh} {et~al.}(2006){Welsh}, {Wheatley}, {Browne}, {Siegmund},
  {Doyle}, {O'Shea}, {Antonova}, {Forster}, {Seibert}, {Morrissey}, \&
  {Taroyan}}]{welsh2006}
{Welsh}, B.~Y., {Wheatley}, J., {Browne}, S.~E., {et~al.} 2006, \aap, 458, 921

\bibitem[{{Wright} {et~al.}(2011){Wright}, {Drake}, {Mamajek}, \&
  {Henry}}]{wright2011}
{Wright}, N.~J., {Drake}, J.~J., {Mamajek}, E.~E., \& {Henry}, G.~W. 2011,
  \apj, 743, 48

\end{thebibliography}

\end{document}